\RequirePackage{fix-cm}
\documentclass[twocolumn,epjc3]{svjour3}
\smartqed  
\RequirePackage{graphicx}
\usepackage{color}
\usepackage{amssymb}
\usepackage{amsmath}
\usepackage{xspace}
\usepackage{abbreviations}
\journalname{Eur. Phys. J. C}
\begin{document}

\title{Lessons for SUSY from the LHC after the first run 
}


\author{I.-A. Melzer-Pellmann\thanksref{e1,addr1}
        \and
        P. Pralavorio\thanksref{e2,addr2}
}

\thankstext{e1}{e-mail: isabell.melzer-pellmann@cern.ch}
\thankstext{e2}{e-mail: pascal.pralavorio@cern.ch}


\institute{DESY, Notkestr. 85, 22607 Hamburg (Germany)\label{addr1}
           \and
           CPPM, Univ. Aix-Marseille and CNRS/IN2P3, 163 avenue de Luminy, case 902, 13288 Marseille cedex 09 (France) \label{addr2}
}

\date{}

\maketitle

\begin{abstract}
A review of direct searches for new particles predicted by Supersymmetry after the first run of the LHC is proposed. This review is based 
on the results provided by the ATLAS and CMS experiments.
\end{abstract}

\tableofcontents
\section{Introduction}
\label{sec:intro}

Ernest Rutherford once said at the beginning of the XX$^{\rm th}$ century ``Theorists play games with their 
symbols while we discover truths about the Universe''. This was before the birth of the Bohr atom and 
quantum mechanics. At the beginning of the XXI$^{\rm st}$ century this sentence can probably be reverted, 
as the theory predictions are so successful that experimental discoveries of the last 40 years fit 
perfectly in the Standard Model (SM) theory framework. In this respect, the systematic exploration of the 
electroweak (EW) scale at the Large Hadron Collider (LHC)~\cite{LHC} could have represented an even greater 
confirmation of the theory predictions since most Beyond Standard Model (BSM) theories, and in particular 
low-energy Supersymmetry (SUSY) expect new particles with masses close to the EW scale. As it will be discussed 
extensively in this article, this is clearly not the case. The direct searches conducted at the first LHC run by 
the general purpose experiments ATLAS~\cite{ATLASDetector} and CMS~\cite{CMSDetector} did not reveal the presence 
of any new particle beyond that of the Standard Model. In contrary, the Standard Model is now fully established 
by the discovery of the Higgs boson~\cite{HIGGSDISC1,HIGGSDISC2}. 

The article is based on currently published results obtained at a center-of-mass energy of $\sqrt{s} = 8\TeV$ for an integrated 
luminosity of 20\fbinv of LHC data. In case they are not available, previous results, obtained with $\sqrt{s}=7\TeV$ 
and 5\fbinv of LHC data, are discussed. The most important analyses are based on 8\TeV LHC data, either from 
ATLAS or CMS (or both). Therefore, even if the fully final result from Run~1 is not yet published for all 
individual search channels, lessons for SUSY discussed here are not expected to change.

This article is organized as follows: First, a brief recap of the SUSY framework used 
for the LHC searches is proposed in Sect.~\ref{sec:SUSY_LHC}. Then the experimental challenges 
faced by the ATLAS and CMS experiments in terms of object reconstruction and background modeling 
are explained in Sect.~\ref{sec:Exp}. The limit setting procedure is also briefly summarized in this section. 
The discussion of the results is split in three different sections, representing the main avenues of the SUSY searches 
at LHC: gluino and squarks of first/second generation at the energy frontier in Sect.~\ref{sec:StrongProd}, 
third-generation squarks in Sect.~\ref{sec:3rdGene} and electroweak SUSY in Sect.~\ref{sec:SUSY_EWK}. All 
assume $R$-parity conservation, and prompt decay of the SUSY particles. Sect.~\ref{sec:RPV_LLP} is devoted 
to escape routes beyond $R$-parity conservation and prompt decays, as well as more exotic SUSY scenarios. Prospects 
with the coming runs of LHC and conclusions are discussed in Sect.~\ref{sec:Prospects} and~\ref{sec:Conclu}, 
respectively. 

\section{SUSY framework for the search at LHC}
\label{sec:SUSY_LHC}

This section provides the reader with the minimum vocabulary and knowledge needed to 
understand the experimental results presented later in this article.

\begin{figure}[htbp]
\begin{center}
\includegraphics[width=\linewidth]{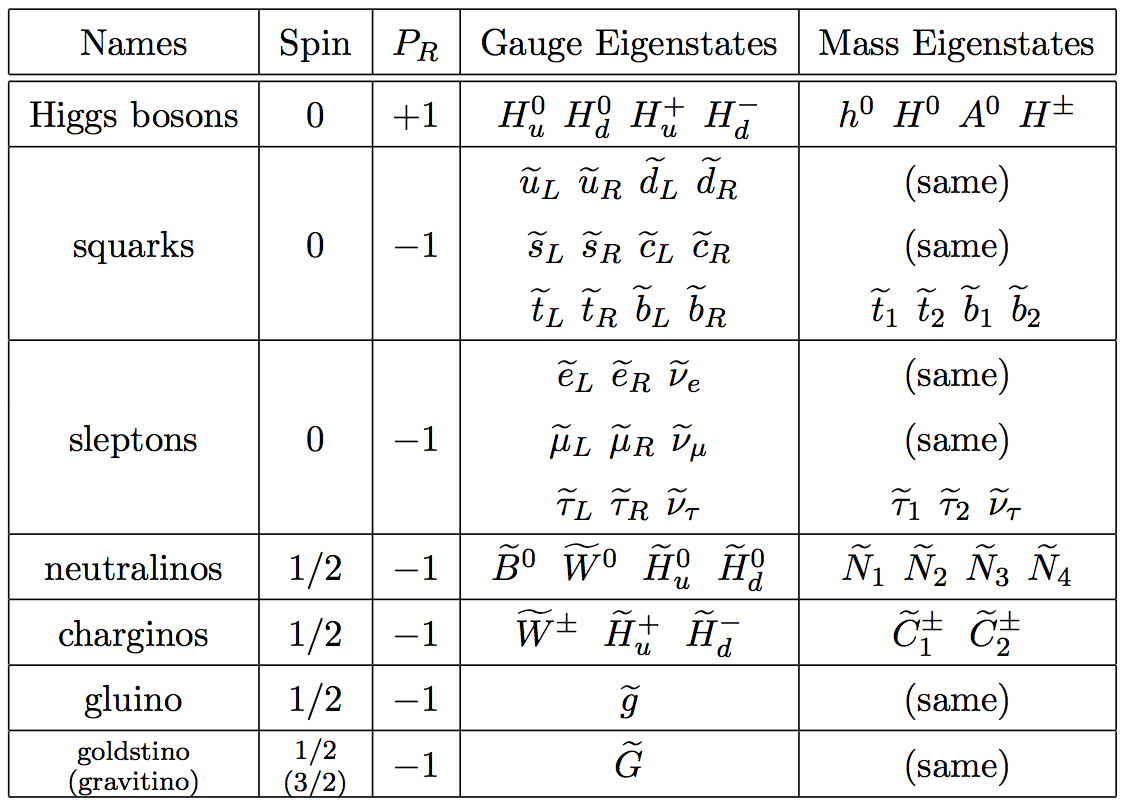}
\end{center}
\caption{SUSY particles in MSSM~\protect\cite{SUSY_Primer}.}
\label{fig:SUSYTheo1}
\end{figure}

As discussed elsewhere in this review, SUSY~\cite{SUSY_Ref1,SUSY_Ref2} can be realized in many different ways. 
Even if the LHC cannot explore the full SUSY phase space, it can probe extensively 
the low-energy (or weak-scale) realization of N=1 SUSY, called the Minimal Supersymmetric 
Standard Model (MSSM)~\cite{SUSY_Primer}. This model predicts new particles, called sparticles, that are superpartners  
of each SM particle in the chiral multiplets, as shown in Fig.~\ref{fig:SUSYTheo1}. A new quantum number, $R$-parity is created and defined 
as $P_R = (-1)^{2s+3B+L}$, where $s$ is the spin, $B$ the baryon number and $L$ the lepton number. It is 
negative/positive for SUSY/SM particles. Therefore, the sparticles have the same quantum numbers as 
their SM partners, except for the spin which differs by half a unit, and $R$-parity. The spectrum 
is characterized by 25 elementary scalars and 10 elementary fermions without counting the SM particles. 

\begin{figure*}
\begin{center}
\includegraphics[width=\textwidth, height=5cm]{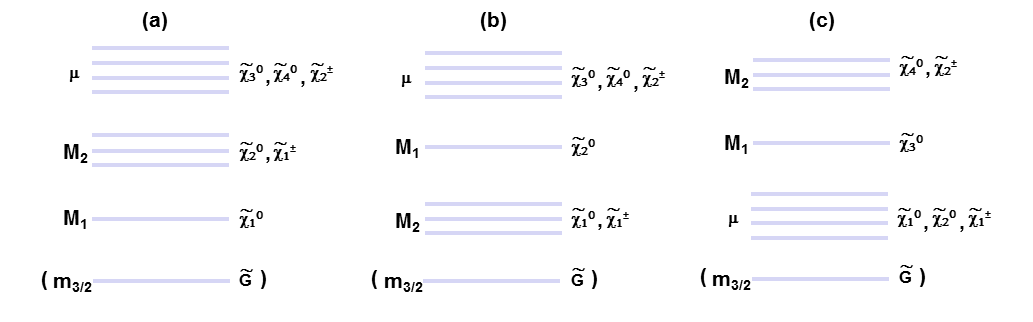}
\end{center}
\caption{Three possible EW SUSY mass spectrum depending of the relative values of $M_1$, $M_2$ and $\mu$ parameters.}
\label{fig:SUSY_EWKinos}
\end{figure*}

For self-containment, we just recap the names and main characteristics of the new particles. 
To generate the masses of the up- and down-type fermions, the SM Higgs sector is extended 
by adding another $SU(2)_L$ complex doublet. Each doublet has a vacuum expectation value (vev) labelled $v_u$ and $v_d$, 
constrained by the SM Higgs vev, $v$ ($v=\sqrt{v_u^2+v_d^2}$). Their ratio are traditionally written $\mathrm{\tan}\beta=v_u/v_d$.
As a result, eight mass eigenstates exist after electroweak (EW) symmetry breaking: three neutral Higgs bosons (\HO, the one with the lightest 
mass, H$^0$ and A$^0$), two charged Higgs bosons (H$^{\pm}$) and three Goldstone bosons (G$^0$, G$^{\pm}$), `eaten' to 
give masses to the \Z and \Wpm bosons. The Higgs boson discovered by ATLAS and CMS is assumed to be the lightest neutral Higgs of the MSSM 
(h$^0$) since it possesses similar properties as the SM one when $m_{\text{A}^0}^2 \gg m_{\Z}^2$ and $\tan \beta>1$ (also called the 
decoupling limit)~\footnote{It could still be H$^0$, implying that h$^0$ is lighter and still to be discovered but this is presently disfavored 
by the data~\cite{SUSY_Higgses}.}.The squarks (\squark) and sleptons (\sLep) are the spin-0 partners of the SM fermions. Similarly, wino (\sWO, \sWpm), 
bino (\sBino) and higgsinos (\sHinouO, \sHinodO, \sHinoup, \sHinodm) are the spin-1/2 superpartners of the electroweak bosons. To complete the list, 
colored gluinos (\gluino) and the gravitino (\gravitino) are the partners of the gluon and graviton. With this setup, the number of fermions and 
bosons is equalized and the Lightest SUSY Particle (LSP) is stable if $R$-parity is conserved. 
Note also that left- and right-handed fermions have two different SUSY partners $\sFer_{L,R}$ that can mix in $\sFer_{1,2}$ 
provided the SM partner is heavy, like in the third generation. Similarly, the wino, bino and higgsino, governed by the gauge eigenstate 
mass terms $M_1$, $M_2$ and $\mu$, mix to give four neutralinos ($\ninozero_{1,2,3,4}$), and four charginos ($\chipm_{1,2}$). 
To simplify the discussion of results in this sector, this article often considers one of the three typical scenario shown in Fig.~\ref{fig:SUSY_EWKinos}. 
Each correspond to a different $\ninoone$ flavor : (a) bino-like, (b)
wino-like or co-NLSP~\cite{GMSB_coNLSP} and (c) higgsino-like. The last one is 
favored by naturalness arguments, as discussed in the next paragraph. 

\begin{figure}[htbp]
\begin{center}
\includegraphics[width=\linewidth]{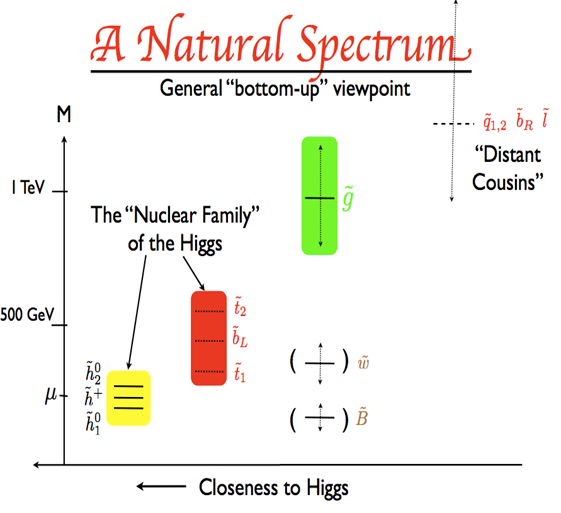}
\end{center}
\caption{Natural SUSY particle mass spectra giving less than 10\% tuning~\protect\cite{SUSY_NatSpectrum}.}
\label{fig:SUSYTheo2}
\end{figure}

When choosing the parameters such that the hierarchy or naturalness problem is solved within the MSSM~\cite{Naturalness_Ref}, stringent 
constraints appear on the masses of the new predicted particles, 
especially the ones which are most closely related to the Higgs, as can be seen in Fig.~\ref{fig:SUSYTheo2}. 
Interestingly, there was no experimental sensitivity to this spectrum before the startup of the LHC. Moreover, by design, the
LHC is perfectly suited to access this particular region of phase space, already with its first run at 
$\sqrt{s}=7-8$\TeV. Searches for hints in this particular spectrum generally shape the analysis strategy 
at the LHC. 

In the MSSM framework, three main theoretical unknowns influence the search direction: the LSP nature, the compression (or not) of the SUSY spectra and the status of
$R$-parity. For the first one, experimental constraints restrict the LSP to be the lightest neutralino (\ninoone) or the almost massless gravitino. 
In the later case, final states are increased compared to the former. The reason is that the Next-to-Lightest SUSY Particle (NLSP) which can be 
any of the SUSY particles (squark, gluino, slepton, chargino or neutralino) will decay to the gravitino and the SM partner of 
the NLSP~\cite{GMSB_FinalStates}~\footnote{The most `natural' situation is that the NLSP is the $\ninoone$ (Fig.~\ref{fig:SUSYTheo2}), whose decay will depend on 
its flavor (bino, wino or higgsino-like of Fig.~\ref{fig:SUSY_EWKinos}) and results in $\ninoone \rightarrow \gamma \gravitino$, 
$\ninoone \rightarrow \gamma/\Z \gravitino$ or $\ninoone \rightarrow \Z/\HO \gravitino$ decays, respectively. The relative proportion 
of $\gamma$, \Z and \HO depends also on others parameters $\theta_{\W}$ or $\mathrm{\tan}\beta$.}. The second MSSM theory unknown is 
the difference ($\Delta M$) between the mass of the highest sparticle produced at the LHC ($M_{\rm SUSY}$) and the 
LSP ($M_{\rm LSP}$), resulting in compressed or open spectra, i.e. hard or soft objects in the final state. The third MSSM theory unknown is the status of $R$-parity. 
In a plain vanilla MSSM scenario $R$-parity is conserved (RPC), but it could well be violated (RPV) or even a continuous symmetry. 
It is important to mention that in the huge MSSM phase space long-lived particles decaying within the detector or even outside often exist. 
This situation could arise from low mass difference between two sparticles in the spectrum, very weak coupling to the LSP, 
very small $R$-parity Yukawa couplings, etc.

\begin{figure}[htbp]
\begin{center}
\includegraphics[height=10cm]{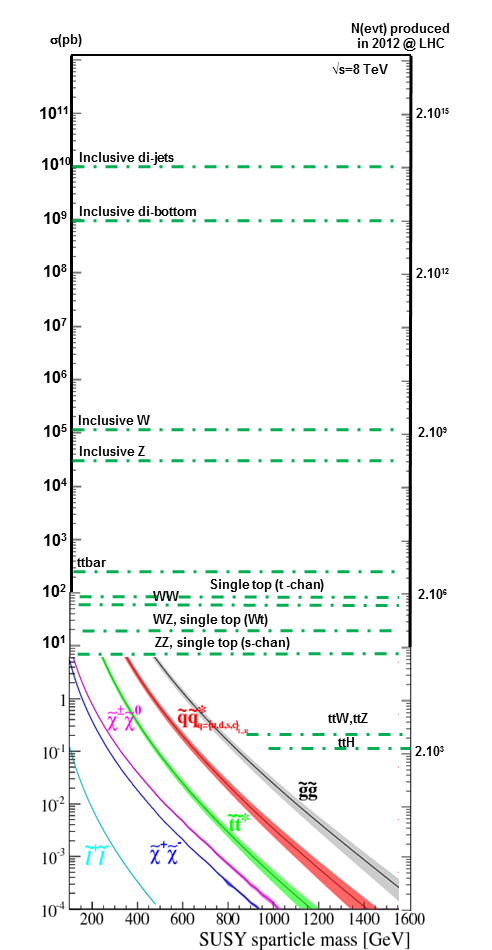}
\end{center}
\caption{Cross sections of several SUSY production channels~\protect\cite{ATLAS_CMS_xs}, superimposed with Standard Model process at 
$\sqrt{s}=8$ TeV. The right-handed axis indicates the number of events for 20\fbinv.}
\label{fig:SUSYxs}
\end{figure}

\begin{figure}[htbp]
\begin{center}
\includegraphics[height=6cm]{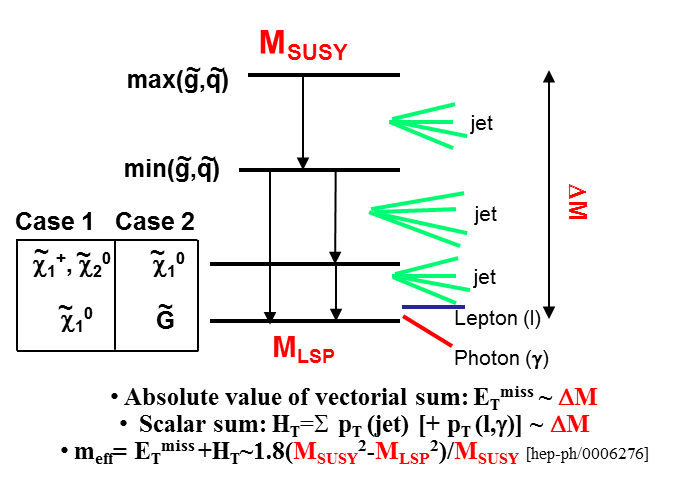}
\end{center}
\caption{Typical decay of a colored SUSY particle at LHC. The two cases shown at the bottom of the SUSY spectrum correspond 
to the two considered LSP types.}
\label{fig:SUSY_LHC}
\end{figure}

\begin{figure*}[htbp]
\begin{center}
\includegraphics[width=\linewidth, height=4cm]{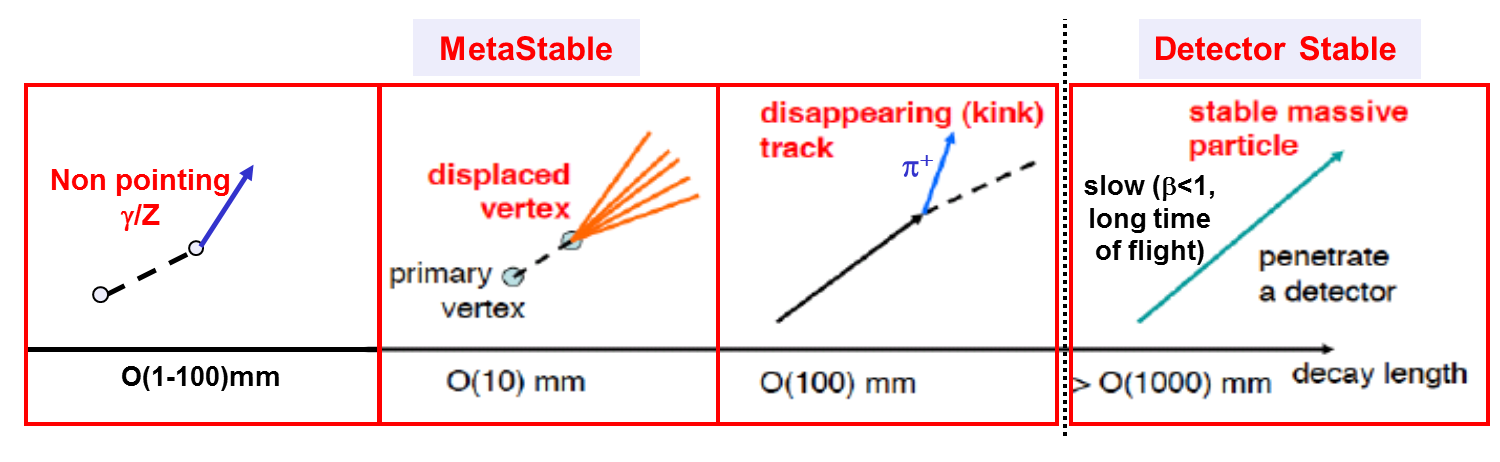}
\end{center}
\caption{Possible signatures from non-prompt sparticle decay.}
\label{fig:SUSY_LLP}
\end{figure*}

Typical SUSY cross sections of pair-produced sparticles at the LHC are given in Fig.~\ref{fig:SUSYxs} -- 
for two different sparticles mass 
degeneracy is assumed~\footnote{Following an ATLAS-CMS agreement~\cite{ATLAS_CMS_xs}, all SUSY cross sections are calculated in the
MSSM at NLO precision in the strong coupling constant, including the resummation
of soft gluon emission at next-to-leading-logarithmic (NLO+NLL) accuracy, using
PROSPINO and NLL-fast~\cite{Beenakker:1996ch,Kulesza:2008jb,Kulesza:2009kq,Beenakker:2009ha,Beenakker:2011fu}.}. 
Since each SM particle and its superpartner belong to the same multiplet, the sparticle decay generally 
involves the SM partner and the LSP. However due to the high number of new particles many different decays are 
possible depending on the sparticle mass spectrum, generating long decay chains. Plain vanilla MSSM searches 
are therefore characterized by pair-produced particles, two LSPs escaping the detection and long decay chains 
involving jets and/or leptons and photons as depicted in Fig.~\ref{fig:SUSY_LHC}. Note that for the gravitino LSP scenario, 
the mass difference with the NLSP is always sizeable in the natural spectrum, O(100) GeV, because the gravitino is 
almost massless. If the masses of all colored particles are too high 
to be produced at LHC, the production may be dominated by chargino-neutralino pair production, resulting in less 
complicated final states, generally containing several leptons. Finally, non-prompt sparticle decays can generate 
striking signatures as displayed in Fig.~\ref{fig:SUSY_LLP}. 

\section{Experimental challenges for SUSY searches at LHC}
\label{sec:Exp}

This section is mainly addressed to non-experts in LHC analyses and
analysers or theorists who want to understand better the 
many experimental facets of a SUSY analysis at LHC. 

Discovering SUSY at the LHC is an extremely challenging task, even within the restricted framework of the MSSM. 
First, every corner of the parameter space needs to be covered, including all possible decay channels which 
provide a high number of final states with different mixtures of reconstructed objects (photon, electron, muon, tau, 
jets, \botq-jets, missing transverse energy). Second, due to the presence of many scalars and weakly interacting particles, 
cross sections are generally extremely tiny with respect to the SM background (cf.~Fig~\ref{fig:SUSYxs}). In the plain vanilla 
MSSM scenario, the few signal events are generally located in the tails of the kinematic distributions, 
requiring challenging trigger, powerful discriminating variables and accurate background modeling in a complicated 
region of the phase space. In other SUSY scenarios where $R$-parity is violated and/or non-prompt decays are possible, 
the experimental challenge generally shifts to taking the best performance of each sub-detector to improve secondary 
vertex reconstruction, timing resolution, jet substructure reconstruction, lepton coverage, etc.
Therefore, SUSY searches provide an excellent way to push the detector and analyser capabilities to their best.

This section is organized as follows. Experimental matters, i.e. LHC data, trigger and detector/object 
performance relevant for SUSY searches, are treated in Sect.~\ref{sec:LHC_Det}. Commonly used discriminating 
variables for the design of the signal regions are then discussed in Sect.~\ref{sec:discri} and methods 
to estimate the remaining background in these signal regions are described in Sect.~\ref{sec:bckg}. 
Finally, the limit setting tools and SUSY models used for interpretations are briefly reviewed in Sect.~\ref{sec:limit} 
and~\ref{sec:interp}, respectively.

\subsection{LHC data and detector performance}
\label{sec:LHC_Det}

After a brief reminder of the main characteristics of the LHC data (section~\ref{sec:LHC}), ATLAS and CMS detectors (section~\ref{sec:ATLAS-CMS}), the object 
and detector performance relevant to SUSY searches are discussed (section~\ref{sec:Obj} and~\ref{sec:Det}). 

\subsubsection{LHC data}
\label{sec:LHC}

\begin{figure*}
\begin{center}
\includegraphics[width=\textwidth, height=5cm]{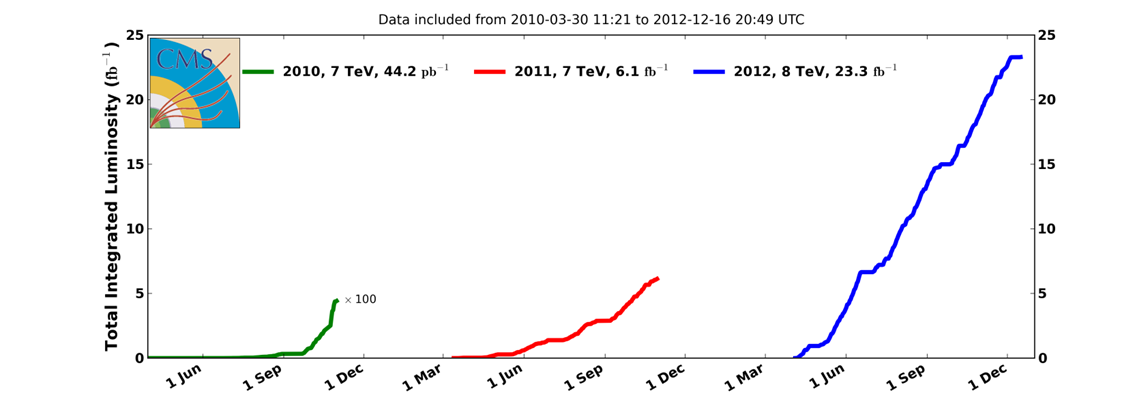}
\end{center}
\caption{LHC luminosity recorded by the CMS experiment}
\label{fig:LHCPerf1}
\end{figure*}

\begin{figure*}
\begin{center}
\includegraphics[width=\textwidth, height=5cm]{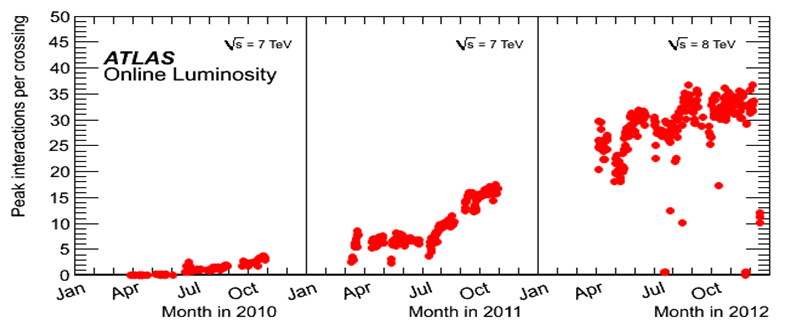}
\end{center}
\caption{corresponding number of pile-up events as a function of time.}
\label{fig:LHCPerf2}
\end{figure*}

The LHC is a particle collider at CERN, `probably the largest and the most complex machine ever constructed by 
humans'~\cite{LHC_Nobel}. It is housed in a 27\km long tunnel $\sim$100\m underground and is ultimately designed to collide 
proton beams at a center-of-mass energy of up to $\sqrt{s}=14$\TeV at a rate of 40\MHz. The first proton-proton run of the 
LHC (Run~1) lasted from March 2010 to December 2012 with $\sqrt{s}=7$\TeV and 8\TeV, and collisions every 50\ns. It was extremely 
successful and provided more than 20\fbinv in 3 years, as shown in Fig.~\ref{fig:LHCPerf1}. Because of the increase of the 
proton density per bunch and the tuning of the beam optics, the number of interactions per beam crossing (pile-up) increased 
regularly during Run~1 to reach more than 30 at the end of 2012; see Fig.~\ref{fig:LHCPerf2}. This increases the complexity of the event 
reconstruction, as discussed later. For practical considerations, unless mentioned otherwise, searches presented 
in this article make use of the the full $\sqrt{s}=8$\TeV dataset.

\subsubsection{ATLAS and CMS detectors}
\label{sec:ATLAS-CMS}

Four large detectors have been built at the collision points. Among these, ATLAS 
and CMS are the two general-purpose experiments. Because of the huge complexity of the detectors that can cope 
with very high collision rate and high pile-up conditions, world-wide collaborations of few thousands 
of physicists and engineers were set-up, giving these projects a flavor of a modern cathedral, dedicated to science. 

The interesting particles are produced over the full solid angle down to small polar angles ($\theta$) with respect 
to the incoming beams (a fraction of a degree corresponding to pseudo\-rapidities of $|\eta|$ up to 5, where 
$\eta=-\ln[\tan(\theta/2)]$) and in the full azimuth $0 \leq \phi \leq 2\pi$. The transverse plane plays a special role for LHC analyses:
the vectorial sum of all particle momenta produced by the collision is null and it is the bending plane in the central part. 
Therefore, selections generally rely on transverse momenta, labelled \PT.

The two detectors are based on two different technologies for the central magnets that are used to bend the charged particle trajectories: 
CMS uses a 4-T superconducting solenoid magnet of 3\m radius, fully containing the trackers and the calorimeters, 
while ATLAS chose a smaller central solenoid (2\T and 1.2\m radius), complemented by outer toroids. 
These choices influence the design of all detector technologies~\cite{CMSvsATLAS}. 

Inner tracking systems measure the momentum of charged particles, which are bent by the magnetic field. To achieve this, 
ATLAS and CMS have designed tracker systems providing a similar geometrical coverage (over $|\eta|<2.4-2.5$). 
They are based on the same silicon detector technology near the interaction vertex, i.e. 
below 50\cm, with a silicon pixel and strip tracker, providing around 10 precision points per track. 
However, they differ considerably at larger radii: 
ATLAS uses straw-tube detectors (TRT), allowing 35 extra measurements per track in the bending plane for $|\eta|<2.0$, 
with drift-time information for momentum measurements and pattern recognition, while CMS extends the silicon-strip technology
up to a radius of $\sim$110\cm. Stand-alone tracker performance is generally better for CMS because of the higher magnetic field.

The electromagnetic (EM) calorimeter is the key system for measuring the kinematics of electrons and photons. CMS chose 
$\sim$75000 scintillating PbWO$_4$ crystals with an excellent energy resolution but extremely low light yield, 
while ATLAS built a granular lead/liquid argon sampling calo\-ri\-meter with $\sim$200k channels. This last technology is robust and well known, 
with poorer energy resolution at low energy, but comparable to CMS in the 0.1-1\TeV energy range. 

Key parameters for the hadronic calorimeters are the coverage up to $|\eta|<5$ for both ATLAS and CMS, the depth in interaction 
length ($\lambda \sim$10) and the sampling fraction, 3 times better in ATLAS than in CMS~\cite{CMSvsATLAS}. 

Finally, the muon spectrometers are also quite different: in ATLAS they can provide an independent and high-accuracy 
measurement of muons over $|\eta|<2.7$ coverage, whereas CMS relies on a combined measurement of muon chambers and inner tracker up to 
$|\eta|<2.4$. 

The ability of ATLAS and CMS to use more than 90\% of the high-quality data 
delivered by the LHC for physics analyses demonstrates the excellent functioning of both experiments.

\subsubsection{Object reconstruction performance relevant to SUSY searches}
\label{sec:Obj}

Because of different detector concepts, the final-state reconstruction differs quite a lot between ATLAS and CMS. 

In ATLAS, electrons, photons, and jets are seeded by calorimeter clusters. Electrons and photons are eventually 
combined with the tracker information. In CMS, all final-state particles are reconstructed with the particle-flow method~\cite{PF},
generally seeded in the inner detector and further combined with the information from all sub-detectors.

Jets are reconstructed using the anti-kt jet clustering algorithm~\cite{AntiKt1,AntiKt2} with a distance parameter of 
0.4 and 0.5 for ATLAS and CMS, respectively. After calibration, pile-up corrections and cleaning, jets are generally 
considered only above \PT$>$ 20 GeV. The jet energy scale uncertainty 
(and the jet energy resolution to a lesser extent) is generally the dominant systematic uncertainty for $R$-parity conserving 
strong production. It is lower than 2\% for $\PT>100\GeV$, degrading to 4\% for jets with $\pt = 20\GeV$~\cite{CMS-JEC,ATLAS-JEC}.

Identifying \botq-jets from light-quark and gluon jets is crucial for all third-generation searches and is possible 
thanks to secondary vertex information provided by the tracker. Dedicated algorithms are used by ATLAS~\cite{ATLAS_BTAG} 
and CMS~\cite{Chatrchyan:2012jua}. Typically with a 60-70\% \botq-jet identification efficiency, a light-quark jet rejection between 100 and 1000 is 
obtained depending on \PT and $\eta$. Similarly, hadronically decaying taus can be separated from light-quark jets with O(10) less 
rejection power than for a \botq-jet assuming a 60\% tau identification efficiency~\cite{CMS-TauPerf}.

Leptons are key ingredients for SUSY sear\-ches targeting compressed spectra, EW production and/or RPV final states. 
Due to trigger requirements, the leading lepton has to be generally above 
20-25\GeV in \PT, but it is possible to lower the \PT down to 6-7 GeV in analyses considering multi-leptons. 
Further separation from jets is obtained by requiring the leptons to be isolated in the calorimeters and/or the tracker. 

A crucial variable for SUSY searches at the LHC is the magnitude of the missing transverse momentum vector (\MET). In ATLAS, it is based on the 
vector sum of transverse momenta of jets, leptons and all calorimeter clusters not associated to such objects (within $|\eta|<5$). 
In CMS, it is based on all particles reconstructed by the particle-flow method which compensate for the lower calorimeter jet energy resolution. 
Because of the hadronic environment, fake \MET can arise from 
jet mismeasurements which can be efficiently removed by rejecting events where a high-energetic jet (or lepton) and \VEtmiss are close-by (with a 
relative angle of $\Delta \phi(j,\MET)<0.4$).
Detector malfunctions and poorly instrumented regions can cause high \MET as well. In some lepton-veto analyses, \HET is considered, i.e.  
the vectorial sum of jets above few tens of GeV in \PT, to decrease the sensitivity to low-energy jets coming from pile-up.

Obviously the very first experimental challenge at the LHC is the trigger. In Run~1, ATLAS and CMS concentrated their efforts on 
single triggers (\MET, multi-jets, electrons and muons), but also allocated a part of the bandwidth to combined triggers (di- and multi-leptons, 
jets+\MET, several central jets, leptons in combination with large hadronic energy, etc.) which are more analysis-specific.

\subsubsection{Detector performance relevant to SUSY searches}
\label{sec:Det}

Performance of specific sub-detectors are crucial to detect non-prompt sparticle decays or heavy stable charged particles which are slowly 
moving ($\beta=v/c < 0.9$). 

In this respect, the tracker provides a lot 
of relevant information: (i) the ionization energy loss (d$E$/d$x$) measured in the silicon detectors, significantly higher for 
low $\beta$ than for minimum ionizing particles, (ii) characteristics of displaced vertices via dedicated algorithms and (iii) for ATLAS a continuous 
outer tracker, the TRT, to identify late decays. Note that RPC searches generally veto long-lived particles decaying in the tracker by imposing impact 
parameter requirements to reject cosmic muons. 

Low-$\beta$ particles or particles coming from long-lived particle decays in the tracker will arrive late in the calorimeter. The excellent timing 
resolution of the ATLAS and CMS EM calorimeters, around 0.3-0.4\ns per cell, is a very powerful tool to discriminate against SM particles. In case of ATLAS, stand-alone 
pointing capability can also be used. Similarly, the excellent control of the calorimeter noise yields sensitivity to late-decaying particles trapped in the detector. 

Finally, the time-of-flight measured in the muon spectrometer can be exploited as well. Stand-alone resolution of the order of 5\% or better can be achieved 
over the whole $\eta$ range of the spectrometer. It can be ultimately improved by combining information with the calorimeter; see for example~\cite{ATLAS_LLP}.

\subsection{Discriminating variables}
\label{sec:discri}

In RPC searches, the main background for SUSY signals is generally caused by processes involving particles with masses close to the weak scale (\W, \Z, H, top), 
or QCD multijet production with real or fake \MET; see Fig.~\ref{fig:SUSYxs}. These processes can have cross sections that are up to ten 
orders of magnitude higher than the SUSY signal. Therefore, finding SUSY at the LHC requires to design signal regions (SR) 
by exploiting at best the main characteristics of the decay cascade: the presence of two stable LSPs, whose momentum is directly 
proportional to \MET, and/or long decay chains involving jets, i.e. large calorimeter activity in the event. The latter is 
efficiently measured by the scalar sum, \HT, of the transverse energy of reconstructed objects (for some analyses, only a part of the objects can be also considered). 
Rectangular cuts can then be applied on \HT and \MET (or \HET). 

\begin{figure}[htbp]
\begin{center}
\includegraphics[width=\linewidth]{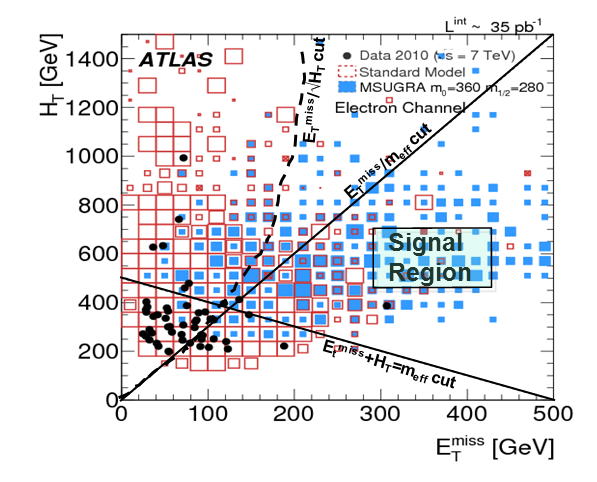}
\end{center}
\caption{Illustration of the variables used in a standard SUSY search for strong production at LHC with a one-lepton, three-jets and \MET final state~\cite{MET_HT_Plane}. 
For completeness the variable $\MET/\sqrt{\HT}$ is also represented in this plane.}
\label{fig:Meff}
\end{figure}

Sensitivity can be improved by taking advantage of the correlation between \HT and \MET 
and by computing $m_{\mathrm{eff}}= \HT + \MET$, called the effective mass~\cite{MEff}. This is illustrated in Fig.~\ref{fig:Meff}. 
The other advantage is that $m_{\mathrm{eff}}$ can be linked to characteristic SUSY parameters like $M_{\mathrm{SUSY}}$, 
the mass of the highest colored object, $M_{\mathrm{LSP}}$, the LSP mass, and their mass difference $\Delta M$; see Fig.~\ref{fig:SUSY_LHC}. 
Typically $m_{\mathrm{eff}}$ will peak at $1.8(M_{\mathrm{SUSY}}^2-M_{\mathrm{LSP}}^2)/M_{\mathrm{SUSY}}$~\cite{ATLASMeff}. 
For open spectra ($\Delta M=M_{\mathrm{SUSY}}-M_{\mathrm{LSP}}> \mathrm{O}(500)\GeV$), this value is well above the SM background which has no correlation 
between \MET and \HT, and therefore peaks at lower values. However, for compressed 
spectra ($\Delta M<500$ GeV), $m_{\mathrm{eff}}$ looses its separation power and selection requirements have to be relaxed, or other discriminant variables 
have to be used. 

An alternative approach, which can also served as a cross-check, is to take advantage of the kinematic distribution of the decay of two heavy 
sparticles (typically gluinos or squarks around 1 TeV). By grouping the reconstructed objects in two hemispheres, two mega-jets 
can be formed and kinematic properties used to distinguish signal from SM background. Two common variables are 
$\aT$~\cite{alphaT}, the ratio of the \pt of the second hardest jet and the invariant mass formed
from the two hardest jets, and {\it Razor}~\cite{Razor}. In the former, SM di-jet events are back-to-back and 
trail off at $\aT=0.5$, whereas a SUSY signal can be asymmetric (causing $\aT>0.5$) because of the presence of the 
LSP in the decay. For {\it Razor}, the idea is to use the transverse and longitudinal information to reconstruct the mass $M_R$ 
of the two mega-jets in the rest frame of the two-jet system ($R$-frame). For signal with open spectra $M_R$ will peak at $M_{\mathrm{SUSY}}$ 
and at $m_{\topq}$ or $m_{\W}$ for \ttbar and WW events. Other quantities related to the $R$-frame (transverse mass, 
Lorentz boost) can be used to increase the discrimination against the background. 

If the presence of lepton(s) is required several types of invariant transverse masses could be considered as discriminant variables. The simplest case corresponds 
to the single lepton channel. There, the leptonic \W-boson background, either coming from \W (+jets) or \ttbar production, is efficiently removed by requiring that
the transverse mass \MT is above the \W mass. This variable is defined as $\MT \equiv \sqrt{2 \MET \; \pt^l (1 - \cos(\Delta \phi))}$, 
where $\pt^l$ is the transverse momentum of the lepton and $\Delta \phi$ the difference in the azimuthal angle between the 
lepton and the \MET direction. The \MT variable has an end-point for backgrounds containing a single leptonically decaying \W 
boson, while signal events contain additional \MET due to the LSPs, leading to an excess at large \MT. When two leptons are considered
the best discriminant variables are the generalized transverse masses \MTt~\cite{Lester:1999tx,Barr:2003rg,Burns:2008va} and \mCT~\cite{mCT1,mCT2}. 
It can be applied for example to search for direct slepton production $\slepp \slepm \rightarrow \lep_1  \lep_2  \ninoone (p_1) \ninoone (p_2)$.
\MTt minimises the larger of the two transverse masses and is defined as
\begin{equation}
\MTt^2 = \min_{ \ponevec + \ptwovec = \VEtmiss } \left\{\max { \left( \MT^2\left(\lep_1,p_1\right),  \MT^2\left(\lep_2,p_2\right) \right)}\right\}
\end{equation} 
while \mCT is a simple combination of the visible decay products and is defined as
\begin{equation}
\mCT^2 = [\ET(\lep_1) + \ET(\lep_2)]^2 - [\ptvec(\lep_1) - \ptvec(\lep_2)]^2 \; ,
\end{equation}
with $\ET = \sqrt{\pt^2 + m^2}$. In both cases -- assuming that leptons are massless -- a signal end-point is defined by 
$(m_{\sleppm}^2-m_{\ninoone}^2)/m_{\sleppm}$ while the \ttbar's endpoint is $(m_{\topq}^2-m_{\W}^2)/m_{\topq} \sim 135\GeV$, 
providing a very powerful discriminating variable for open spectra, i.e. $m_{\sleppm}-m_{\ninoone}>$ O(100 GeV). This approach can be applied 
to all sparticle direct production cases (where the sparticle decays as SM partner and the LSP). As for 
$m_{\mathrm{eff}}$, if SUSY is discovered this will provide a way to measure sparticle masses. Other possibilities exist for 
mass constraining variables and are described in Ref.~\cite{DiscriList}.

\subsection{Background modeling}
\label{sec:bckg}

As just discussed, the signal regions are generally located in extreme regions of the phase space which are not necessarily 
well understood and described in simulation. As a consequence, the number of remaining background events 
in the signal region needs to be estimated as precisely as possible (to increase the sensitivity to the SUSY signal), 
preferably without relying only on simulation. Several methods using data in the background determination are possible, e.g. a
semi data-driven approach using background enriched `control' regions (CR) in data, or even a fully data-driven 
approach. 

The fully data-driven approach is particularly suited for background processes with very large cross sections and fake 
\MET-like dijet production, fake lepton/photon in leptonic/photonic channels and long-lived particle searches. The first type of background is 
estimated with the jet smearing method, where the jet response for well-measured and badly-measured jets is estimated in dedicated 
samples (see more details in~\cite{JetSmearingMethod}). The second type of background is estimated by counting the number of leptons/photons 
passing a loose leptons/photons selection but failing a tight one. With the measure of the true and fake lepton/photon efficiency to pass 
or fail the tight criteria from dedicated orthogonal samples, it is possible to estimate the number of fake lepton/photon in the tight selection.
Finally, background to long-lived particles is estimated with a template method.

\begin{figure}[htbp]
\begin{center}
\includegraphics[width=\linewidth]{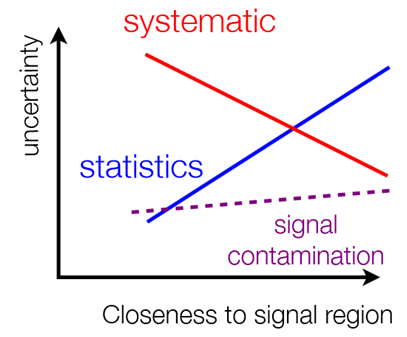}
\end{center}
\caption{Parameters entering in the design of a control region.}
\label{fig:CR}
\end{figure}

The semi data-driven approach is particular suited for SM
processes with large cross section, as top-quark, W-, or Z-boson production. The definition of the control region is a trade-off 
between kinematic requirements as similar as possible to the signal region to minimize systematics, the highest 
achievable purity and minimization of the contamination from potential signal. This is illustrated in Fig.~\ref{fig:CR}. 
The simulation is normalized to the event yield in the control region ({\it scale} factor), and the background in the signal
region is estimated by extrapolating the background level via simulation from the control region to the signal region ({\it transfer} factor). 

Determination from Monte Carlo simulations only are generally adequate for backgrounds that are expected to be very 
small in the signal region (e.g.~diboson production for strong SUSY searches), or for rare processes with very small 
cross sections (e.g.~\ttH or triboson production).

Monte Carlo (MC) simulated event samples therefore play a central role in SUSY analyses. They allow not only to develop and validate the analysis procedure
but also, in many cases, to evaluate the SM backgrounds, and calculate the acceptance and efficiency 
of the studied signal samples. SUSY analyses rely heavily on the progresses made last 20 years in the calculations 
and simulation of high $Q^2$ processes in hadronic collisions. Simulated samples of top (including \ttW and \ttZ) and \W- or \Z-boson production 
with multiple jets are produced with multi-parton generators such as ALPGEN~\cite{ALPGEN}, SHERPA~\cite{SHERPA} or MADGRAPH~\cite{MADGRAPH}, with (in some cases)
up-to six additional partons in the matrix element. The next-to-leading-order (NLO) generators MC@NLO~\cite{MCatNLO} and
POWHEG~\cite{POWHEG1,POWHEG2,POWHEG3} are generally used for top or diboson production. Parton shower and fragmentation processes are simulated 
with either HERWIG~\cite{HERWIG} or PYTHIA~\cite{PYTHIA}.

All SM backgrounds are then fed into a {\sc Geant4}-based~\cite{GEANT4} model of the CMS or ATLAS detector. Due to the large 
amount of signal points, the signal events are usually processed with a fast detector simulation~\cite{CMS_fastsim,ATLAS_fastsim}. 
The effect of multiple proton-proton collisions from the same or different bunch crossings is included in all simulations by 
overlaying minimum bias events onto hard-scattering events. The distribution of the number of interactions per bunch crossing is 
reweighted in the simulation to match the data.
 
Systematics are often estimated by comparing two generators of the same type, by using different parton distribution functions,
and by changing the factorization, renormalization and matching scales. The choice of the generator is made case-by-case 
and is analysis dependent: RPC strong production searches with a large number of jets generally use multi-parton generators, 
whereas EW production searches preferably use next-to-leading order generators.

\subsection{Limit extraction}
\label{sec:limit}

After all cuts have been applied, the number of data events $n$ is counted in the signal region ('cut-and-count' method). It is compared to 
the expected number of SM events to enter the signal region. For that, a likelihood function for observing $n_{\mathrm B}$ background events in the 
signal region is constructed as the product of Poisson probability distributions for event counts in the signal region 
and each of the main control regions, and of constraints for systematic uncertainties on the expected yields, called nuisance 
parameters~\footnote{Nuisance parameters are modelled by a Gaussian probability density function with a width given by the size of the uncertainty.}. The Poisson
probability density functions also include free parameters, for example to scale the expected contributions from
the major backgrounds. The free parameters and nuisance parameters are adjusted to maximize the likelihood function. 
The result of the likelihood fit includes therefore a set of estimates and uncertainties for the background in the SR~\footnote{The procedure is checked by building `validation regions' 
in between the control and the signal and comparing with data.}. The significance of an excess is the probability that a background-only 
experiment is more signal-like than observed and is computed following the \textit{CLs} prescription~\cite{CLS}. 

If no excess is observed, limits at 95\% confidence level (CL) are set. Note that to obtain more constraining limits, some 
analyses make use of the discriminant variable shape. For each public result, efficiency maps are provided to allow theorists 
to test their own models. Following an ATLAS-CMS agreement, the numbers quoted for exclusion refer to the observed limit minus one standard deviation.

\subsection{SUSY models for interpretation}
\label{sec:interp}

The sensitivity of the SUSY searches are estimated by three complementary approaches, given it is not possible to cover 
the entire parameter phase space. 

First, constrained SUSY models are tested, where boundary conditions at a high energy scale reduce the number of parameters to a few making it realistic to scan systematically. 
Benchmark models are MSUGRA/CMSSM~\cite{MSUGRA1,MSUGRA2,MSUGRA3,MSUGRA4,CMSSM}, minimal GMSB~\cite{GMSB} and AMSB~\cite{AMSB1,AMSB2} models. Each model corresponds to 
a particular SUSY breaking messenger: gravity for the former and the latter and gauge bosons for GMSB. Similarly the LSP is generally the 
bino-like \ninoone for MSUGRA/CMSSM, the wino-like \ninoone for AMSB and always the gravitino for GMSB.

Second, we have topological or simplified models~\cite{SMS1,SMS2,SMS3} where only a few SUSY particles are involved, while the masses 
of all other SUSY particles are set to multi-TeV values, out of range at the LHC. The cascade decay of the remaining
particles to the LSP, typically with zero or one intermediate step, is only characterized
by the particle masses. These models are particularly suited for direct sparticle production.

Finally, results can also be interpreted in phenomenological MSSM (pMSSM)~\cite{PMSSM} models where the number of MSSM
parameters is reduced to 19 by assuming the absence of new sources of flavor changing neutral
currents and CP violation and universality of the first and second generation. By sampling a limited number of pMSSM parameters,
the sensitivity of the searches to more `realistic' configurations of SUSY particle
masses and branching ratios can be assessed.

\section{Gluino and first/second generation of squarks}
\label{sec:StrongProd}

In the MSSM, TeV-scale squarks and gluinos produced in pp collisions will decay promptly in long decay chains containing mainly quark and gluon jets and the LSP. SUSY events are 
therefore characterized by multiple energetic jets as well as transverse missing energy (\MET) originating from the undetected LSP energies; see Fig.~\ref{fig:SUSY_LHC}. 
Depending on the sparticle present (or not) in between the squarks/gluinos and the LSP, charged lepton(s) and/or photons could also appear in the cascade. This section 
summarizes the present status of searches for gluinos and first/second generation squarks when the \ninoone is the LSP (Sect.~\ref{sec:Strong_Chi10}) and when the gravitino is 
the LSP (Sect.~\ref{sec:Strong_gravitino}).

\subsection{SUSY models with $\ninoone$ as LSP}
\label{sec:Strong_Chi10}

To improve the sensitivity to these models, searches are usually divided in lepton veto (Sect.~\ref{sec:LepVet}) and leptonic (Sect.~\ref{sec:LepTonic}) searches. 
The former target more inclusive or generic scenarios while the latter are generally more optimal for specific models. In both cases, requiring that some jets are originating 
from a \botq quark can increase the sensitivity (Sect.~\ref{sec:bjets}). A summary is proposed in Sect.~\ref{sec:summary-strong}.

\subsubsection{Lepton-veto searches}
\label{sec:LepVet}

The preferred gluino (squark) decay modes are 
$\gluino \rightarrow \quark \squark$  ($\squark \rightarrow \quark \ninoone, \quark \gluino$) which generate signatures with two to $\ge$10 jets. Low jet multiplicities probe 
squark-squark (two jets or more), squark-gluino (three jets or more) or gluino-gluino (four jets or more) production. Additional jets compared to the tree 
level processes originate from initial- and final-state radiation jets (ISR/FSR) or from the presence of a top quark in the decay chain. In the last case, an increase of sensitivity 
is possible by requiring the presence of one or several \botq-tagged jets (see Sect.~\ref{sec:bjets} for more details). 

\begin{figure}[htbp]
\begin{center}
\includegraphics[width=\linewidth]{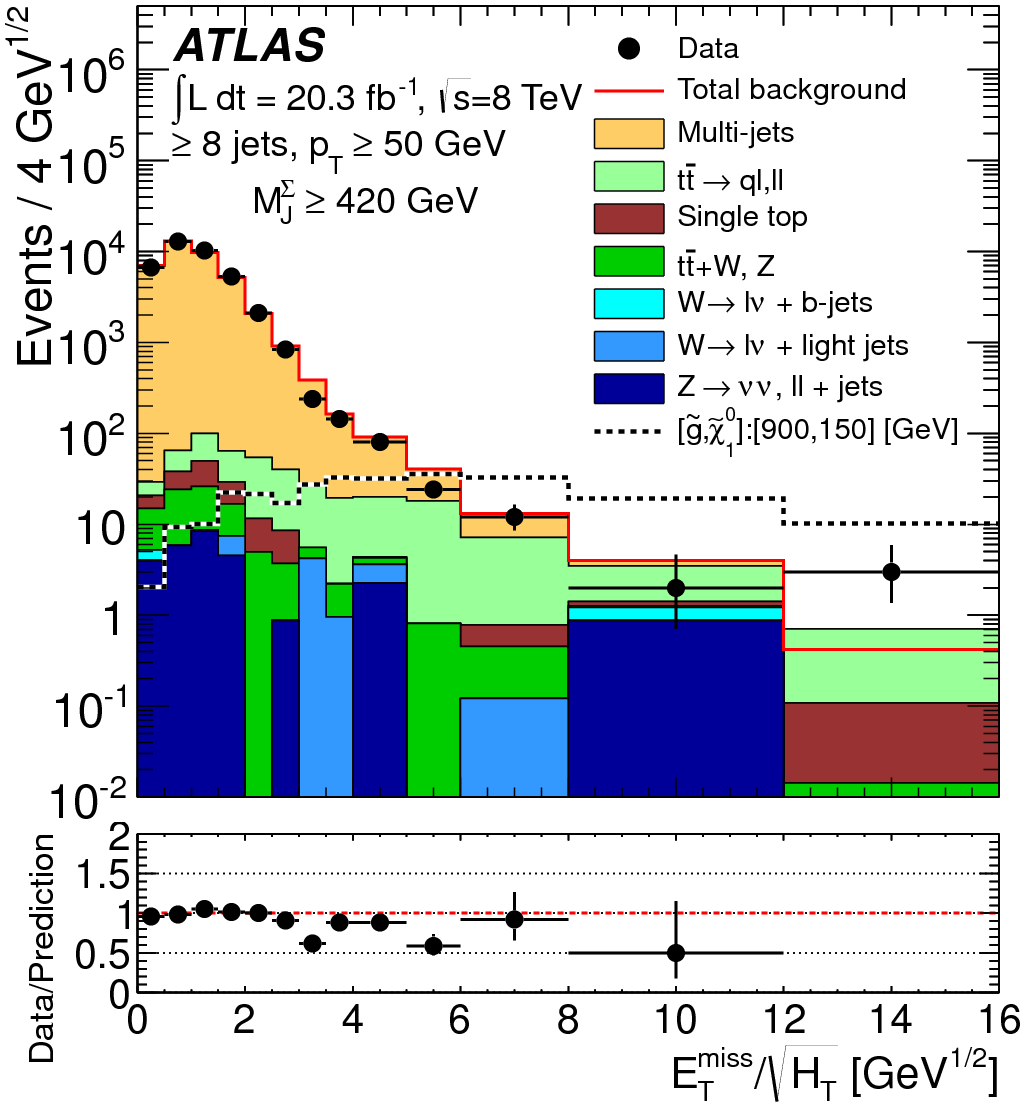}
\end{center}
\caption{Example of a discriminating variable: the $\MET$ significance from~\protect\cite{ATLAS_Strong_0l7j}.}
\label{fig:0lStrong_Discri}
\end{figure}

Several optimizations are possible depending on the discriminating variables chosen, and this section only discusses the already published results based on rectangular cuts 
on \MET and \HT~\cite{CMS_Strong_0lMETjets}, \MHT and \HT~\cite{CMS_Strong_0l_MHT} on one side and \MET significance~\cite{ATLAS_Strong_0l7j} on the other side. 
As an example Fig.~\ref{fig:0lStrong_Discri} shows the \MET significance distribution. The SM background, composed of $\W$+jets, $\Z(\rightarrow \nu \nu)$+jets, \ttbar and 
QCD multi-jets, peaks at low $\MET$ significance value. A typical SUSY signal, where gluinos of mass 900\GeV are pair produced and decay each to a \ttbar pair 
and a LSP of mass of 150\GeV, leads to much higher values. The signal region is defined as $\MET/\sqrt{\HT}>4\GeV^{1/2}$. 

In all lepton-veto analyses, the challenge is to properly estimate backgrounds that are poorly modeled by Monte Carlo simulations: the QCD multi-jet background is obtained with 
a jet smearing method or a template method for the \MET significance search. $\Z(\rightarrow\nu \nu)$+jets relies on close-by Standard Model process like $\gamma$+jets or 
$\Z(\rightarrow\mu \mu)$+jets samples to get an estimate of $\Z(\rightarrow\nu \nu)$. $\W$+jets and \ttbar are estimated by designing control regions close to the signal regions -- 
requiring one lepton for example. Note that other searches with \meff or Razor as discriminating variables will become available soon. Searches with \aT are only available for half 
the luminosity of 2012 data~\cite{CMS_Strong_alphaT}. They are generally more powerful but will not change the overall picture given in the rest of this section.

\begin{figure}[htbp]
\begin{center}
\includegraphics[width=\linewidth]{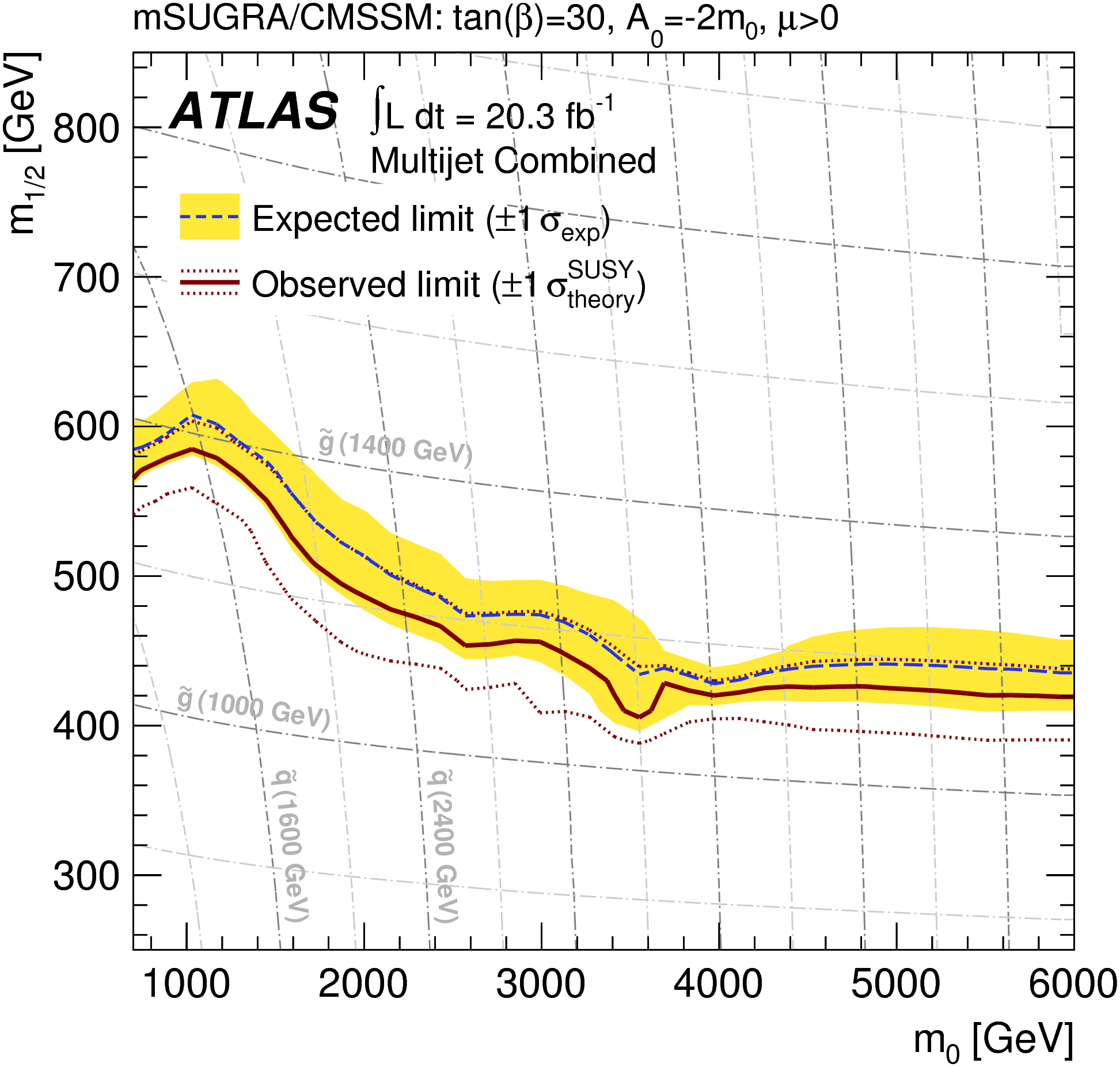}
\end{center}
\caption{Typical exclusion limit at 95\% CL from the lepton-veto inclusive search in the MSUGRA/CMSSM scenario~\protect\cite{ATLAS_Strong_0l7j}.}
\label{fig:0lStrong_MSUGRA}
\end{figure}

All these sear\-ches are particularly efficient for open spectra where the mass difference between the LSP and the gluino/squark 
is large ($\Delta M> \mathrm{O}(500)\GeV$), providing high-energetic jets. This is for example the case in the constrained SUSY model MSUGRA/CMSSM where the two most relevant 
parameters, the universal scalar and fermion masses at Grand Unified Theory (GUT) scale, $m_0$ and $m_{1/2}$, are varied to construct a grid of 
points~\footnote{The other fixed parameters ($\mathrm{\tan}\beta$, $A_0$ and the sign of $\mu$) are chosen to accommodate a 126\GeV Higgs mass.}. 
Figure~\ref{fig:0lStrong_MSUGRA} shows the limits obtained with the $\MET$ significance search. Squark and gluino masses at the EW scale are 
proportional to the $m_0$ and $m_{1/2}$ parameters and shown by the isolines which indicate that 
gluino and degenerate squark masses below 1.2\TeV are excluded. For equal degenerate squark and gluino masses a limit of 1.3\TeV is reached. 

\begin{figure}[htbp]
\begin{center}
\includegraphics[width=\linewidth]{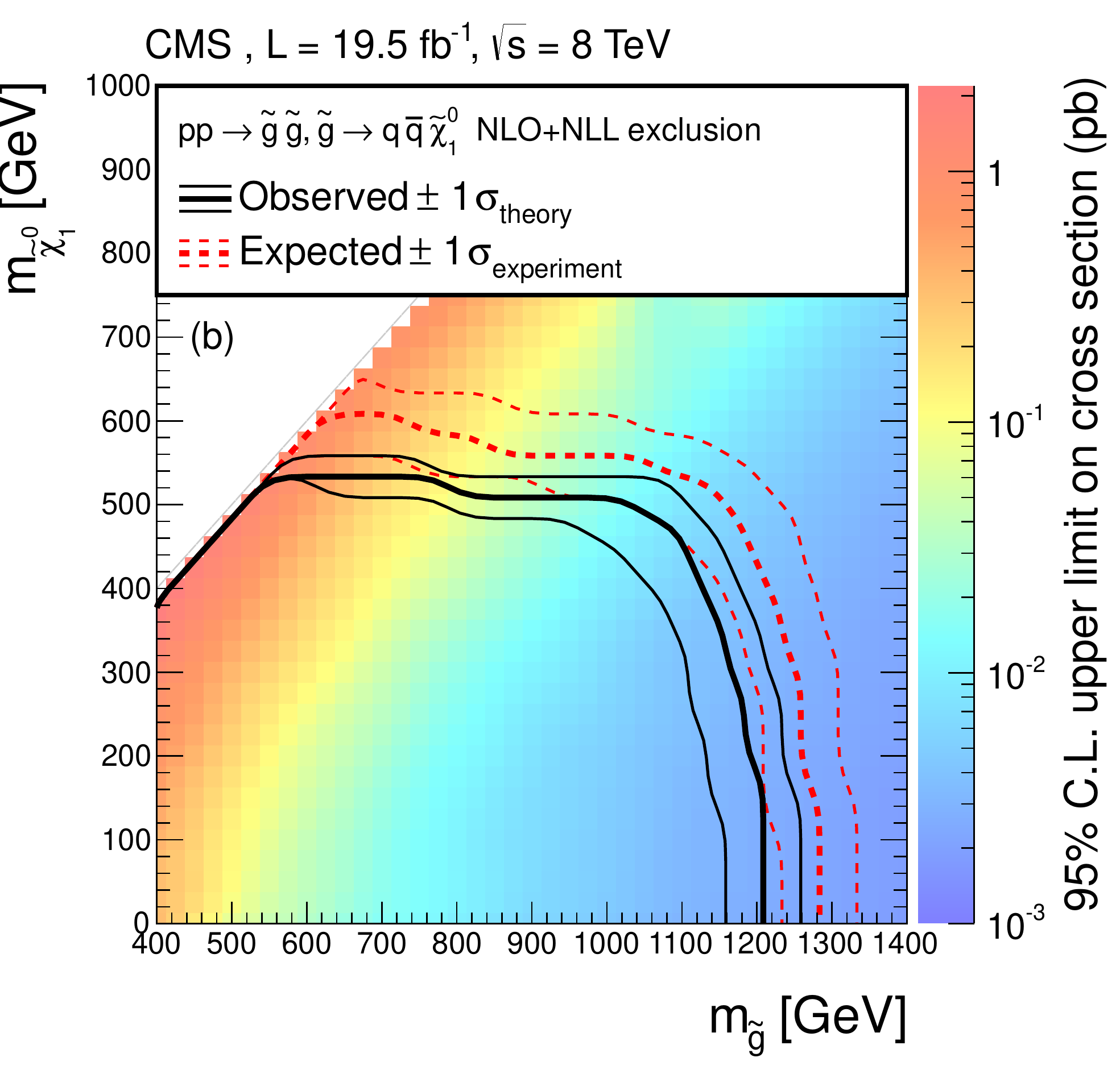}
\end{center}
\caption{Exclusion limits at 95\% CL on the gluino-gluino production in the gluino-LSP mass plane~\protect\cite{CMS_Strong_0lMETjets}. 
The gluino always decays as $\gluino \rightarrow \quark \aQuark \ninoone$ and all other SUSY particles are decoupled.}
\label{fig:0lStrong_gluinos}
\end{figure}

\begin{figure}[htbp]
\begin{center}
\includegraphics[width=\linewidth]{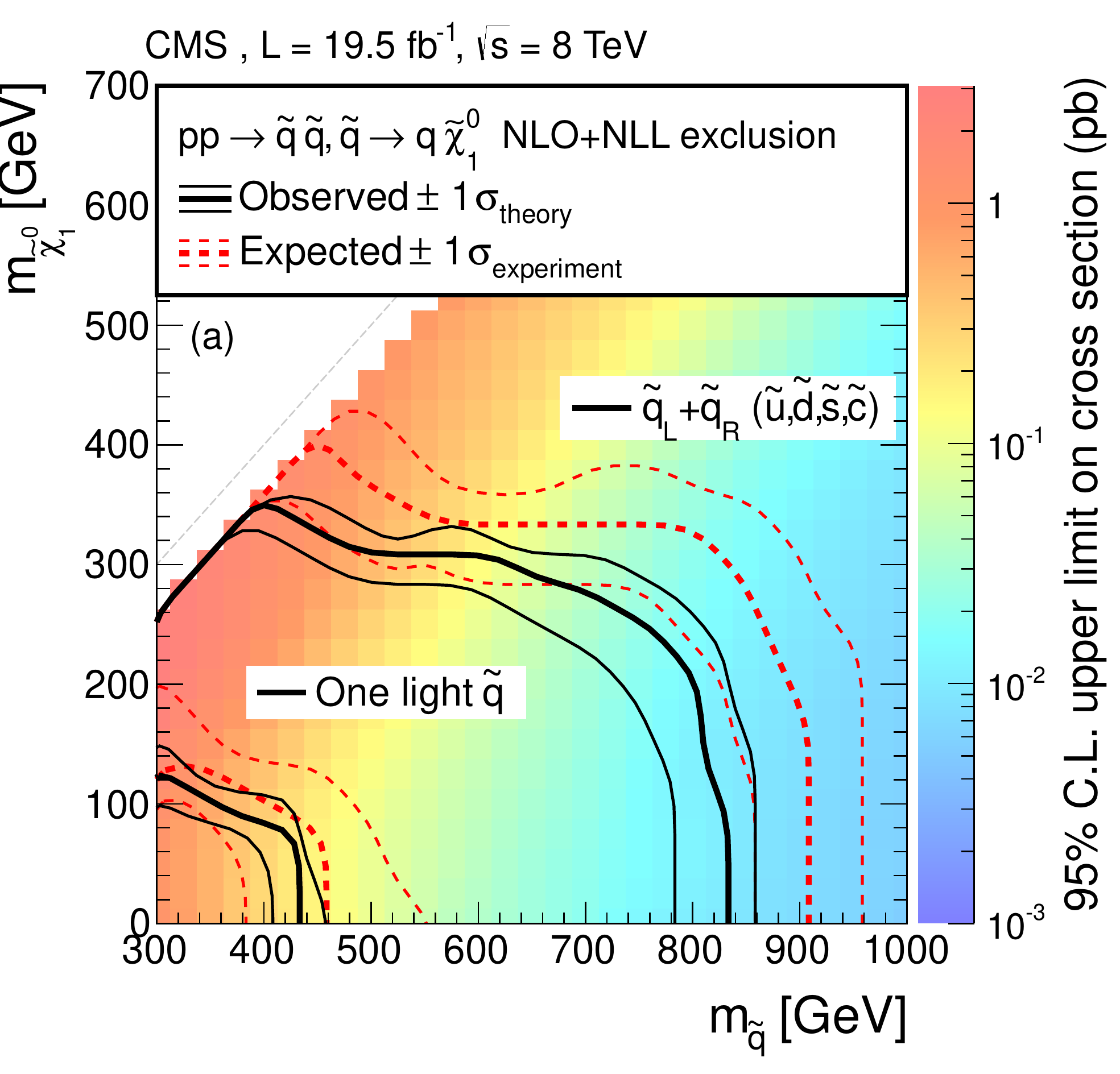}
\end{center}
\caption{Exclusion limits at 95\% CL on the squark-squark production in the squark-LSP mass plane~\protect\cite{CMS_Strong_0lMETjets}. 
The squark always decays as $\squark \rightarrow \quark \ninoone$ and all other SUSY particles are decoupled.}
\label{fig:0lStrong_squarks}
\end{figure}

Another way to represent the power of the lepton veto searches is to use 
simplified models where $\gluino \rightarrow \quark \aQuark \ninoone$ decays are enforced. Here again a limit of 1.2\TeV on the gluino mass is obtained (for massless LSPs), 
see Fig.~\ref{fig:0lStrong_gluinos}. However, for more compressed spectra the limits degrade and LSP masses cannot be excluded beyond 550\GeV. A similar situation occurs when 
considering mass-degenerate light flavor squarks forced to decay as $\squark \rightarrow \quark \ninoone$; see Fig.~\ref{fig:0lStrong_squarks}. Compared to the gluino situation, 
the limits are degraded to 800\GeV for squark masses (again in the case of massless LSPs), and LSP masses can not be excluded beyond 300\GeV. 
These limits are reduced to 400\GeV and 100\GeV, respectively, when only one light squark 
is considered. 

Overall these results represent an increase of about one order of magnitude compared to the previous limits from Tevatron and LEP. From lepton-veto analyses, 
the strongest limits are obtained for the gluino in open SUSY spectra, and they exclude a large part of the favored region from naturalness (cf. Fig.~\ref{fig:SUSYTheo2}).

\subsubsection{Leptonic searches}
\label{sec:LepTonic}

Requiring one or two isolated leptons (electron or $\mu$) on top of multi-jets and \MET allows to probe other regions of parameter space, and especially more compressed mass spectra.
The lepton generally comes from \W leptonic decays originating from chargino, top or slepton decay. 
Experimental challenges drastically change: lepton triggers can be exploited and requirements on jet kinematics can be reduced. Lowering cuts on \MET and \HT is possible 
since the multi-jet QCD background is naturally suppressed by the presence of isolated lepton(s). Very soft leptons (in the 6-25\GeV range) are also considered to probe the 
compressed gluino-chargino-LSP case~\cite{ATLAS_SoftLepton}. Finally, other variables exists like the transverse mass \MT, 
which efficiently reduces \ttbar and $\W$+jets backgrounds by requiring $\MT > m_{\W}$, as discussed in Sect.~\ref{sec:discri}. This allows to compensate the loss due to 
the leptonic branching ratio(s) when comparing with lepton-veto analyses. 
While this works well for single-lepton analyses to be competitive with lepton-veto analyses, this is generally not sufficient when two leptons are considered~\footnote{This 
final state could be very useful to determine SUSY parameters but is generally not for discovery.}. 
In that case, it is more advantageous to consider two leptons of same-sign since this signature is almost not produced by SM processes and appears naturally 
in many SUSY decays. The two reasons are : $i)$ gluinos are Majorana particles and produced in pairs, therefore if leptons are present in each leg they have 50\% probability to 
be of same-sign, $ii)$ multi-\W final states occur frequently through top and chargino decays and leptonic \W decay will ensure the presence of two same-sign leptons in most cases.
For a same-sign dilepton analysis, the main background is caused by the rare $\topq \aTop X$ ($X=\Higgs,\Z,\W$) SM processes, fake leptons and mis-measured lepton charge because of the process 
$\lep \rightarrow \lep \gamma \rightarrow \lep \lep \alep$ where $\alep$ inherits most of the energy of the original lepton.
 
\subsubsection{Multi \botq-tagged jet searches}
\label{sec:bjets}

\begin{figure}[htbp]
\begin{center}
\includegraphics[width=\linewidth]{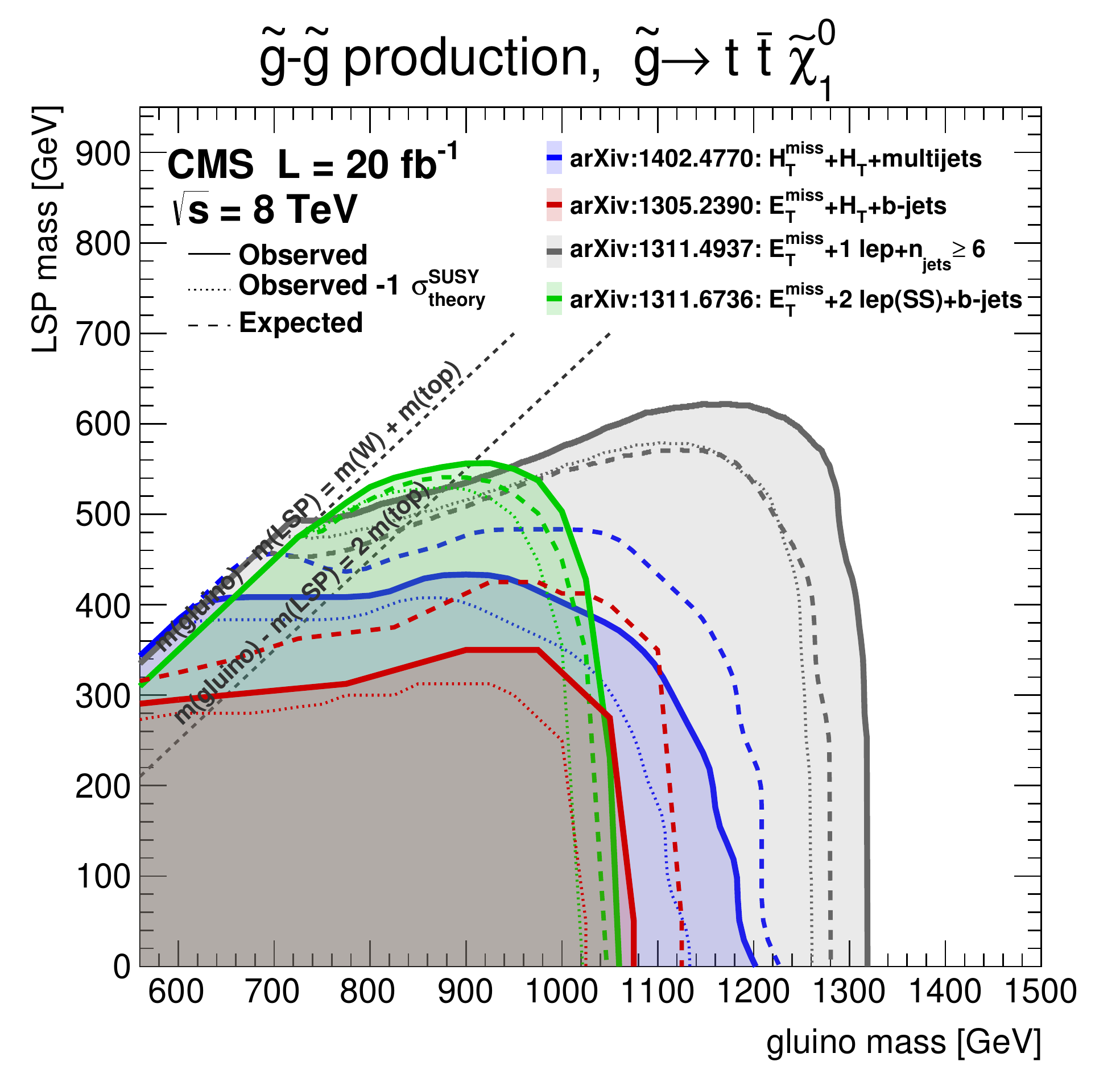}
\end{center}
\caption{Exclusion limits at 95\% CL on the gluino-gluino production in the gluino-LSP mass plane for a simplified model where gluino decays promptly 
via off-shell top squark and a LSP. The other sparticle masses are assumed to be decoupled. Four different analyses are shown.}
\label{fig:StrongGtt}
\end{figure}

As for leptons, identifying \botq-tagged jets in the multi-jet final states can be a precious help, especially together with leptons. This is particularly
true for the decay $\gluino \rightarrow \topq \sTop \rightarrow \topq \aTop \ninoone$ favored by the natural mass spectrum. This will 
provide 4 tops+\MET final states. Several dedicated analyses have been designed to reach this striking final state and obtain extra sensitivity compared 
to lepton-veto analyses described in Sect.~\ref{sec:LepVet}. Reducing the dominant $\topq \aTop \rightarrow \Wp \Wm \botq \aBot$ background is possible when considering 
$i)$ a single isolated lepton and at least five jets, two or three of which are identified as \botq-tagged jets~\cite{CMS_Strong_1lepton}, $ii)$ two same-sign 
leptons with one, two or three \botq-tagged jets~\cite{CMS_Strong_2lSS}~\footnote{Three \botq-tagged jets without a lepton is also considered but no public results 
exist yet at $\sqrt{s}=$8\TeV.}. 
The best sensitivity is obtained by the former, which can exclude gluino masses up to 1.3\TeV for LSP masses below 600\GeV, assuming a 100\% branching ratio for the decay 
$\gluino \rightarrow \topq \sTop \rightarrow \ttbar \ninoone \ninoone$. The same-sign dilepton analysis allows one to probe the compressed spectra part when one top 
is off-shell. The relative strengths of the different analyses for this model can be judged from Fig.~\ref{fig:StrongGtt}. Similar results are obtained for 
$\gluino \rightarrow \botq \sbL \rightarrow \bbbar \ninoone \ninoone$. 

\subsubsection{Summary}
\label{sec:summary-strong}

Ultimate limits from the first LHC run on gluino and squark masses when the $\ninoone$ is the LSP will be obtained by combining all lepton-veto and leptonic analyses -- many results are still to come. 
However, it is already fair to say that the most important constraint from the LHC experiments is the exclusion of the gluino 
mass below 1\TeV for open spectra. This fairly generic limit excludes a great part of the favored region from naturalness (cf. Fig.~\ref{fig:SUSYTheo2}). Since the gluino mass 
is governed by only one parameter ($M_3$), and enters in the top squark and EWKino masses through loop corrections, this has the general effect to pull up the whole natural 
spectrum~\cite{GluinoAfterRunI}. Compressed spectra are still poorly explored, but more work is still going on to exploit monojet-like final states provided by ISR/FSR. Extra 
sensitivity will also be brought by analyses based on reduced lepton and jet thresholds -- this is possible when the peak luminosity decreases at the end of a LHC run. Another 
important information is coming from the LSP mass constraint which can reach up to 600\GeV in leptonic analyses, though it is less generic than the gluino limit. Finally, squark-mass constraints 
are more model-dependent and single squark-mass limits are at most of 500\GeV.


\subsection{SUSY models with \gravitino as LSP}
\label{sec:Strong_gravitino}

SUSY scenarios where the gravitino is the LSP generate a variety of final states, driven by the NLSP-gravitino coupling.
While some of these final states are common with Sect.~\ref{sec:Strong_Chi10} -- when the NLSP is the gluino or a squark, some need
the development of new dedicated analyses -- when NLSP is a slepton, chargino or neutralino. The most natural solution is $\ninoone$ as the NLSP, leading 
to final states with extra $\gamma$, \Z or \HO, depending on the SUSY parameters. 

\begin{figure}[htbp]
\begin{center}
\includegraphics[width=\linewidth]{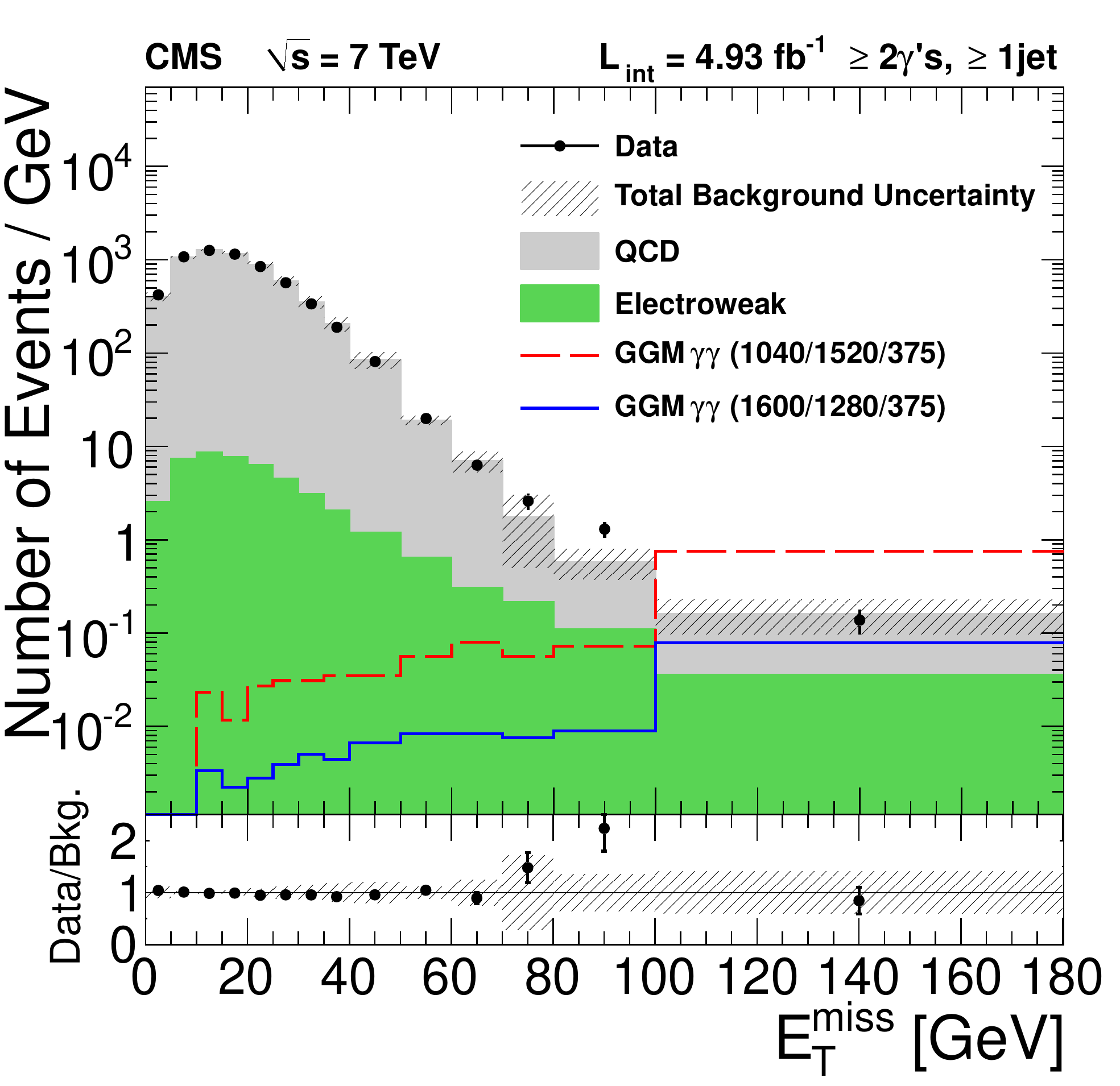}
\end{center}
\caption{$\MET$ distribution in the diphoton analysis~\protect\cite{CMS_2Photon_2011}.}
\label{fig:StrongGMSB_MET}
\end{figure}

\begin{figure}[htbp]
\begin{center}
\includegraphics[width=\linewidth]{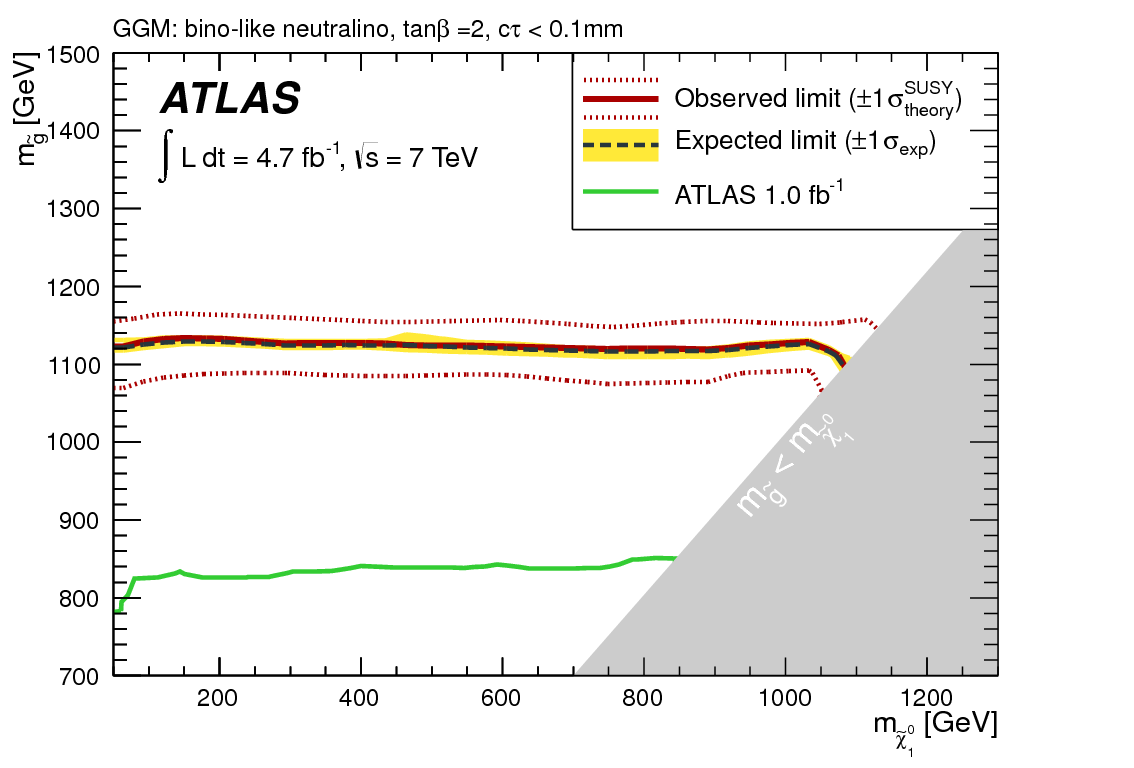}
\end{center}
\caption{Exclusion limits at 95\% CL on the gluino-gluino production in the gluino-LSP mass plane from the diphoton+\MET analysis~\protect\cite{ATLAS_2Photon_2011}. 
The simplified model assumes a bino-like NLSP and a gravitino LSP. The other sparticle masses are assumed to be decoupled.}
\label{fig:StrongGMSB_Limit}
\end{figure}

Among all possible final states, the ones containing photons, jets and \MET from the gluino/squark cascade could have escaped searches described in 
Sect.~\ref{sec:Strong_Chi10}. In this case, the most dangerous background is caused by multi-jet or $\gamma$+jets events when the jet is mimicking a photon. 
This background type can be drastically reduced with a good photon/jet rejection -- measured to be $O(10^4)$ in the $H\rightarrow \gamma \gamma$ channel 
-- and by requiring a high value for \MET. The other type of background are SM electroweak processes, especially $\W(\rightarrow \text{e} \nu_{\text{e}})\gamma$, where 
the electron is reconstructed as a photon and true \MET is caused by the neutrino. In both cases, data-driven methods are used for the background estimate. 
As an example, the distribution of \MET, the main discriminating variable for the $\gamma \gamma$ searches is shown in Fig.~\ref{fig:StrongGMSB_MET}. 
A high sensitivity is observed for the SUSY signal at high \MET. Assuming a bino-like $\ninoone$ NLSP, 
Fig.~\ref{fig:StrongGMSB_Limit} shows that it is possible to exclude gluino masses below 1\TeV regardless of the NLSP mass~\cite{ATLAS_2Photon_2011}.

\begin{figure}[htbp]
\begin{center}
\includegraphics[width=\linewidth]{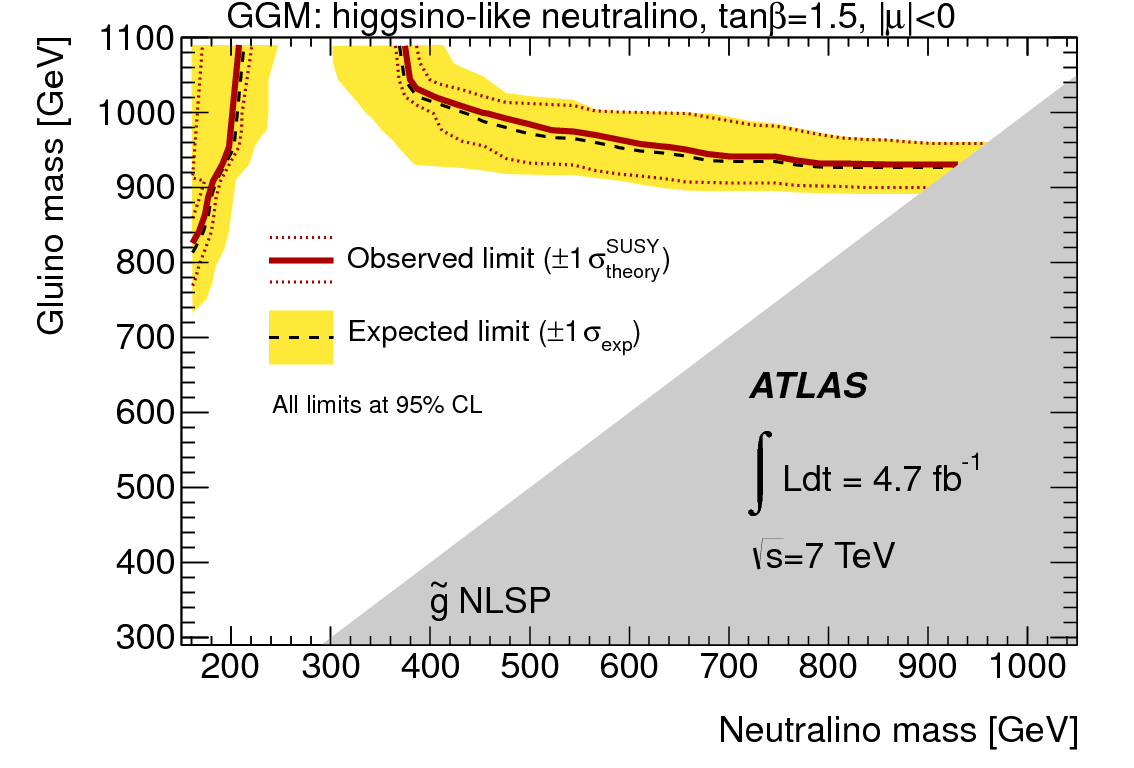}
\end{center}
\caption{Exclusion limits at 95\% CL on the gluino-gluino production in the gluino-LSP mass plane from the $\gamma+\botq+\MET$ analysis~\protect\cite{ATLAS_Photon+b_2011}. 
The simplified model assumes a higgsino-like NLSP. The other sparticle masses are assumed to be decoupled.}
\label{fig:StrongGMSB_Limit_Higgsino}
\end{figure}

When the $\ninoone$ is higgsino-like, the preferred solution from naturalness arguments, $\ninoone \rightarrow \HO (\rightarrow \botq \aBot) \gravitino$ final 
states is expected. No analysis presently attempts to search for a 4\botq+\MET final state. However assuming that $\ninoone$ is also partly bino-like, the 
signature will be $\gamma+\botq+\MET$ and has been searched for. A second photon and a lepton veto are applied to remain orthogonal with other searches and remove 
final states with leptonic \W decays~\cite{ATLAS_Photon+b_2011}. In this case also the gluino mass limit reaches around 1\TeV, as shown in Fig.~\ref{fig:StrongGMSB_Limit_Higgsino}. 

All other NLSP cases are generally already well covered by the searches discussed in Sect.~\ref{sec:Strong_Chi10}~\cite{GMSB_FinalStates}. However, it is possible to 
gain a bit in sensitivity by searching for one or two $\tau$+jets+\MET final states~\cite{ATLAS_taus,CMS_taus}, especially when \sTau is the NLSP or co-NLSP with 
other sleptons. In this case, it is possible to combine with a 2lepton+jets+\MET search~\cite{ATLAS_SoftLepton} and even a 3lepton+jets+\MET search~\cite{CMS_3l_GMSB}. 
In all cases, the gluino mass limit is always above 1\TeV.

As for models with the \ninoone as LSP, SUSY scenarios where the \gravitino is the LSP provide strong 
constraints on sparticle masses and particularly on the gluino mass which is generally excluded 
below 1\TeV whatever the NLSP nature and mass are. Less stringent limits are obtained for squarks. It is worth to note that 
these conclusions are based on 7\TeV results -- presently only 
one 8\TeV result is available.


\section{Third generation of squarks}
\label{sec:3rdGene}

As already discussed in Sect.~\ref{sec:SUSY_LHC}, naturalness predicts light third-generations squarks. Another motivation for the third-generation 
squarks to be the lightest colored sparticles is that the squark mass eigenstates ($\squark_1, \squark_2$) 
depend on orthogonal combinations of the gauge eigenstates ($\squark_{\text R}, \squark_{\text L}$), e.g. for the lighter squark given by 
$\squark_1=\squark_{\text L} \cos(\Theta_{\squark}) + \squark_{\text R}\sin(\Theta_{\squark})$. The off-diagonal elements
of the mass matrix $\Theta_{\squark}$ are proportional to the mass of the SM partner particle, the Higgs-related parameters $\mu$ and $\tan\beta$.
Therefore, the mass of the $\sTop_1$, predominantly $\sTop_{\text R}$, can be small due to the large top quark mass and the $\sBot_1$ mass is 
expected to be light for large $\tan \beta$. For small $\tan\beta$ the $\sBot_{\text L}$ is still expected to be small due to the effects of the 
large top Yukawa coupling as it is part of the doublet containing $\sTop_{\text L}$.

The following two sections review the status of the searches for models with \ninoone being the LSP (Sect.~\ref{sec:stop_neut}), and 
the \gravitino being the LSP (Sect.~\ref{sec:stop_grav}).

\subsection{SUSY models with \ninoone as LSP}
\label{sec:stop_neut}

If the third-generation squarks are lighter than glu\-inos, they are likely to appear in gluino decay chains. But if the gluino masses are too 
heavy to be produced at the LHC energy, searches for direct third-generation squark pair production might be the only way to observe them, 
even though these branching ratios are more than one order of magnitude lower than those for gluino-gluino production (see Fig.~\ref{fig:SUSYxs}).

\begin{figure}[htbp]
\begin{center}
\includegraphics[width=\linewidth]{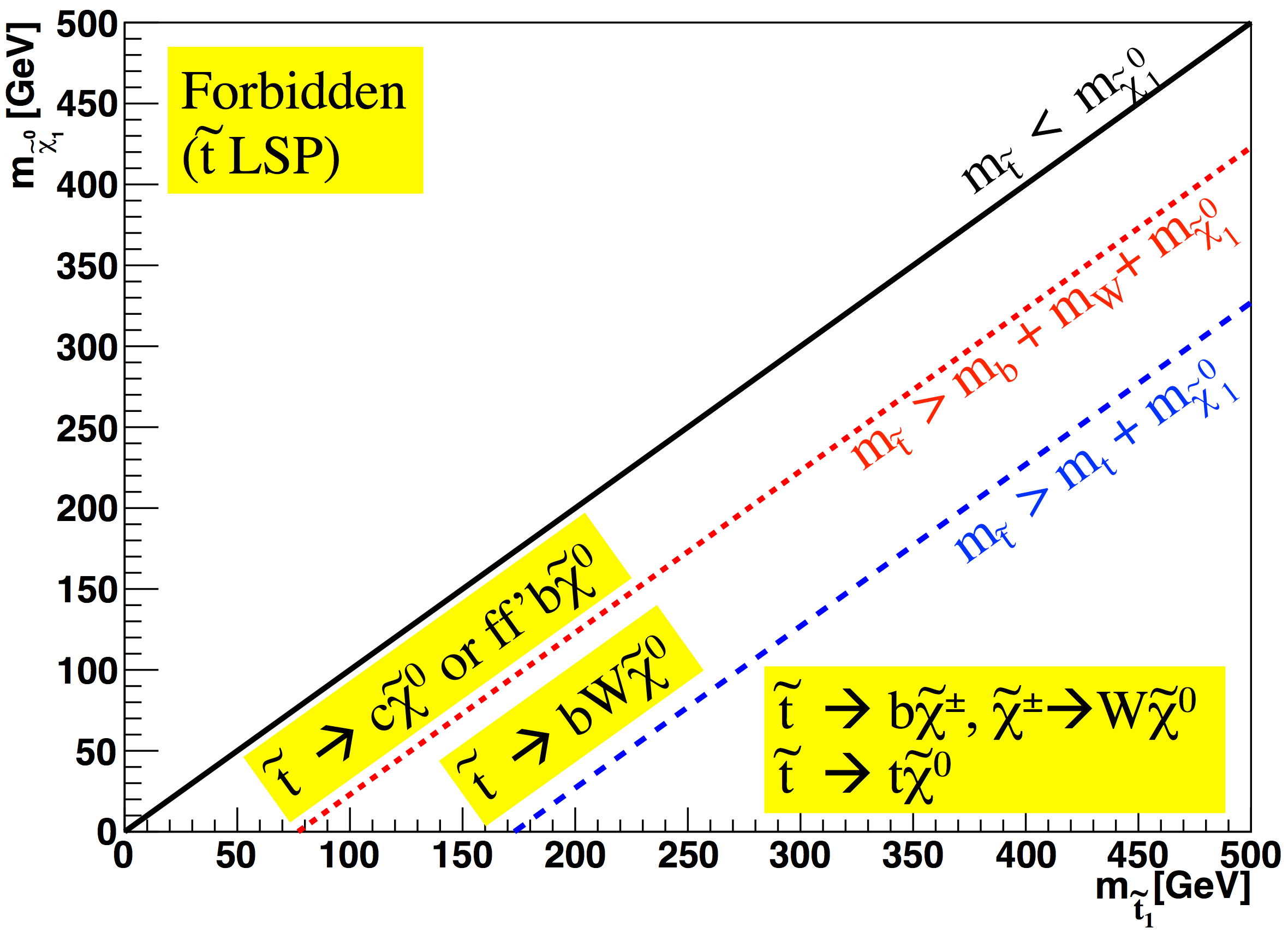}
\end{center}
\caption{Overview of the possible top squark decays depending on the mass of the $\sTop_1$ and the \ninoone.}
\label{fig:stop_plane}
\end{figure}

\paragraph{}
For top squarks, the possible decays and therewith connected search strategies depend on the masses of the accessible particles. Assuming
that the $\sTop_1$, the $\chipmone$ and the $\ninoone$ are the only accessible SUSY particles, the possible decays in the $m_{\sTop_1}$ -- 
$m_{\ninoone}$ parameter plane are displayed in Fig.~\ref{fig:stop_plane}.

The most obvious decay chain is given for $m_{\sTop_1} > m_{\topq} + m_{\ninoone}$: here the top squark will decay either like 
$\sTop \rightarrow \topq \ninoone$ or $\sTop \rightarrow \botq \chipmone$, where $\chipmone \rightarrow \Wpm \ninoone$, with 
cross sections depending on the \chipmone mass. In both cases this gives two \W bosons, resulting in 
a large probability to have (at least) one electron or muon in the final state. Hence, the searches requiring zero or one 
lepton in the final state are the strongest. Zero-lepton searches are expected to have a slightly higher reach in the 
$\sTop_1$ mass for low \ninoone masses, while one-lepton analyses tend to reach closer to the diagonal line at $m_{\sTop_1} - m_{\ninoone} = m_{\topq}$. 
The region around this line is very difficult to cover with current searches, as the kinematics of the decay are similar to the SM top decay kinematics.
An option to test this parameter space exists for cases where the $\sTop_2$ is not too heavy either. A decay chain to
search for would be $\sTop_2 \rightarrow \Z \sTop_1$. Requiring same-flavor dileptons from the \Z boson decay in 
addition to a lepton from the $\sTop_1 \rightarrow \topq \ninoone$ decay provide a powerful background rejection~\cite{ATLAS_stop2}.
 
When the $\sTop_1$ is lighter than the top, in the region defined by $m_{\botq} + m_{\W} + m_{\ninoone} < m_{\sTop_1} < m_{\topq} + m_{\ninoone}$ 
the top squark will decay 
as $\sTop \rightarrow \botq \chipmone$, with the subsequent decay $\chipmone \rightarrow \Wpm \ninoone$. This can best be tested with an analysis 
requiring two leptons and two \botq-tagged jets, which also has sensitivity to the three-body decay $\sTop \rightarrow \botq \Wpm \ninoone$,
which becomes important for high \chipmone masses. Also a one-lepton search has sensitivity in this area, as shown below.

For $m_{\sTop_1} < m_{\botq} + m_{\W} + m_{\ninoone}$, $\sTop_1$ is expected to decay to c\ninoone. This case is best tested
with a monojet analysis which can contain a charm-tag as well, which is not yet published. The results of top squark searches are discussed 
in Sect.~\ref{sec:stop}.

\paragraph{}
Searches for the decay of the bottom squark via $\sbone \rightarrow \botq \ninoone$, are usually performed in zero-lepton analyses
requiring two \botq-tagged jets, as no prompt leptons are expected from \botq decays. If $m_{\sBot_1} > m_{\topq} + m_{\ninoone}$, the decay 
$\sBot_1 \rightarrow \topq \chimone$, with $\chimone \rightarrow \W^- \ninoone$, is open as well, and searches with lepton signatures are 
again 
advisable, e.g. a same-sign dilepton search is well suited due to the low SM background in this channel.
For \sBot decays to a bottom quark and \ninotwo, the \ninotwo can decay with a certain probability to a \Z or Higgs boson and \ninoone. The 
additional boson could be tagged to further reduce the background. The results of bottom squark searches are discussed in Sect.~\ref{sec:sbottom}.

\subsubsection{Search for direct top squark production}
\label{sec:stop}

The first search~\cite{CMS_Stop_1l} discussed here is focused on the direct production of two top squarks, with two possible decay 
modes of the top squark: $\sTop \rightarrow \topq \ninoone$ and $\sTop \rightarrow \botq \chipmone$, with $\chipmone \rightarrow
\Wpm \ninoone$, for which a one-lepton final state is a favorable final state.

The dominant \ttbar background, where both \W bosons decay leptonically, 
is reduced by the quantity \MTtW, which is defined as the minimum mother particle mass compatible with
all transverse momenta and mass-shell constraints~\cite{Bai:2012gs}. This variable is similar to the \MTt (cf. Sect.~\ref{sec:discri}). \MTtW 
has by construction an endpoint at the top-quark mass in case of a
di-leptonic top-quark decay, where one lepton is not identified or lies outside the acceptance of the analysis. In the search for 
the decay $\sTop \rightarrow \topq \ninoone$, the dilepton \ttbar background is further suppressed by requiring that three 
of the jets in the event are consistent with the $\topq \rightarrow \botq \W \rightarrow \botq \quark \aQuark$ decay chain.

\begin{figure}[htbp]
\begin{center}
\includegraphics[width=\linewidth]{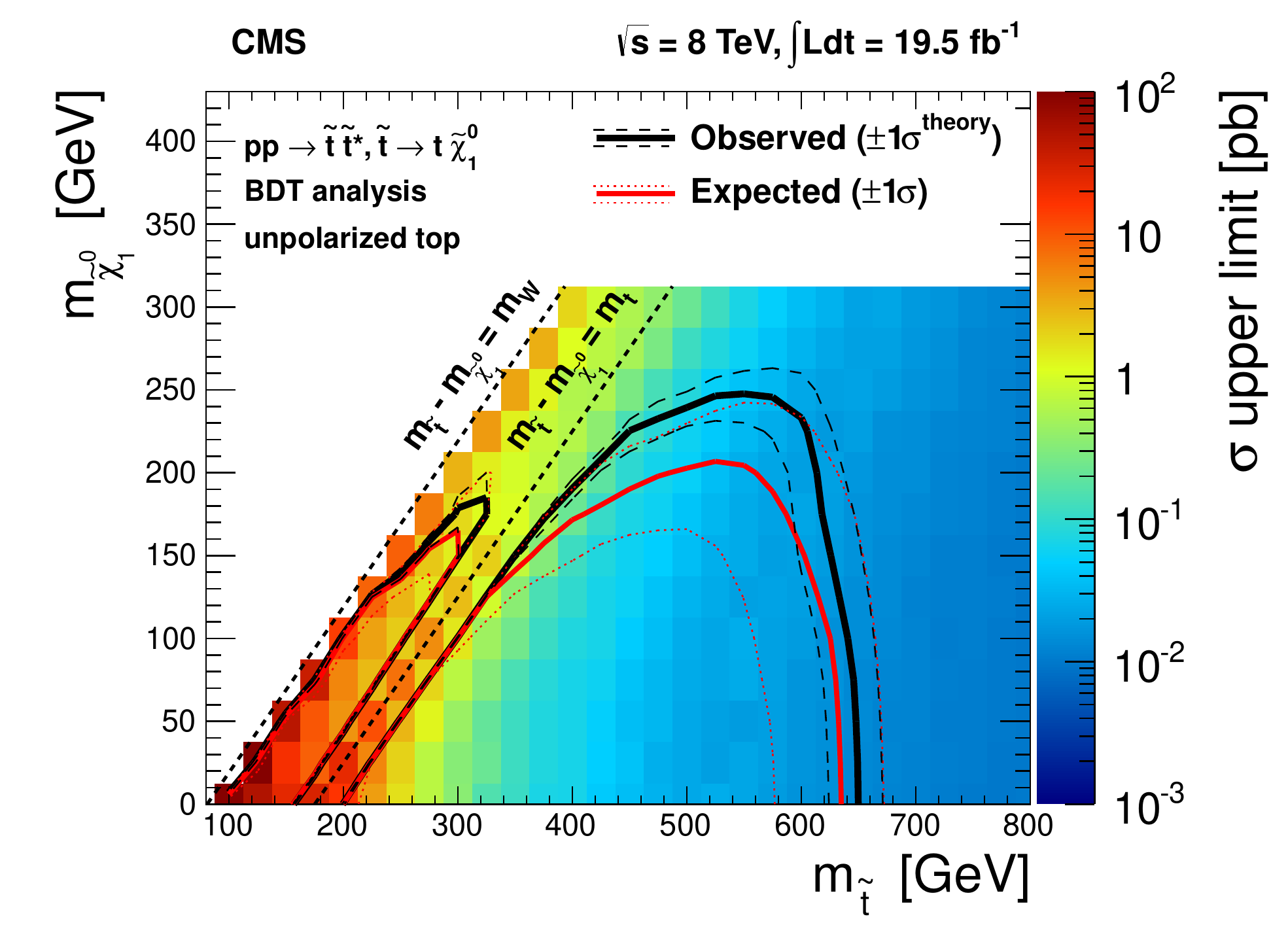}
\end{center}
\caption{Exclusion limits at 95\% CL on the top squark pair production in the top squark-LSP mass plane for a simplified model assuming 
$\sTop \rightarrow \topq \ninoone$~\protect\cite{CMS_Stop_1l}. All other sparticles are decoupled.}
\label{fig:SUSY_stop_1l1}
\end{figure}

The exclusion limits are shown in Fig.~\ref{fig:SUSY_stop_1l1} for the simplified model describing
 $\sTop \rightarrow \topq \ninoone$. Here the top is unpolarized, where a maximum limit of $620\GeV$ 
for the top squark mass and of $225\GeV$
for the \ninoone mass can be set. In case of 100\% right-handed tops one would expect leptons with larger \pt leading to a larger acceptance, 
and hence to an extension of the limit at high masses by 25--50\GeV. Accordingly, a 100\% left-handed top would reduce the limit by the same
amount. Also, one has to take into account that the simplified model assumes a probably too optimistic branching fraction of 100\%. 
If the branching fraction would be reduced to 60\% with no possibility to detect other decay chains, 
the excluded limit would drop to $m_{\sTop} < 500\GeV$ and $m_{\ninoone} < 125\GeV$.

\begin{figure}[htbp]
\begin{center}
\includegraphics[width=\linewidth]{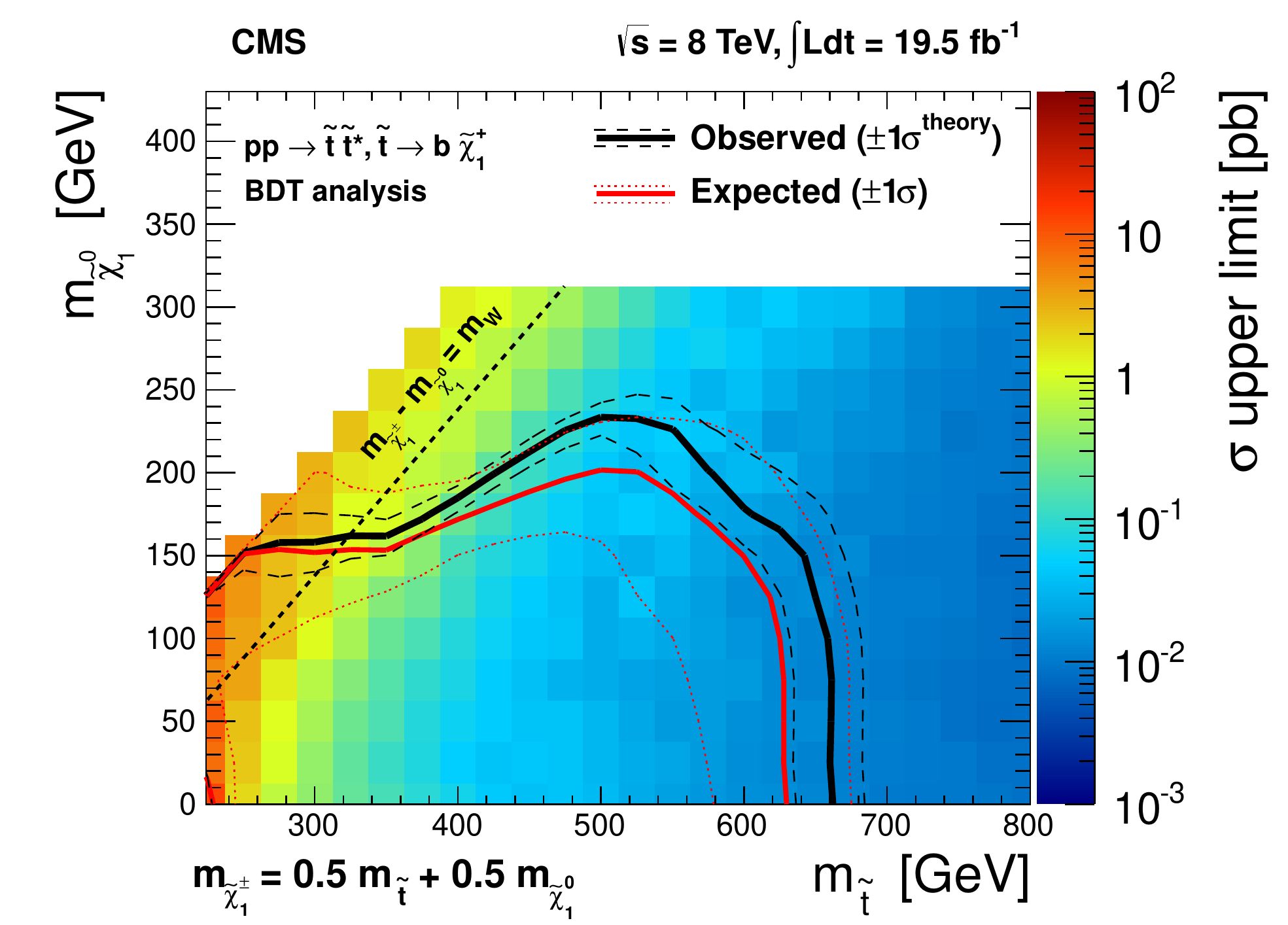}
\end{center}
\caption{Exclusion limits at 95\% CL on the top squark pair production in the top squark-LSP mass plane for a simplified model assuming $\sTop \rightarrow \botq \chipmone$, 
with $\chipmone \rightarrow \Wpm \ninoone$. The \chipmone mass parameter is $x=0.5$ (see text)~\protect\cite{CMS_Stop_1l}. All other sparticles are decoupled.}
\label{fig:SUSY_stop_1l2}
\end{figure}

Effects of similar size are also observed for the simplified model describing $\sTop \rightarrow \botq \chipmone$, with $\chipmone \rightarrow
\Wpm \ninoone$.
Fig.~\ref{fig:SUSY_stop_1l2} 
shows this model for the unpolarized chargino, left-right symmetric $\W \ninoone \chipmone$ coupling, and the mass parameter
of the \chipmone set to $m_{\chipmone} = x m_{\sTop} + (1-x) m_{\ninoone}$ with $x=0.5$. For a larger mass parameter (with $x=0.75$) the excluded top 
squark and \ninoone mass is shifted up by about 25--50\GeV, while for a lower mass parameter ($x=0.25$) the limit becomes slightly weaker.
For right-handed charginos and right-handed $\W \ninoone \chipmone$ couplings the limit is up to 50\GeV stronger, while it is 
weaker for other combinations of polarizations of chargino and $\W \ninoone \chipmone$ couplings.
Searches for top squarks can also be performed in 2-lepton final states \cite{ATLAS_stop_2l} and leads to slightly less stringent results, 
which are compatible with the single-lepton search results.

\paragraph{}
In summary, top squarks around 600\GeV can be excluded for \ninoone masses lower than 200\GeV, assuming that \chipmone and \ninoone 
are the only lighter SUSY particles, in which case the branching fractions for the investigated decays are large. This result excludes for the 
first time a large part of the phase space allowed by naturalness. Here it should be mentioned that the limit is not so strong for models with
rich EWKino and slepton spectra with masses below the top squark mass. In such models many more decay chains may open up, leading partly to very 
soft objects in the final state and therefore deteriorating the acceptance and hence the achievable limit.

\subsubsection{Search for direct bottom squark production}
\label{sec:sbottom}

\begin{figure}[htbp]
\begin{center}
\includegraphics[width=\linewidth]{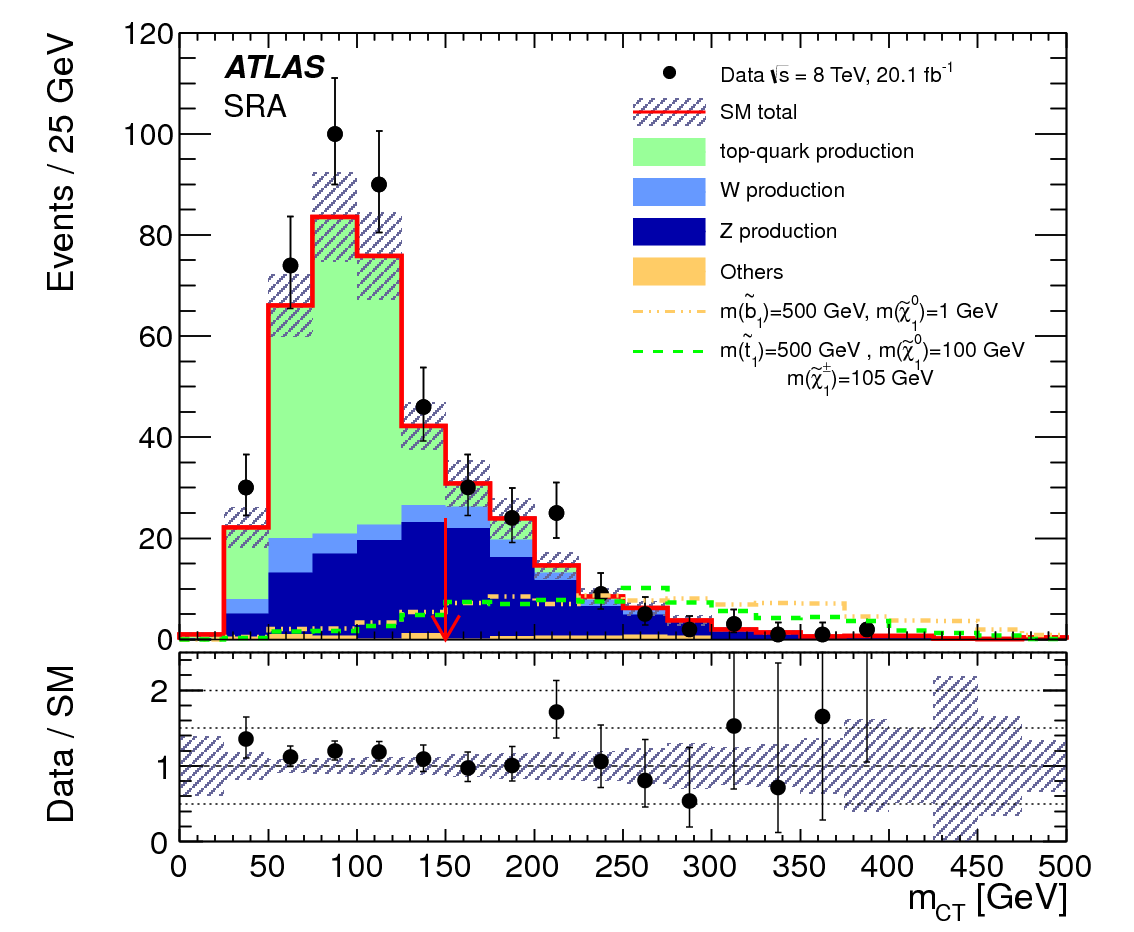}
\end{center}
\caption{The \mCT distribution with all selection criteria applied except for the \mCT thresholds~\protect\cite{ATLAS_Sbottom_2b}.}
\label{fig:SUSY_sbottom_0l1}
\end{figure}

Assuming $\sBot_1 \rightarrow \botq \ninoone$, the search for direct bottom squark production can be performed by requiring a lepton-veto, two \botq-tagged jets 
and a moderate amount of missing energy due to the LSPs ($\MET > 150\GeV$)~\cite{ATLAS_Sbottom_2b}. Further discriminating variables are the 
minimum angle \DP between the \MET vector and either of the three highest-\pt jets, which is expected to be larger for signal than for background from 
multi-jet events, and a requirement on the contransverse mass \mCT (cf. Sect.~\ref{sec:discri}), which is displayed in Fig.~\ref{fig:SUSY_sbottom_0l1}.

\begin{figure}[htbp]
\begin{center}
\includegraphics[width=\linewidth]{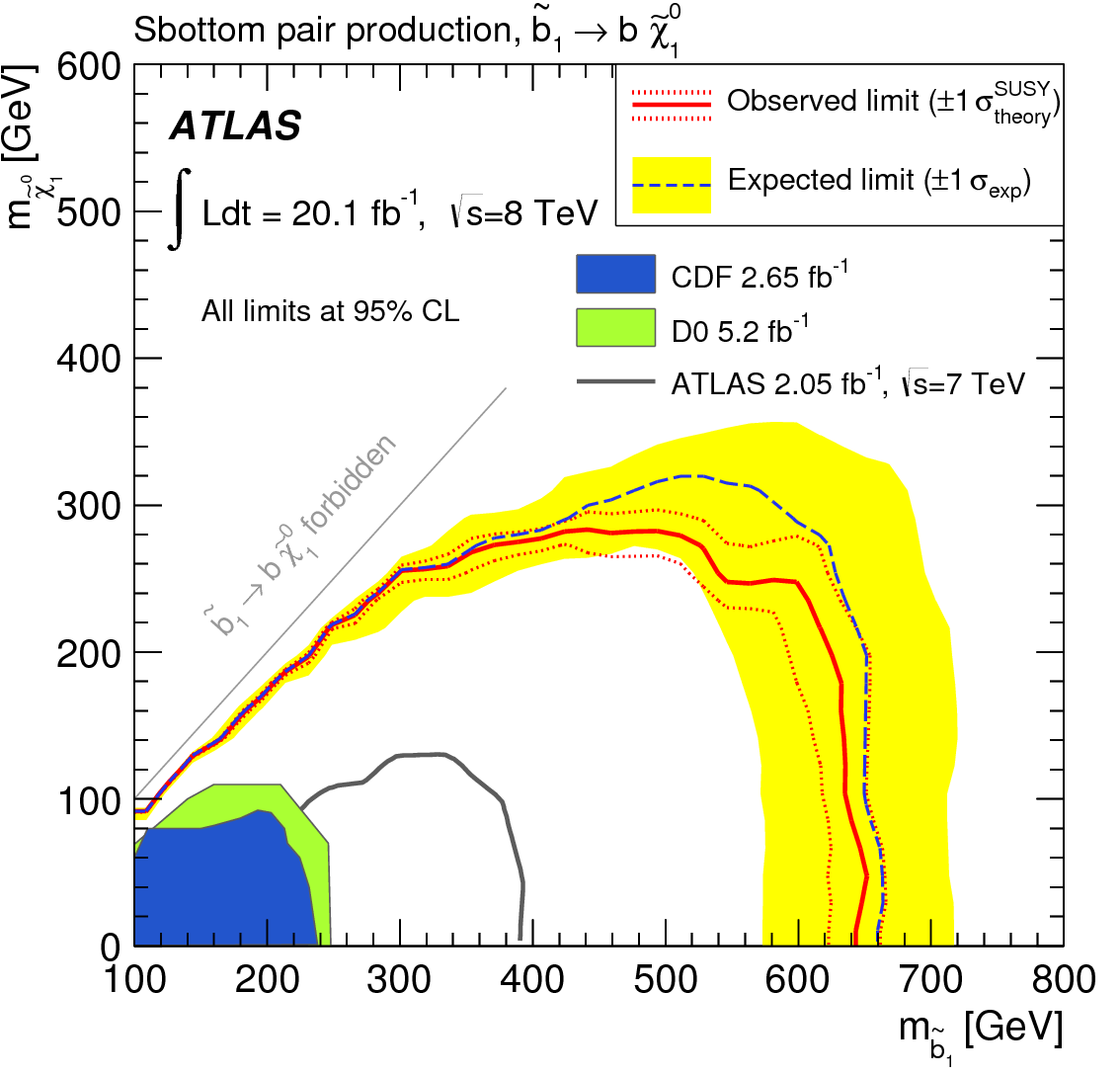}
\end{center}
\caption{Exclusion limits at 95\% CL on the bottom squark pair production in the bottom squark-LSP mass plane for a simplified model assuming 
$\sBot \rightarrow \botq \ninoone$~\protect\cite{ATLAS_Sbottom_2b}. All other sparticles are decoupled.}
\label{fig:SUSY_sbottom_0l2}
\end{figure}

Fig.~\ref{fig:SUSY_sbottom_0l2} shows the exclusion limit for direct bottom squark pair production. Bottom squark masses 
up to 620\GeV and \ninoone masses up to 260\GeV are excluded at 95\% CL. Up to bottom squark masses of 300\GeV, mass differences of at least 50\GeV
between \sBot and \ninoone can be excluded. Again, these limits correspond to a branching fraction of 100\%. If a lower 
branching fraction of 60\% and no possibility to detect other decay chains is assumed, the limit drops to $m_{\sBot} < 520\GeV$ and 
$m_{\ninoone} < 150\GeV$. Note that this search is also sensitive to the direct production of top squarks with subsequent decay 
to $\sTop \rightarrow \botq \chipmone$, with $\chipmone \rightarrow \Wpm \ninoone$. The sensitivity of this search is comparable to 
the one-lepton search discussed above for small chargino-neutralino mass differences of a few GeV, where the leptons are too soft to be detected.

The results of the same-sign dilepton search~\cite{CMS_Strong_2lSS}, discussed in Sect.~\ref{sec:LepTonic}, can also be
interpreted in terms of a direct bottom squark search, where the bottom squarks are pair-produced and then each decay as 
$\sBot \rightarrow \topq \chimone$, with $\chimone \rightarrow \W^- \ninoone$ (and charge conjugate, respectively). Here, same-sign
dileptons can originate from leptonic top-quark and \W-boson decays. Bottom squark 
masses below 500\GeV can be excluded for chargino masses up to 350\GeV and neutralino masses up to 150\GeV, if $m_{\ninoone}/m_{\chipmone} = 0.5$.
Slightly higher neutralino masses, up to 180\GeV can be excluded for $m_{\ninoone}/m_{\chipmone} = 0.8$.

In summary, directly produced bottom squark have been searched for the first time in the region predicted by naturalness. 
Bottom squark masses below 600\GeV up to \ninoone masses of 250\GeV can be excluded assuming large branching fractions to the 
examined final states.

\subsection{SUSY models with \gravitino as LSP}
\label{sec:stop_grav}

\begin{figure}[htbp]
\begin{center}
\includegraphics[width=\linewidth]{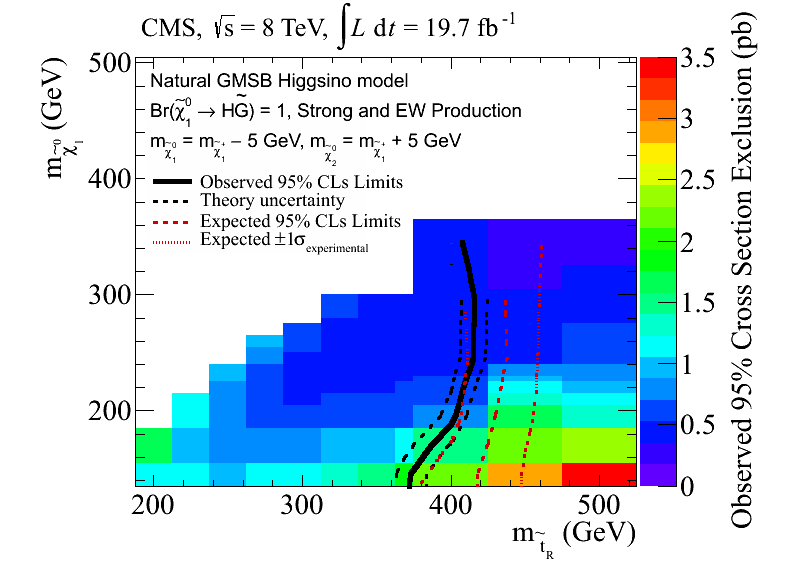}
\end{center}
\caption{Exclusion limits at 95\% CL on the top squark pair production in the top squark-LSP mass plane for a simplified model assuming 
$\sTop_{\mathrm R} \rightarrow \chipone \botq$. Further assumptions are that \chipone is higgsino-like, \ninoone is 
the NLSP decaying as $\ninoone \rightarrow \HO \gravitino$ and \gravitino is the LSP~\protect\cite{CMS_Stop_GMSB}.}
\label{fig:stop_GMSB1}
\end{figure}

GMSB models where only the $\sTop_{\mathrm R}$ and the higgsinos are accessible~\cite{CMS_Stop_GMSB} provide a perfect example of a natural scenario. 
The lightest chargino and the two lighter neutralinos are almost pure higgsinos and therefore nearly mass-degenerate which correspond to 
scenario (c) of Fig.~\ref{fig:SUSY_EWKinos}. The following decay is therefore considered $\sTop_{\mathrm R} \rightarrow \chipone \botq 
\rightarrow f f' \ninoone \botq \rightarrow f f' \HO (\rightarrow \gamma \gamma, \bbbar) \gravitino \botq$ where f and f$'$ are low-energetic quarks 
or leptons. SM background events are suppressed by requiring the invariant mass of two photons to be within the Higgs mass window, exploiting the 
sidebands for the background estimation. 
In addition, two \botq-tagged jets are required, originating from the top squark decay. As shown in Fig.~\ref{fig:stop_GMSB1}, top squark mass below 
360 to 410\GeV are excluded, depending on the higgsino mass. Note that the direct $\chipone\chimone$ production channel can have a similar final state 
when \botq-tagged jets are coming from one of the \HO. Considering this decay therefore increases the sensitivity for low \ninoone mass and top squark mass 
above 300 \GeV, where the $\chipone\chimone$ production cross section dominates over direct top squark production.

\begin{figure}[htbp]
\begin{center}
\includegraphics[width=\linewidth]{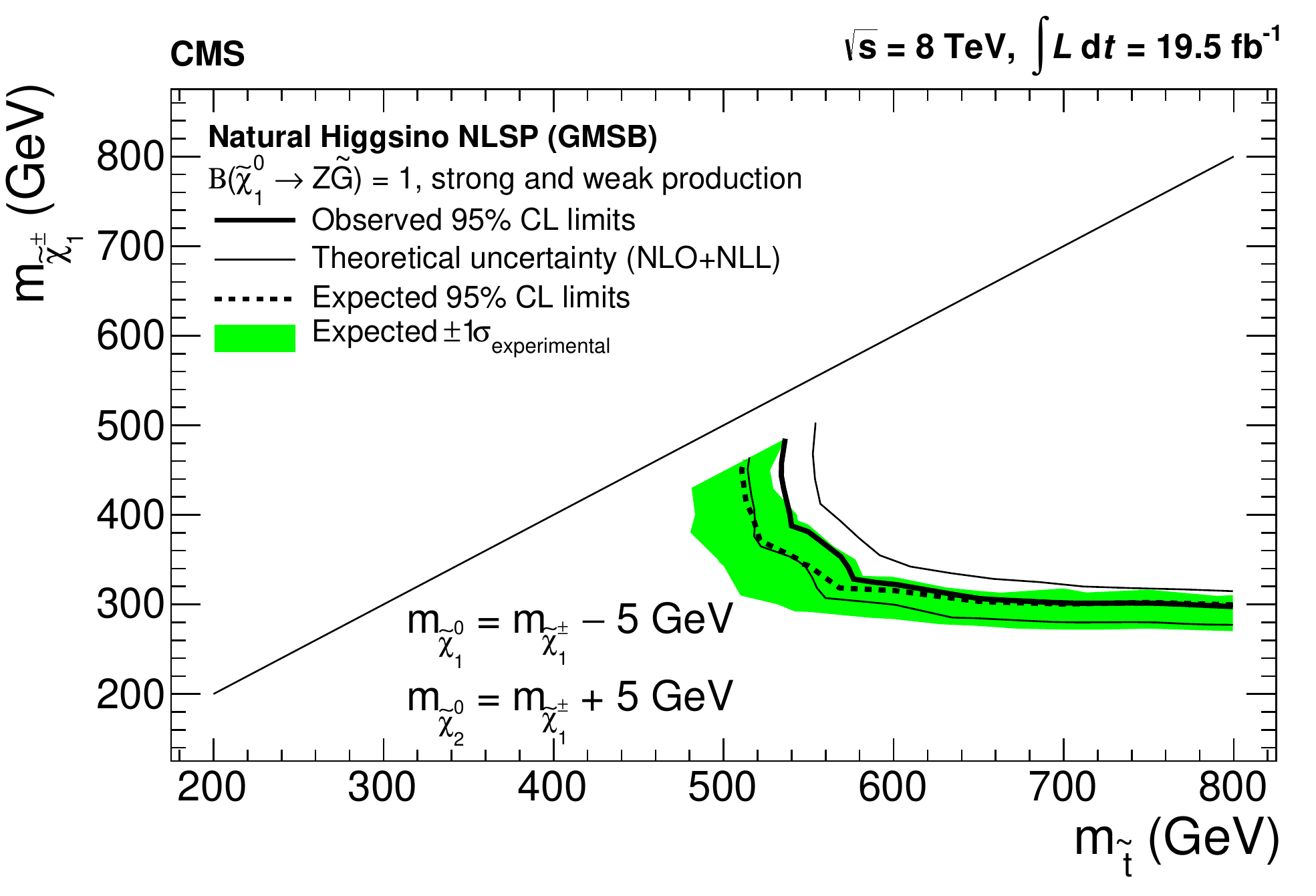}
\end{center}
\caption{Exclusion limits at 95\% CL on the top squark pair production in the top squark-LSP mass plane for a simplified model assuming 
$\sTop_{\mathrm 1} \rightarrow \chipone \botq$. Further assumptions are that \chipone is higgsino-like, \ninoone is the NLSP decaying as 
$\ninoone \rightarrow \Z \gravitino$ and \gravitino is the LSP~\protect\cite{CMS_3l_GMSB}.}
\label{fig:stop_GMSB2}
\end{figure}

For the top squark decay considered, other final states can occur as well. First the two Higgs bosons could decay as $\HO \rightarrow \Z \Z \rightarrow \lep \lep \lep \lep$ 
instead of  $\gamma \gamma$/\bbbar. Second the lightest neutralino could also decay as $\ninoone \rightarrow \Z (\rightarrow \lep \lep) \gravitino$ giving $\Z\HO$ or 
even $\Z\Z$ final states. In all of these cases, the multilepton analysis~\cite{CMS_3l_GMSB}, described in Sect.~\ref{sec:Strong_gravitino}, is particularly sensitive.
Models with a branching ratio of 100\% for $\ninoone \rightarrow \Z \gravitino$ and a branching ratio of 50\% for each \ninoone decay are considered. 
As shown in Fig.~\ref{fig:stop_GMSB2}, for a branching ratio of 100\% to Z bosons and \gravitino, top squark masses below 510\GeV 
can be excluded for \chipmone masses of up to 450\GeV. The limits for the other two cases are weaker. 

In summary, the mass reach of GMSB top squark searches is about 200\GeV weaker than for models with \ninoone as LSP. No dedicated bottom squark search in the GMSB model is 
published yet, but similar limit should be obtained since the final state $\sBot_{\mathrm 1} \rightarrow \ninoone \botq$ is very similar to that of the top squark 
($\sTop_{\mathrm 1} \rightarrow \chipone \botq \rightarrow f f' \ninoone$) when the low-energetic fermions f and f$'$ are not reconstructed.

\section{ElectroWeak SUSY sector}
\label{sec:SUSY_EWK}

In natural SUSY, many weakly interacting particles are expected to be close to the EW scale. Searches for neutral and charged Higgses 
with positive $R$-parity are presented in Sect.~\ref{sec:SUSYHiggs}. Searches for partners of the Higgses and 
electroweak gauge bosons, called electroweakinos (EWKinos) are discussed in Sect.~\ref{sec:EWKinos}. Finally limits on 
sleptons are discussed in Sect.~\ref{sec:Sleptons}.

\subsection{SUSY Higgses}
\label{sec:SUSYHiggs}

As already mentioned in Sect.~\ref{sec:SUSY_LHC}, the Higgs boson discovered is assumed to be the lightest neutral Higgs of the MSSM (h$^0$). 
It is worth to note that a value of 126\GeV is close to the upper mass bound possible for h$^0$ in MSSM and requires O(1\TeV) top squark mass or a 
fine-tuned value of top squark mixing. This creates a tension with the natural SUSY spectrum where top squark mass should be 600\GeV maximum. Results 
of searches are extensively discussed in~\cite{SUSY_Higgses} of this review. This section therefore proposes only a short summary.

Extra neutral and charged Higgses, which preferentially couple to the most massive down-type fermions, are actively searched. At tree level, their masses only 
depend on $\mathrm{tan}\beta$ and $m_{\text{A}^0}$~\footnote{For charged Higgses $m_{\text{H}^{\pm}}^2=m_\text{A}^2+m_{\W}^2$ 
at tree level. Other SUSY parameters enter via radiative corrections and are fixed to particular benchmark values, chosen to exhibit certain MSSM features.}.
At LHC, neutral Higgses are produced singly or accompanied by \botq-jet(s) and decay 
via $\tau^+ \tau^-$, $\bbbar$ and more marginally $\mu^+ \mu^-$ final states. Charged Higgses with lower masses than the top quark will 
predominantly appear in the top decay via t $\rightarrow$ bH$^{\pm}$. When charged Higgses have higher masses than the top quark, they will be 
produced in association with top and bottom quarks. In both cases, they mainly decay via H$^{\pm} \rightarrow \tau^{\pm} \nu$. Results from searches favor neutral and 
charged SUSY Higgses with masses higher than h$^0$, even if no model-independent limits exist yet.

It worth to mention that more intricate searches are also investigated. For example, searches for a topology in which a H$^{0}$ decays via a cascade of lighter
charged and neutral Higgs bosons~\footnote{H$^0 \rightarrow \W^{\mp} $H$^{\pm} \rightarrow \W^{\mp} \Wpm \HO \rightarrow \W^{\mp} \Wpm \bbbar$} have been performed 
by ATLAS~\cite{ATLAS_HiggsCascade}.

\subsection{EWKinos}
\label{sec:EWKinos}

As discussed in Sect.~\ref{sec:SUSY_LHC}, neutralino and chargino masses are obtained by mixing gauge eigenstates. The sensitivity to the three typical scenarios 
shown in  Fig.~\ref{fig:SUSY_EWKinos} and corresponding to (a) bino-like, (b) wino-like and (c) higgsino-like \ninoone 
are now reviewed both for $\ninoone$ (Sect.~\ref{sec:EWK_Chi10}) and the gravitino (Sect.~\ref{sec:EWK_gravitino}) being the LSP. 

\subsubsection{SUSY models with \ninoone as LSP}
\label{sec:EWK_Chi10}

At LHC, most efforts concentrate on processes involving the two lightest neutralinos ($\ninoone$, $\ninotwo$) and the lightest chargino ($\chipmone$). 
Assuming that the EWKinos are the lightest sparticles of the spectrum (Fig.~\ref{fig:SUSYTheo2}), the main production occurs via the s-channel exchange of
a virtual gauge boson. EWKinos then naturally decay as $\ninotwo \rightarrow \Z/\text{h}^{0(*)} \ninoone$ 
and $\chipmone \rightarrow \W^{(*)} \ninoone$. Given the low values of cross sections compared to SM backgrounds, searches are conducted most of the times 
for leptonic decays of \Z and \W, giving 1-4 leptons+\MET final states. Note that an excess in these channels could well be the only SUSY signal at LHC 
if colored sparticles are too heavy or decay through intricate chains.

At LHC, scenario (a) with bino-like $\ninoone$ and wino-dominated $\ninotwo$ and $\chipmone$ is the most favorable scenario 
especially when the mass difference between $\ninotwo$ and $\chipmone$ and the LSP allows for on-shell \Z, \HO and \W. 
The highest cross section is coming from $\chipmone \ninotwo$ production, covered by a 3-lepton+$\MET$ search. Most sensitive signal regions require the three leptons to 
be electrons or muons (the leading one should have $\pt>25\GeV$) and the invariant mass of the two same-flavor opposite-sign leptons ($m_{\text{SFOS}}$) to be close 
to the \Z-boson mass. Further discrimination is obtained by selecting ranges of $\MET$ and $\MT$ (formed with $\MET$ and the lepton not forming the SFOS lepton pair).
The search sensitivity is driven by the ability to reduce and control the $\W\Z$ background. Assuming mass degeneracy between $\chipmone$ and $\ninotwo$, stringent limits 
are obtained on EWKinos : $m_{\ninotwo,\chipmone}<340\GeV$ for LSP masses lower than 70\GeV are excluded~\cite{ATLAS_EWK_3l}~\footnote{Note that 
with no mass degeneracy these upper bounds can be significantly lower.}. Recent efforts were made to cover the case 
where \Z and/or \W are not on-shell and where \HO is present in the decay. This is done by considering bins of $m_{\text{SFOS}}$ outside the \Z-mass and 
requiring the presence of tau-lepton(s). In this case the reducible background coming from jets or photons faking leptons is also of importance and could dominate 
over $\W\Z$, altering the sensitivity to this more compressed EWKino mass spectrum. However, together with the increase in luminosity and $\sqrt{s}$, this provides 
considerable improvement over 7 TeV results, even if the compressed scenario case still has poor sensitivity. 

\begin{figure}[htbp]
\begin{center}
\includegraphics[width=\linewidth]{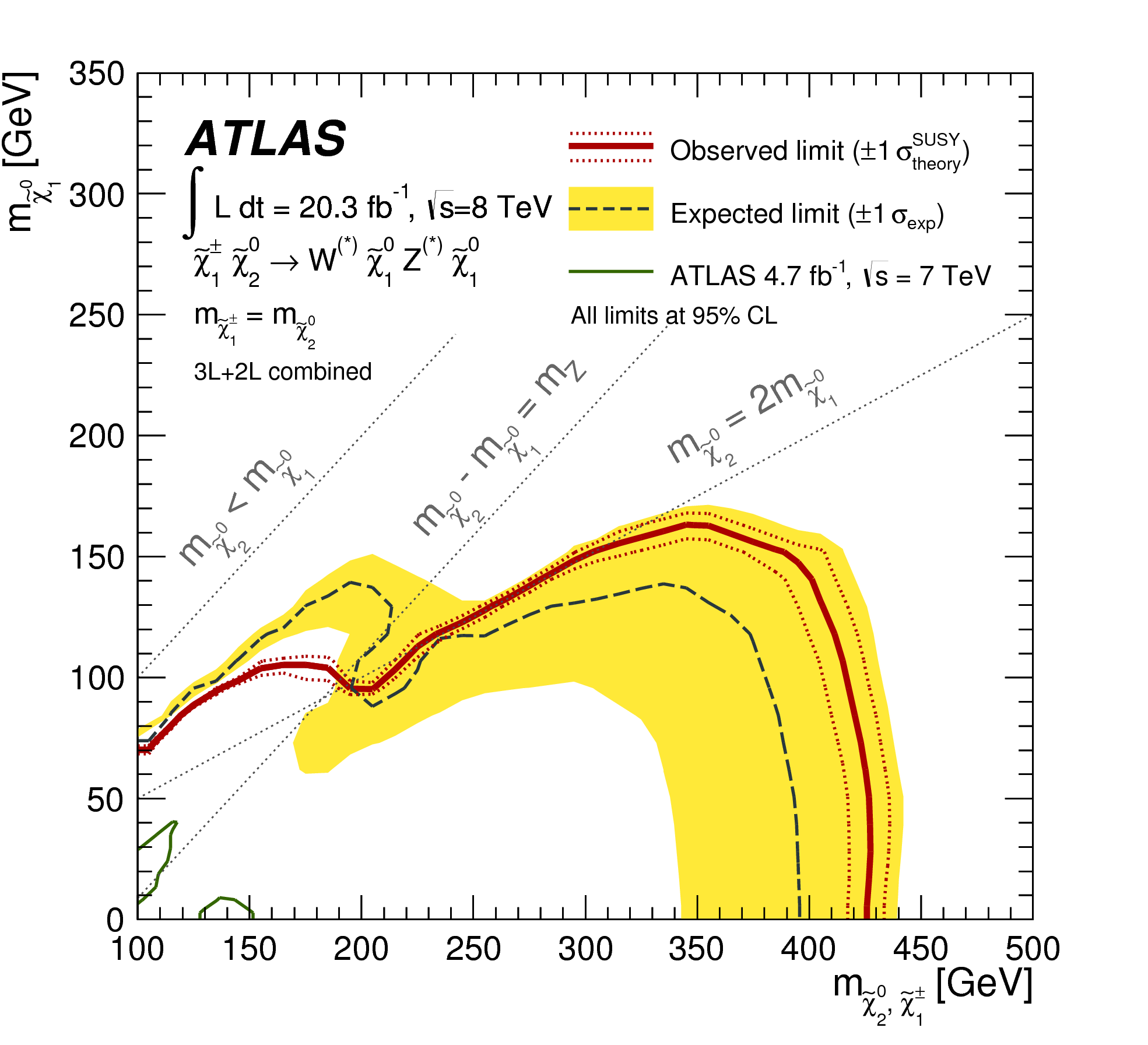}
\end{center}
\caption{Exclusion limits at 95\% CL on the \chipmone \ninotwo production in the \chipmone-LSP mass plane~\protect\cite{ATLAS_EWK_2l} for a simplified model assuming 
\chipmone and \ninotwo are mass degenerate and decay via $\chipmone \rightarrow W^{\pm(*)} \ninoone$ and $\ninotwo \rightarrow Z^{0(*)} \ninoone$. 
All other SUSY particles are decoupled.}
\label{fig:SUSY_EWK}
\end{figure}

The 3-lepton+$\MET$ final state is less favorable for scenario (a) where $M_1 \ll \mu < M_2$ since the $\chipmone\ninotwo$ production cross section is divided by 3 due 
to the lower coupling of the higgsino to SM gauge bosons. Similarly scenario (b) suffers from the too large $\chipmone$-$\ninotwo$ mass difference. Finally, no sensitivity is 
expected from scenario (c) because of the closeness of $\chipmone$, $\ninotwo$ and $\ninoone$, resulting in too soft objects in the final state. To partially recover the 
sensitivity, other channels with 2-lepton+$\MET$ final state targeting the search for $\chipone \chimone \rightarrow \Wp (\rightarrow \lep^+ \nu) \Wm (\rightarrow 
\lep^- \nu)\ninoone \ninoone$ and $\chipone \ninotwo \rightarrow \Wp (\rightarrow \quark \quark') \Zz (\rightarrow \lep^- \lep^+)\ninoone \ninoone$ are being developed~\cite{ATLAS_EWK_2l}. 
In scenario (a) with massless \ninoone, the former excludes chargino masses in the range $120 <m_{\chipmone}<160$ \GeV while the latter extends further the
\chipmone/\ninotwo mass limit to 410 \GeV; see Fig.~\ref{fig:SUSY_EWK}. Other modes like $\chipone \ninoone \rightarrow \Wp (\rightarrow \lep^+ \nu) \ninoone \ninoone$ and $\ninotwo \ninoone \rightarrow 
\Zz (\rightarrow \lep^+ \lep^-) \ninoone \ninoone$ are not yet explored due to the very low cross section and overwhelming inclusive \W and \Z cross section. Finally, 
even in the most favorable scenario, (a), $\ninoone \ninoone$ and $\ninotwo \ninotwo$ productions are heavily suppressed at production level -- O(1~fb) for 100\GeV $\nino$ 
mass -- and cannot be searched for at LHC even in dedicated monojet analyses~\cite{CMS_Monojet,ATLAS_Monojet}. 

\subsubsection{SUSY models with \gravitino as LSP}
\label{sec:EWK_gravitino}

\begin{figure}[htbp]
\begin{center}
\includegraphics[width=\linewidth]{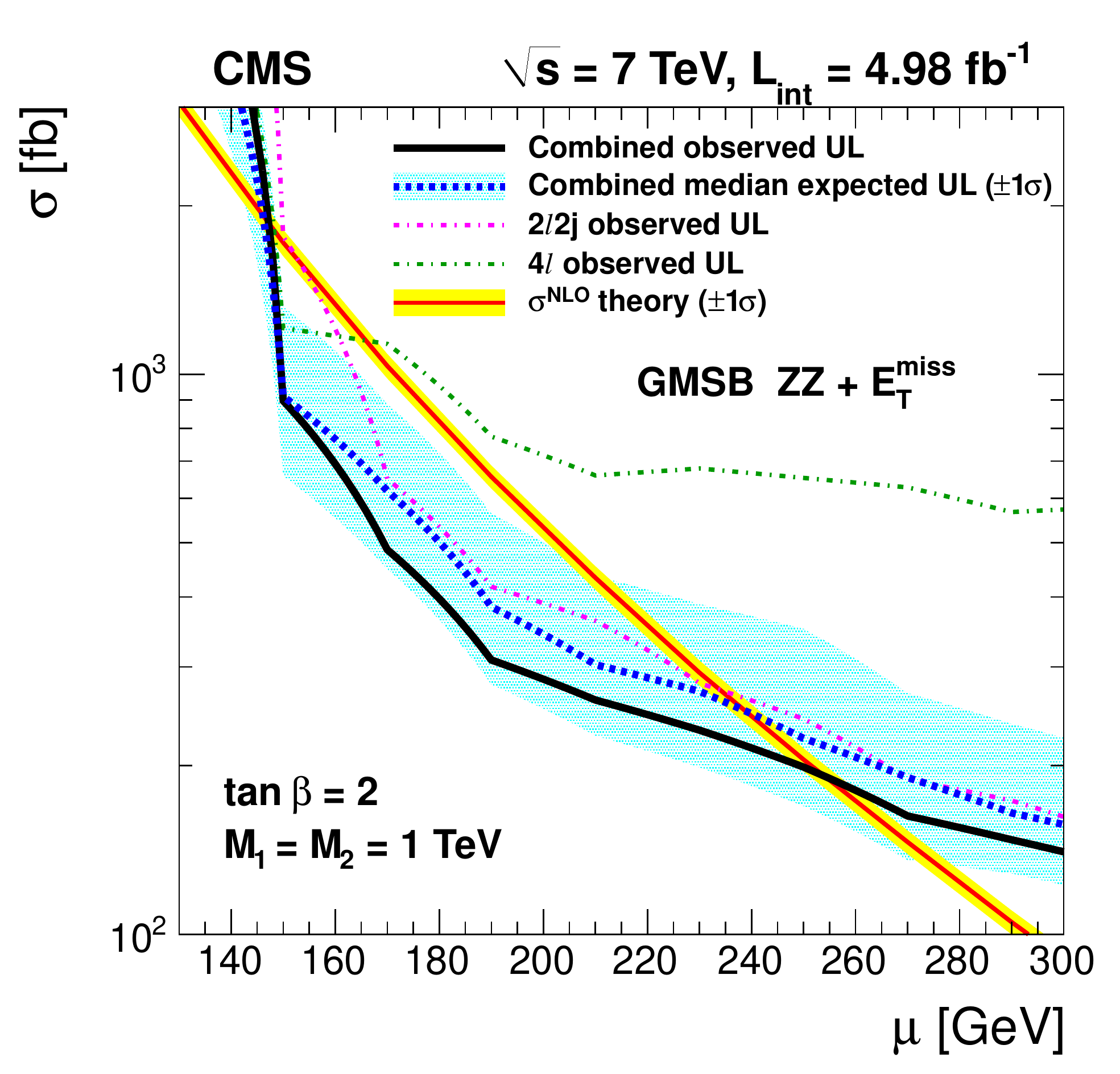}
\end{center}
\caption{Exclusion limits at 95\% CL on the production cross section of higgsino like EWKinos \ninoone, \ninotwo,\chipmone. The decay of the NLSP 
$\ninoone \rightarrow \Z^{0(*)} \gravitino$ is forced and all other SUSY particles are decoupled~\protect\cite{CMS_EWK}.}
\label{fig:SUSY_EWK_GMSB}
\end{figure}

As for the strong production, new final states and search possibilities can emerge when the gravitino is the LSP. 
In scenario (a), the final state will contain two additional photons, reducing drastically the 
background from gauge bosons and therefore increasing the reach in mass. Reinterpreting the two 2-photons+$\MET$ analysis described in Sect.~\ref{sec:Strong_gravitino}, 
and considering $\chipmone \ninoone$ production which have the highest cross section, exclude $m_{\chipmone}<550\GeV$ independently of 
the $\ninoone$ mass~\cite{CMS_2Photon_2011}. Similarly scenario (b) implies 
$\chipmone \rightarrow \Wp \gravitino$ and $\ninoone \rightarrow \gamma/\Z \gravitino$ and it could be searched for in the 1-lepton+1-photon channel, but no 
publication exists yet. Finally, scenario (c) is accessible thanks to the decay $\ninoone \rightarrow \gamma/\Z/\text{h}^0 \gravitino$ where the branching ratios to 
$\gamma$, \Z and h$^0$ depend primarily on $\tan\beta$ and on the mass difference between \ninoone and \gravitino. In case of \Z-rich higgsino (low $\tan\beta$ value 
and positive $\mu$), final states with 4-leptons+$\MET$ or 2-leptons+2-jets will provide interesting sensitivity, as shown in Fig.~\ref{fig:SUSY_EWK_GMSB} for the 2011 LHC
data~\cite{CMS_EWK}. Mixed Z/\HO (higher $\tan\beta$ value) scenarios can be covered by 2leptons+2b+$\MET$ final states. \HO-rich higgsino scenarios (low $\tan\beta$ value and 
negative $\mu$) can be covered by 4b+$\MET$. 

More complicated situations can occur beyond the three scenarios discussed in Fig.~\ref{fig:SUSY_EWKinos}. For example, 
if $M_1$ and $\mu$ are approximately equal and the NLSP is a bino/higgsino admixture, large branching ratios to photons and Higgs bosons are generated. Then the 1b+photon+$\MET$ 
final state, described in Sect.~\ref{sec:Strong_gravitino}, can exclude  
$\ninoone$ masses between 200 and 400\GeV, see Fig.~\ref{fig:StrongGMSB_Limit_Higgsino}.

\paragraph{}
In conclusion, EWKino searches provide presently much weaker constraints on the natural SUSY scenario than the strong production searches. Limits are 
still model-dependent and rely on many assumptions. More results are still 
expected in the near future, and ultimately the limits should be set in the pMSSM to understand better how complete the current searches are. Given that, 
it is fair to say that a complete exploration of this sector is still to come and will greatly benefit from the high luminosity program of LHC.

\subsection{Sleptons}
\label{sec:Sleptons}

Sleptons ($\sEl$, $\sMu$, $\sTau$ and $\sNu$) are governed by 5 parameters: masses of the left-handed and right-handed $\sEl$/$\sMu$, which are assumed to be mass degenerate in the MSSM, 
masses of the left-handed and right-handed staus and the stau mixing angle. \sNu masses can be related to the 
charged slepton parameters. From naturalness arguments O(1\TeV) slepton masses are expected: the very low slepton production cross section, see Fig.~\ref{fig:SUSYxs}, 
will therefore prevent their discovery. However, searching for O(500\GeV) sleptons could be achievable with high luminosity. The reason is that these very low cross 
sections with respect to EWKino production, are largely recovered 
by the more favorable branching ratio BR($\slepp \rightarrow \lep^+ \ninoone$)=100\% compared to leptonic branching ratios of \W and \Z. Lepton searches rely almost 
entirely on the very powerful discriminant variables like \MTt or \mCT and electron(s) or muon(s) in the final states (presently no sensitivity is obtained for staus). 
Like in the direct bottom squark search (Fig.~\ref{fig:SUSY_sbottom_0l1}), the edge of slepton events with $m_{\sLep}-m_{\ninoone}>$O(100\GeV) appears far above the top 
and $\W\W$ background ones. 

\begin{figure}[htbp]
\begin{center}
\includegraphics[width=\linewidth]{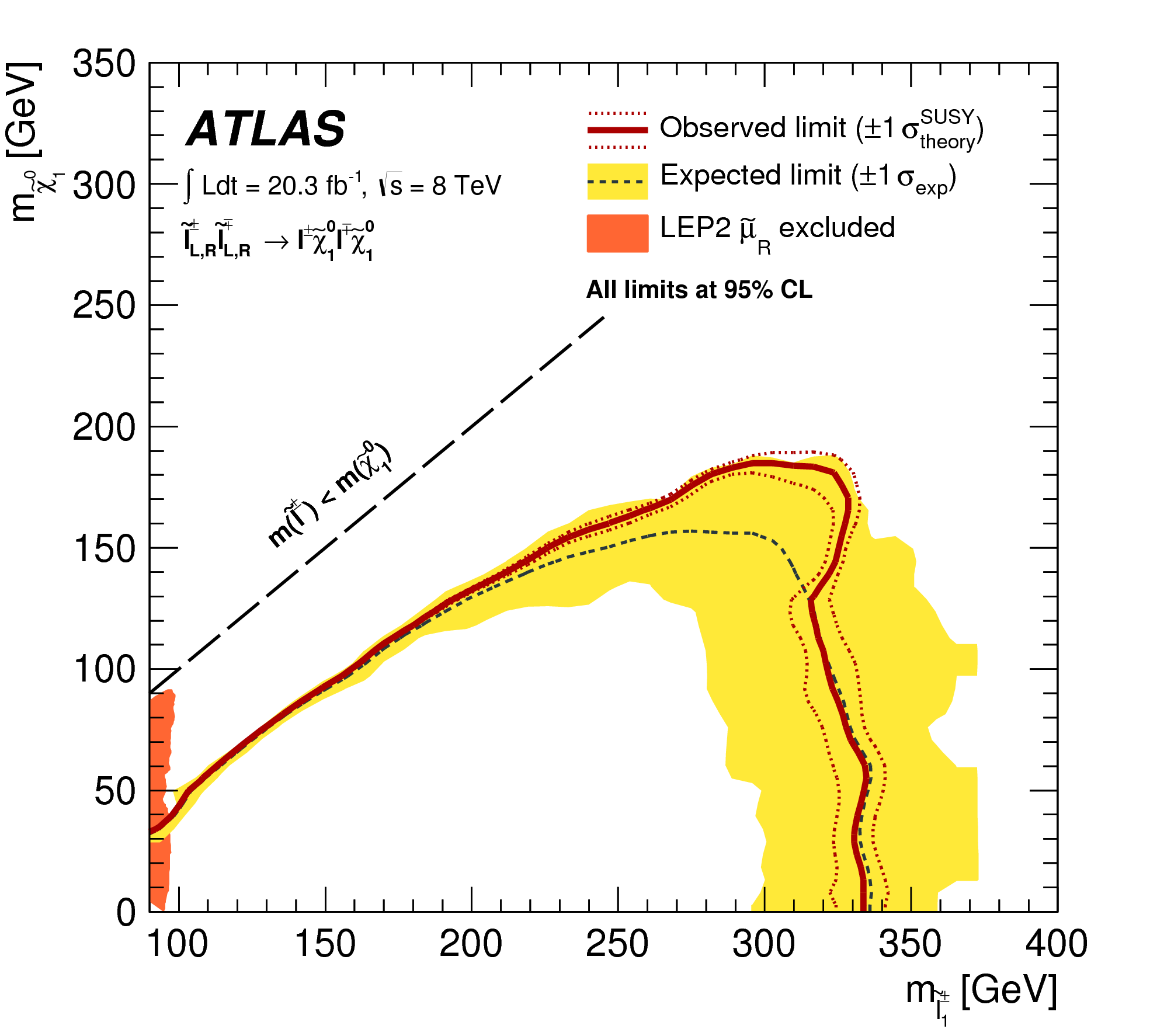}
\end{center}
\caption{Exclusion limits at 95\% CL on the slepton-slepton production in the slepton-LSP mass plane. The slepton decay
$\sLep \rightarrow \lep \ninoone$ is forced and all other SUSY particles are decoupled~\protect\cite{ATLAS_EWK_2l}.}
\label{fig:SUSY_Slepton}
\end{figure}

The most promising final state is 2-lepton$+\MET$, coming from $\slepp \slepm \rightarrow \lep \alep \ninoone \ninoone$. Mass degenerate left- and 
right-handed selectrons and smuons are excluded below 325 \GeV masses for massless neutralinos, largely exceeding the LEP limit, as shown in Fig.~\ref{fig:SUSY_Slepton}~\cite{ATLAS_EWK_2l}. 
Upper limits for left- and right-handed slepton masses are also obtained and give 250 \GeV and 300 \GeV for massless neutralinos.

Direct slepton production can also be searched when the gravitino is the LSP. Various final states can be considered depending on the nature of the NSLP generating final 
states very similar to the EWKino searches. A particular interesting one corresponds to the NSLP slepton scenario where right-handed \sEl/\sMu decay to an 
electron/muon and a right-handed \sTauR gives a tau and a gravitino. This generates a multilepton final state which lepton multiplicity depends on the tau 
decay~\cite{CMS_3l_GMSB}. \sTauR masses can then be excluded up to 200\GeV.

To be complete, a very favorable situation arises if the sleptons are interleaved between $\ninotwo$/$\chipmone$ and $\ninoone$ in scenario (a). Two cases can be envisaged : 
assume that all sleptons are mass-degenerate (a1) and consider that only $\sTau$ and $\sNuTau$ are light (a2). In both 
cases, the EWKinos will couple to the sleptons generating 2-leptons+$\MET$ and 3-lepton+$\MET$ final states with higher $\sigma \times$ BR than in the direct slepton 
or direct EWKino case discussed previously in this section. If the slepton masses are exactly in between $\ninotwo$/$\chipmone$ and $\ninoone$, mass limits reached in 
Fig.~\ref{fig:SUSY_EWK} are increased by a factor 2 to 3 for scenario (a1) and remain similar for scenario (a2)~\cite{ATLAS_EWK_3l}.

\section{Escape routes: $R$-parity violation, long-lived particle searches and beyond MSSM signatures}
\label{sec:RPV_LLP}

Given the absence of signal from plain vanilla MSSM signatures, it is of paramount importance to look at scenarios where 
$R$-parity is violated (Sect.~\ref{sec:RPV}) and/or sparticle decays are not prompt (Sect.~\ref{sec:LLP}). In both 
cases, the stringent limits discussed in Sect.~\ref{sec:StrongProd},~\ref{sec:3rdGene} and~\ref{sec:SUSY_EWK} generally do not apply. Pushing 
further this idea, sensitivity to signatures appearing in scenarios beyond MSSM are also discussed in Sect.~\ref{sec:Oth}.

\subsection{$R$-parity violation searches}
\label{sec:RPV}

$R$-parity conservation implies pair production of the superpartners and requires the lightest superpartners to be
stable, leading to typical missing transverse energy signatures in the detector. If $R$-parity is not conserved these 
constraints do not exist anymore and dedicated searches need to be performed.

In RPV scenarios, the current limits of the proton decay can be met if only either $B$ or $L$ (and not 
both) is violated and the violation is sufficiently small~\cite{RPV-Review}. Such models can also accomodate 
non-zero neutrino masses and neutrino oscillations. The RPV superpotential $W_{\mathrm {RPV}}$ includes 
three trilinear terms parameterized by the 48 Yukawa couplings $\lambda_{ijk}$, $\lambda_{ijk}'$, $\lambda_{ijk}''$:
\begin{equation}
\label{eq:RPV}
W_{\mathrm {RPV}} = \frac{1}{2} \lambda_{ijk} L_i L_j \bar{E_k} + \lambda_{ijk}' L_i Q_j \bar{D_k} + \lambda_{ijk}'' \bar{U_i} \bar{D_j} \bar{D_k} \; ,
\end{equation}
where $i,j,k$ are generation indices, $L$ and $Q$ the $SU(2)_L$ doublet superfields of the leptons and quarks, and 
$\bar{E}$, $\bar{D}$ and $\bar{U}$ the $SU(2)_L$ singlet superfields of the charged leptons and the up- and down-type
quarks.

The nature of the LSP -- which is neutral and colorless in $R$-parity conserving models -- might be different in RPV
models and might be charged and/or carry color as well.

Searches for models with leptonic RPV interactions ($\lambda_{ijk} \neq 0$ or $\lambda'_{ijk} \neq 0$)   
are discussed in Sect.~\ref{sec:RPV-leptonic} and \ref{sec:RPV-semi-lep}, respectively, while the 
quark RPV interactions ($\lambda_{ijk}''  \neq 0 $) are reviewed in Sect.~\ref{sec:RPVmultjet}. Because of present 
constraints on RPV couplings, the values considered are generally in the range O($10^{-2}-10^{-5}$). If the phase space 
for the LSP decay is very small ($\lambda<10^{-5}$), it might also be long-lived. Such cases are covered in Sect.~\ref{sec:LLP}.

\subsubsection{Search for leptonic RPV interactions ($\lambda_{ijk} \neq 0$)}
\label{sec:RPV-leptonic}

With leptonic RPV interactions, multi-lepton final states are expected, which is particularly favorable at LHC where the QCD 
background overwhelmingly dominates. The plethora of models considered is discussed 
in the same order as for the RPC models, starting with strong production, then focussing on the third-generation,
EWKino production and, finally, slepton production. 

\begin{figure}[htbp]
\begin{center}
\includegraphics[width=\linewidth]{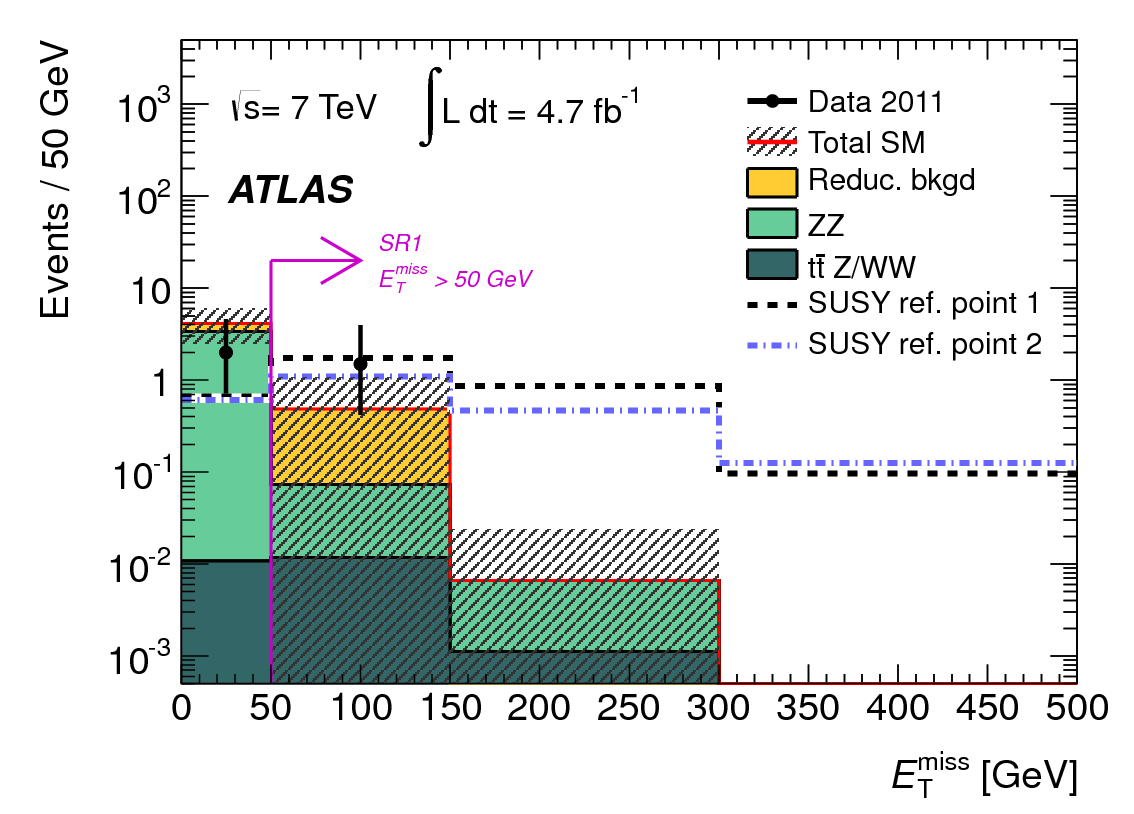}
\end{center}
\caption{\MET distribution for events with at least four leptons and no \Z-boson candidates. 
``SUSY ref. point 1'' is a simplified model point defined by $\m_{\chipmone} = 500\GeV$ and $m_{\ninoone} = 300\GeV$, 
while ``SUSY ref. point 2'' is a MSUGRA/CMSSM model point defined by $m_{1/2} = 860\GeV$ and $\tan \beta = 37$~\protect\cite{ATLAS_RPV_4l_7TeV}.}
\label{fig:RPV-4l}
\end{figure}

\paragraph{}
As first example, a search with four or more leptons (electrons or muons) in the final state is discussed~\cite{ATLAS_RPV_4l_7TeV}. A 
non-zero coupling of $\lambda_{121}$ is chosen as a representative model. To veto low-energy resonances, the invariant mass of any 
opposite-sign same-flavor pair must be above 20\GeV and outside a window around the \Z-boson mass. Two signal regions according to different 
signal scenarios are defined. The first one requires $\MET > 50\GeV$, to be sensitive to models with missing energy originating from neutrinos. 
As an illustration, the \MET distribution is displayed in Fig.~\ref{fig:RPV-4l}. The other one is tuned
to scenarios with a large multiplicity of high-\pt objects originating from heavier sparticles, by requiring 
$\meff > 300\GeV$. 

The strong production case is considered by looking at a full model, taken from Ref.~\cite{Desch:2010gi}, and tested in a MSUGRA/CMSSM 
parameter plane ($m_{1/2}, \tan \beta$), for $m_0$, $A_0$ both zero, $\mu$ positive, and $\lambda_{121}=0.032$ at the unification scale. 
In this model, the $\sTau_1$ is the LSP and decays through a virtual slepton or sneutrino as $\sTau_1 \rightarrow \tau \mathrm{e} \mu \nu_{\mathrm{e}}$ 
or $\sTau_1 \rightarrow \tau \mathrm{e} \mathrm{e} \nu_{\mu}$. Values of $\m_{1/2}$ below 820\GeV are 
excluded for $10 < \tan\beta < 40$. Note, however, that weak processes
contribute to the SUSY pair production, dominating for $m_{1/2}> 600\GeV$. Therefore, 
a corresponding gluino mass below 1\TeV is excluded in this model. 

\begin{figure}[htbp]
\begin{center}
\includegraphics[width=\linewidth]{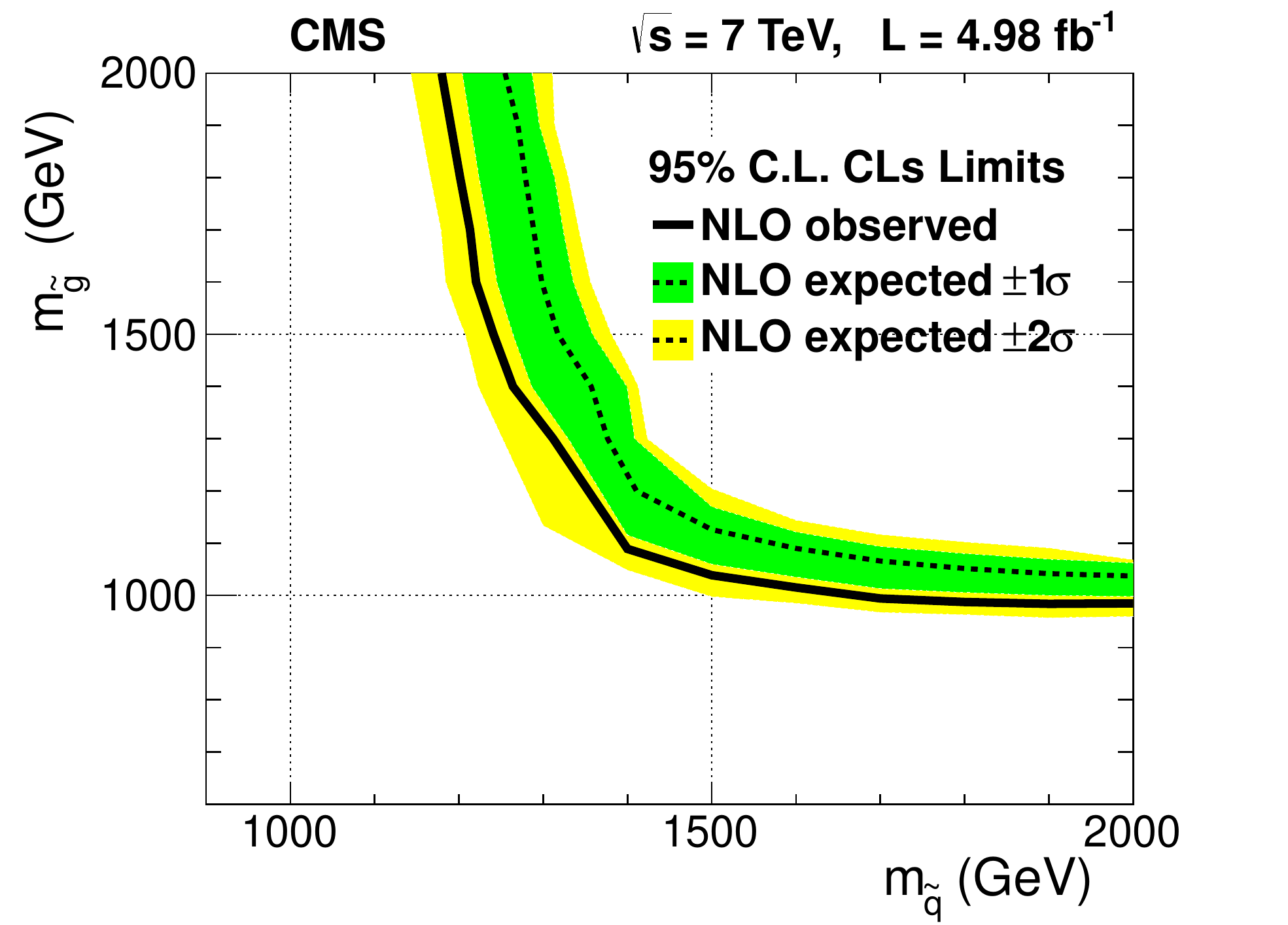}
\end{center}
\caption{Exclusion limits at 95\% CL in the squark and gluino mass plane for a GMSB RPV model with $\lambda_{123}=0.05$~\protect\cite{CMS_RPV_multilepton_7TeV}.}
\label{fig:RPV-CMS}
\end{figure}

Multi-lepton final states can also arise when gravitino is the LSP and all right-handed sleptons are flavor degenerate -- 
known as slepton co-NLSP scenario. Pair-produced gluinos and squarks eventually decay to the 
\ninoone, which further decays to a slepton and a lepton, with the (right-handed) slepton decaying further to another lepton and the 
gravitino, yielding four leptons in the final state. Gluino masses below 1.2\TeV and squark masses below 1\TeV can be excluded
with a 7\TeV analysis~\cite{CMS_RPV_multilepton_7TeV}, assuming an RPV coupling of $\lambda_{123}=0.05$ as shown in Fig.~\ref{fig:RPV-CMS}.

\paragraph{}
Leptonic RPV interactions can also arise in a search for direct top squark production~\cite{CMS_RPV_stop}, 
where each top squark decays to a top quark and an intermediate on- or off-shell bino which decays
further through leptonic RPV interactions ($\lambda_{122}$ or $\lambda_{233}$), via 
$\nino_1^{0*} \rightarrow l_i + \nu_j + l_k$ or $\nino_1^{0*} \rightarrow \nu_i + l_j + l_k$,
where the indices $i,j,k$ refer to the ones in Eq.~(\ref{eq:RPV}).
The signature of direct top squark pair production with RPV decays is different from the one of
RPC models which implies a large amount of missing energy.
In the RPV search, three or more isolated leptons (including hadronic $\tau$ candidates) and one or more \botq-tagged 
jets are required, but only low \MET. Instead $m_{\mathrm{eff}}$ is used as discriminating variable.

The limits are extracted in the bino--top squark mass plane, and found to be approximately independent of the bino mass. 
Stop masses below 1020\GeV can be excluded for bino masses of 200 to 1300\GeV for a non-zero $\lambda_{122}$, 
and top squark masses below 820\GeV for a non-zero $\lambda_{233}$.

\paragraph{}
The results of the 4-lepton+\MET search~\cite{ATLAS_RPV_4l_7TeV}, discussed at the beginning of this section, can also be interpreted in a
simplified model where the lightest chargino and neutralino are the only sparticles with masses below the 
TeV scale. The pair-produced charginos decay each into a \W boson and bino-like \ninoone as in scenario (a) of Fig.~\ref{fig:SUSY_EWKinos}. The LSP
then decays through a virtual slepton or sneutrino as $\ninoone \rightarrow \mathrm{e} \mu \nu_{\mathrm{e}}$ or 
$\ninoone \rightarrow \mathrm{e} \mathrm{e} \nu_{\mu}$ with a branching fraction of 50\% each. The width of the 
\ninoone is fixed to 100\MeV to ensure prompt decays. Choosing the best expected limit for each of the model points, 
with the $\sqrt{s}=7$\TeV data chargino masses up to about 500\GeV are excluded for 
LSP masses between 100 and 540\GeV in the simplified model. 
  
\begin{figure}[htbp]
\begin{center}
\includegraphics[width=\linewidth]{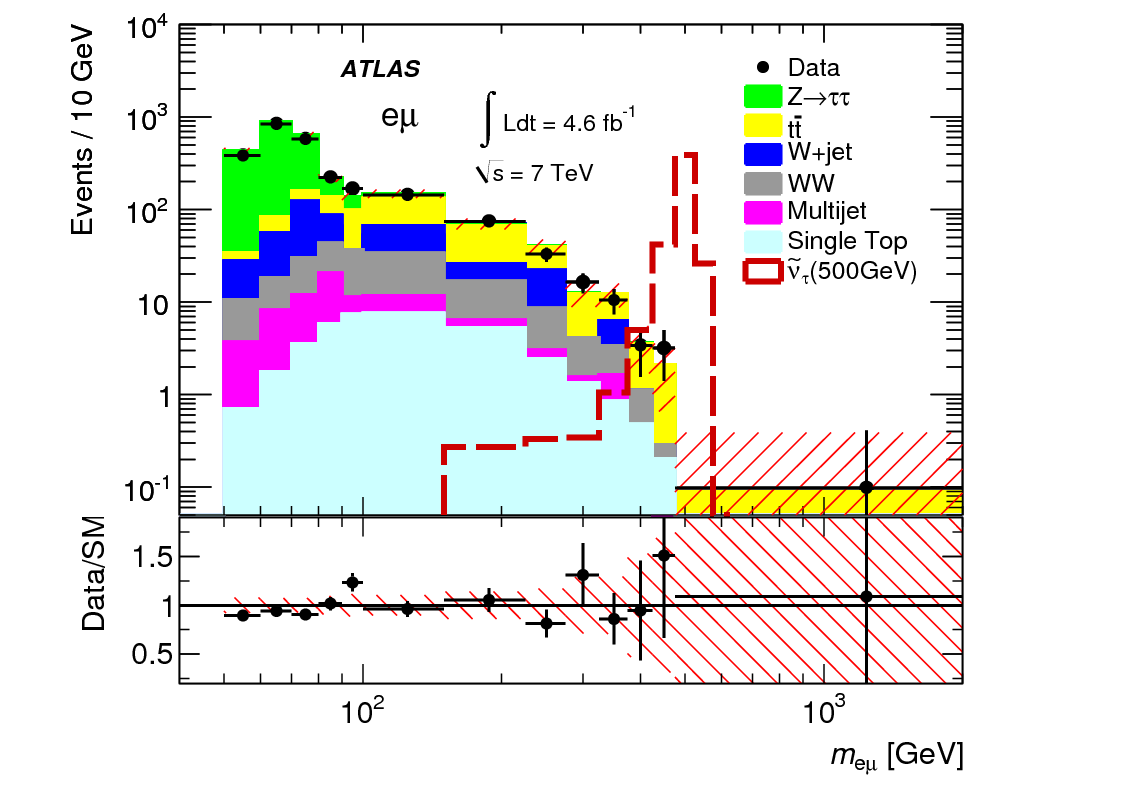}
\end{center}
\caption{Observed and predicted invariant e$\mu$-mass distribution. Signal simulations are shown for sneutrino mass 
of 500\GeV, $\lambda'_{311}=0.11$, and $\lambda_{132}=0.07$~\protect\cite{ATLAS_RPV_sneut}.}
\label{fig:RPV-sneut}
\end{figure}

\paragraph{}
In RPV SUSY, single sneutrinos can be produced via $\lambda_{311}'$ coupling and then decay through $\lambda_{ijk}$ 
couplings to lepton pairs of different flavor.
Searches for such scenarios have been performed in all possible combinations of different-flavor dilepton 
selections~\cite{ATLAS_RPV_sneut}. An example for a possible signal that is 
compatible with current exclusion limits on the strength of the RPV interactions from precision low-energy 
experiments~\cite{Chemtob:2004xr}, is given in Fig.~\ref{fig:RPV-sneut} for the e$\mu$ channel. Sneutrino masses of up to 1.6\TeV are excluded 
in the e$\mu$ selection (for $\lambda_{311}'=0.11$ and
$\lambda_{132} = 0.07$), where the mass resolution is better than in the channels including hadronically decaying 
tau leptons. The latter lead to sneutrino mass exclusion limits of the order of 1.1\TeV for the same
RPV interaction strength.

\paragraph{}
To summarize, as leptonic RPV interactions usually lead to signatures with many leptons, most scenarios are well covered
with the current searches, and often result in sparticle mass limits that are stronger than those of RPC searches.

\subsubsection{Search for semi-leptonic RPV interactions ($\lambda_{ijk}'  \neq 0$)}
\label{sec:RPV-semi-lep}

Such signatures are specifically covered by the HERA experiments~\cite{RPV_ZEUS,RPV_H1}, 
which put stringent limits on the coupling between the first 
and second generation due to the nature of the unique e$^\pm$p accelerator. Therefore, 
LHC searches focus more on the third generation, as detailed below.
 
\paragraph{}
Signatures of models where the top squarks are light, while the other squarks and gluinos are decoupled, 
resemble those of third-generation leptoquarks. Trilinear RPV operators allow
the lepton-number-violating decay
$\sTop_1 \rightarrow \tau \botq$ with a coupling $\lambda'_{333}\neq 0$, resulting in
the same final state as from third-generation leptoquark decay, with similar kinematics. They can be tested by
dedicated searches for $\tau$ and \botq quarks in the final state~\cite{CMS_RPV_stop_taub}. With 7\TeV data, top squarks up to
525\GeV are excluded assuming a simplified model with a branching ratio of 100\% for $\sTop_1 \rightarrow \tau \botq$.

\begin{figure}[htbp]
\begin{center}
\includegraphics[width=\linewidth]{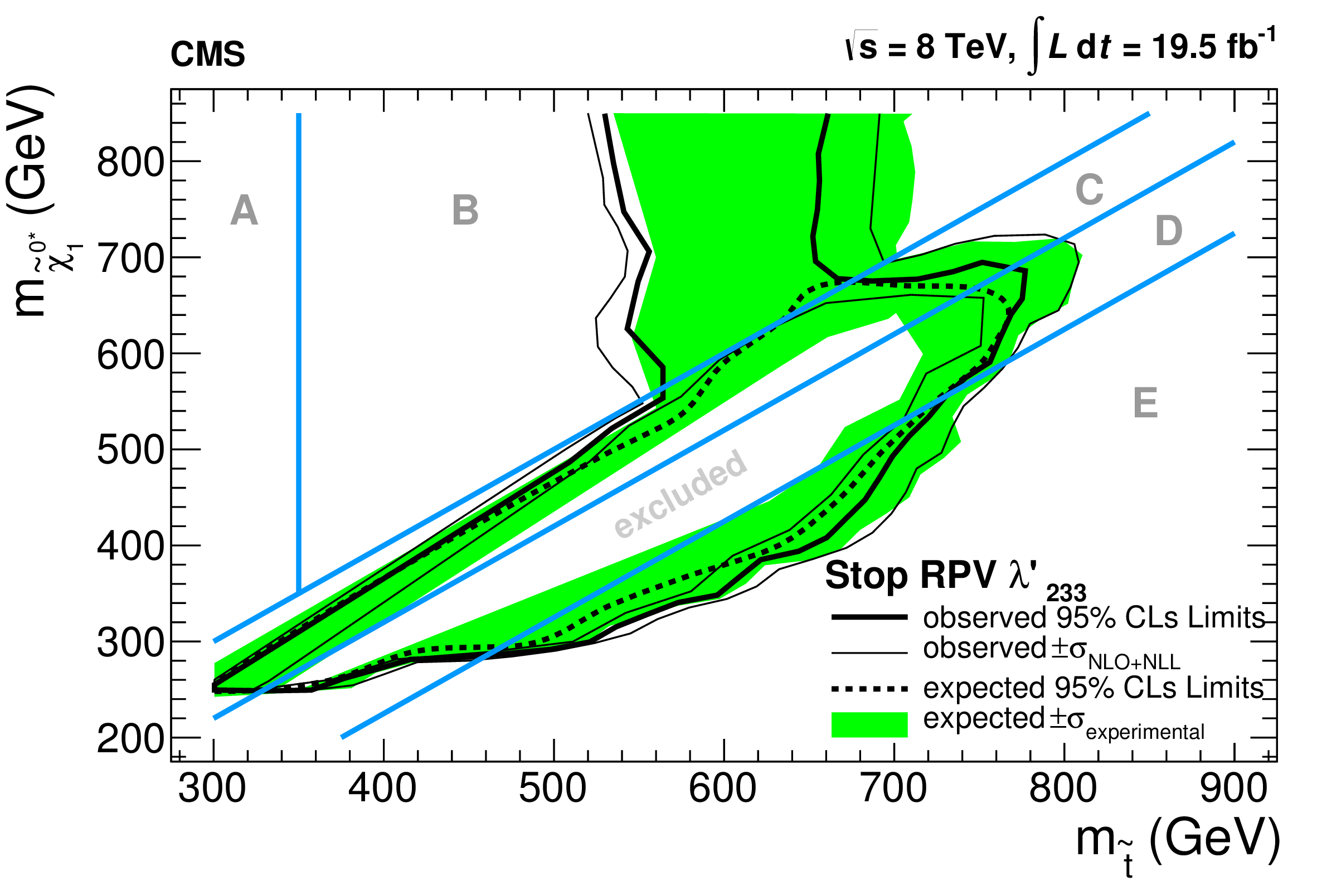}
\end{center}
\caption{Exclusion limits at 95\% CL on the for RPV top squark decay in the top squark-LSP mass plane with non-zero $\lambda_{233}'$ coupling . 
The region inside the curve is excluded. The different regions, A, B, C, D, and E mark different kinematic 
regions with different top squark decay products~\protect\cite{CMS_RPV_stop}.}
\label{fig:RPV_stop}
\end{figure}

\paragraph{}
The top squark search~\cite{CMS_RPV_stop}, discussed in Sect.~\ref{sec:RPV-leptonic}, 
can also be exploited for semileptonic RPV interactions ($\lambda_{233}'$), 
via $\nino_1^{0*} \rightarrow l_i + \quark_j + \quark_k$ or  
$\nino_1^{0*} \rightarrow \nu_i + \quark_j + \quark_k$, where the indices $i,j,k$ refer to the ones in Eq.~(\ref{eq:RPV}).
Limits are set for $\lambda_{233}'$, where different kinematic regions 
lead to different allowed decays, ranging from two-body decays (for $m_{\topq} + m_{\ninoone} < m_{\sTop_1}$) to four-body decays
(e.g. for $m_{\ninoone} > m_{\sTop_1} > 2 m_{\topq}$). The resulting excluded region is shown in Fig.~\ref{fig:RPV_stop}.

\paragraph{}
In summary, the signatures of semi-leptonic RPV interactions are similar those expected in leptoquark decays. 
The LHC experiments can complement these searches with
analyses including the third-generation, where the most stringent limits up to now could be achieved.

\subsubsection{Search for quark RPV interactions ($\lambda_{ijk}''  \neq 0 $)}
\label{sec:RPVmultjet}

The quark RPV interactions described by $\lambda_{ijk}''$ can be well tested with multijet resonance searchess~\cite{Multijets1,Multijets2}. 
We discuss here an analysis searching for three-jet resonances~\cite{CMS_RPV_6jet}, which tests two different RPV
Yukawa couplings. While one search is inclusive, testing $\lambda_{122}''$, the other one 
requires at least one jet of each resonance decay to be \botq-tagged and is sensitive to $\lambda_{113}''$ and $\lambda_{223}''$.
The jet-ensemble technique~\cite{JetEnsemble1,JetEnsemble2} is used to combine the six highest-\pt jets into all 
possible unique triplets.

\begin{figure}[htbp]
\begin{center}
\includegraphics[width=\linewidth]{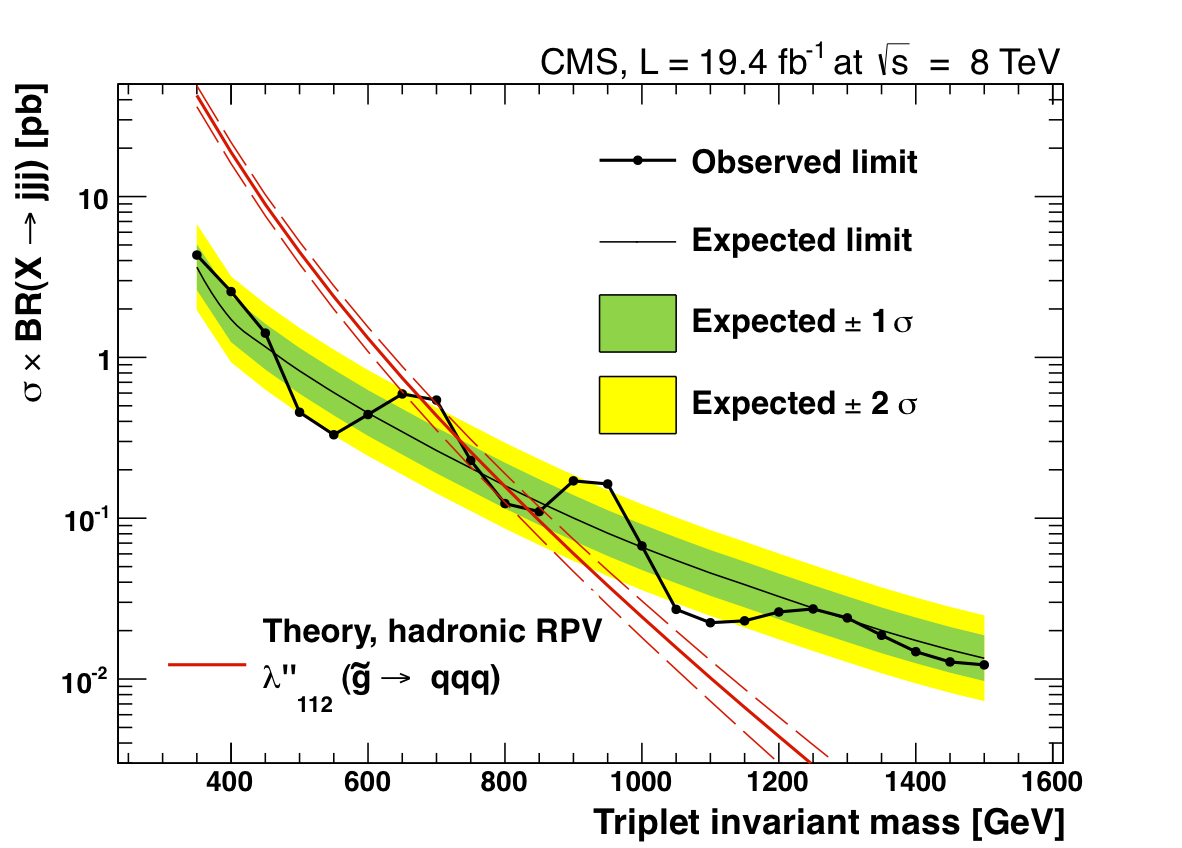}
\end{center}
\caption{Exclusion limits at 95\% CL on the production cross section of pair-produced gluinos for the inclusive RPV multijet 
search~\protect\cite{CMS_RPV_6jet}.}
\label{fig:RPV-multijets1}
\end{figure}

Limits are set on the gluino pair-production cross section times the branching fraction
as function of the gluino mass as shown in Fig.~\ref{fig:RPV-multijets1}. Gluinos with masses below 650\GeV decaying to light-flavor 
jets can be excluded. Decays including heavy-flavor jets can be excluded for even larger gluino masses, between 
200 and 835\GeV.

Two other approaches exist. The first one counts the number of six-jets events above a given \pt threshold to search for high mass gluinos and the second 
one takes profit of the large boost of the low-mass gluinos~\cite{ATLAS_RPV_6jet_7TeV}. With 7\TeV data, gluino masses up to 666\GeV and 255\GeV 
can be excluded respectively. The limit of the first approach is expected to reach 1\TeV with 8\TeV data.

\paragraph{}
Another scenario is given by a gluino decay to two quarks and one \ninoone, which then further decays through a $\lambda_{ijk}''$ interaction
to three quarks, leading to final states with ten jets when gluino pair production is assumed. A search for such scenarios is currently
performed, but it has not yet been published.

\begin{figure}[htbp]
\begin{center}
\includegraphics[width=\linewidth]{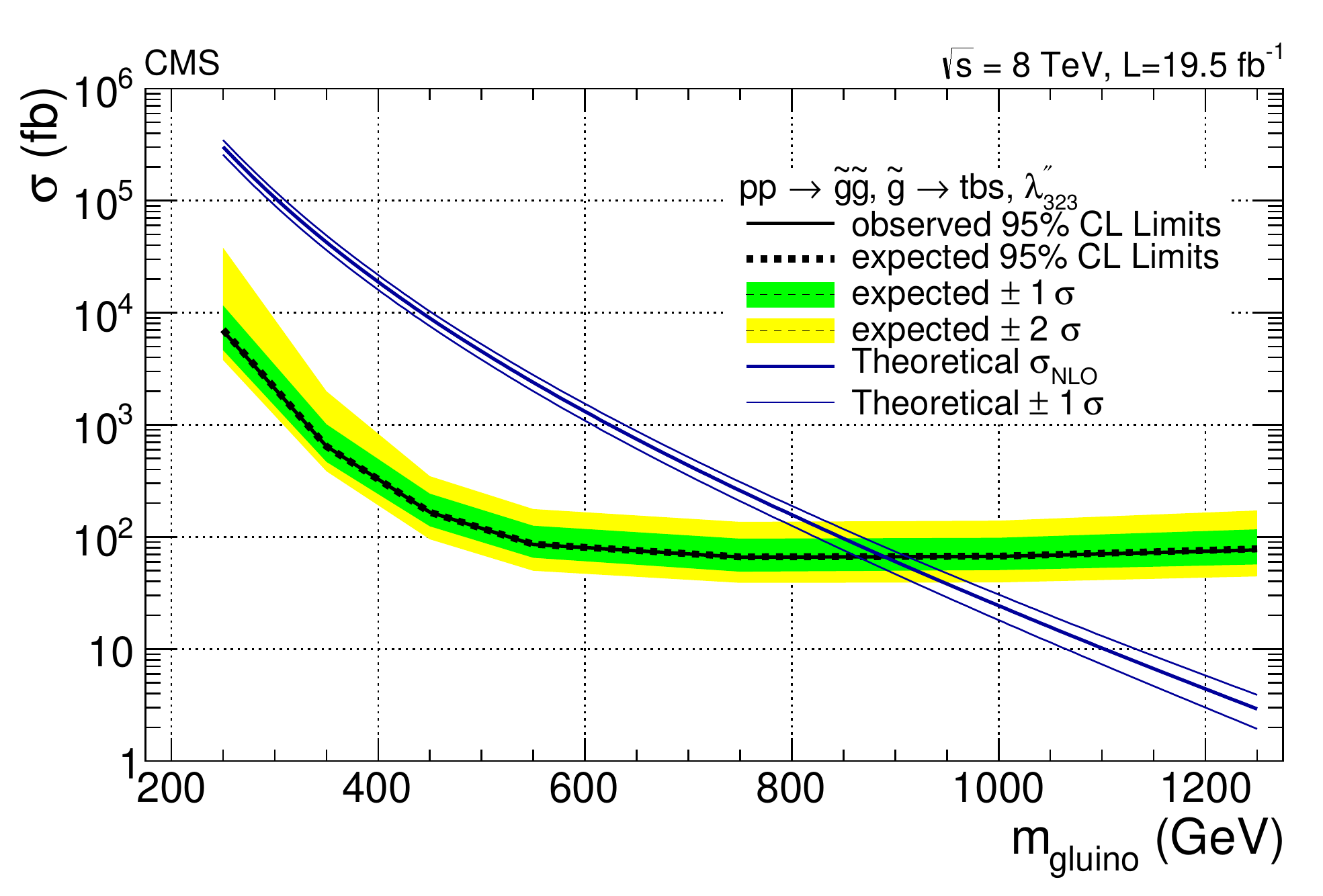}
\end{center}
\caption{Exclusion limits at 95\% CL on the production cross section of pair-produced gluinos for the same-sign dilepton analysis 
including heavy-flavor jets~\cite{CMS_Strong_2lSS}.}
\label{fig:RPV-multijets2}
\end{figure}

\paragraph{}
The results of the same-sign dilepton search~\cite{CMS_Strong_2lSS},
discussed in Sect.~\ref{sec:LepTonic}, 
can also be interpreted in a RPV model, where gluinos are
pair-produced and decay to three quarks via $\gluino \rightarrow \topq \botq \strange (\aTop \aBot \aStrange)$, 
testing the $\lambda_{323}''$ coupling. In this decay 50\% of the \W bosons are expected to be same-sign when both \W bosons decay leptonically. 
As shown in Fig.~\ref{fig:RPV-multijets2}, gluino masses up to 860\GeV can be excluded. 

\paragraph{}
In summary, a plethora of models with different RPV interactions exists, and those with the most striking
signatures have been tested up to now, excluding a large phase space, but leaving holes for more complicated signatures still
to be found with more data. This still leaves a large parameter space for not-yet-detectable RPV SUSY. 

\subsection{Long-lived particle searches}
\label{sec:LLP}

As no metastable particles are present in the Standard Model, long-lived particle searches are generally free of SM background. In turn, they require a deep 
understanding of the detector performance, which represents the only background, as discussed in Sect.~\ref{sec:Det}.
Metastable particles appear generally in the SUSY GUT theory framework~\cite{SUSYGUT1,SUSYGUT2,SUSYGUT3,SUSYGUT4,SUSYGUT5}. They arise in three main situations:
very low mass difference ($\leq \mathrm{O}(1\GeV)$) between a SUSY particle and the LSP in RPC models, very weak $R$-parity 
violation, i.e. $\lambda, \lambda'$ or 
$\lambda'' \leq \mathrm{O}(10^{-5})$, or very weak coupling to the gravitino in GMSB models. Depending on the SUSY mass spectrum, the metastable 
particle can be colored (squarks and gluinos) or not (sleptons, lightest chargino or neutralino). The experimental signatures probed at LHC are 
now reviewed by going from the left to the right of Fig.~\ref{fig:SUSY_LLP}. 

\begin{figure}[htbp]
\begin{center}
\includegraphics[width=\linewidth]{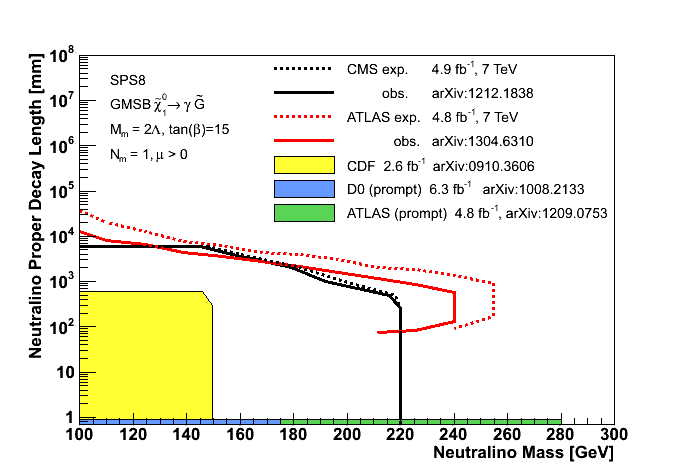}
\end{center}
\caption{Exclusion limits for non-pointing photon searches at 95\% CL in mass--lifetime plane of the \ninoone, assumed to be bino-like NLSP.}
\label{fig:NonPoint}
\end{figure}

Non-pointing photons arise in GMSB models where the NLSP is the lightest neutralino with bino-like flavor, i.e. scenario (a) of Fig~\ref{fig:SUSY_EWKinos}. If the coupling strength 
with the gravitino is weak, the \ninoone lifetime is in the range O(0.1-100\ns) accessible by the experiments, provided the \ninoone mass is 
close to the EW scale. In the EM calorimeter, non-pointing photons exhibit a singular geometric shape for the energy deposit and a late arrival. 
Results of the search are shown in Fig.~\ref{fig:NonPoint} 
in the \ninoone lifetime-mass plane~\footnote{Weaker limits from converted photons reconstructed by the tracker as a pair of electron-positron 
are not shown~\cite{CMS_NonPoint1}.}. The stronger limits obtained by ATLAS~\cite{ATLAS_NonPoint} for long lifetimes are explained by the stand-alone 
pointing capability of its calorimeter, whereas at short lifetimes CMS exploits better the correlation
between \MET and photon energy~\cite{CMS_NonPoint2}. Non-pointing photons are not the only possibility in GMSB models. NLSP \ninoone will give 
non-pointing Z or Higgs. Slepton or squark/gluino NLSPs will give non-pointing leptons or jets. However, in all of these cases, no public 
results exist yet.

\begin{figure}[htbp]
\begin{center}
\includegraphics[width=\linewidth]{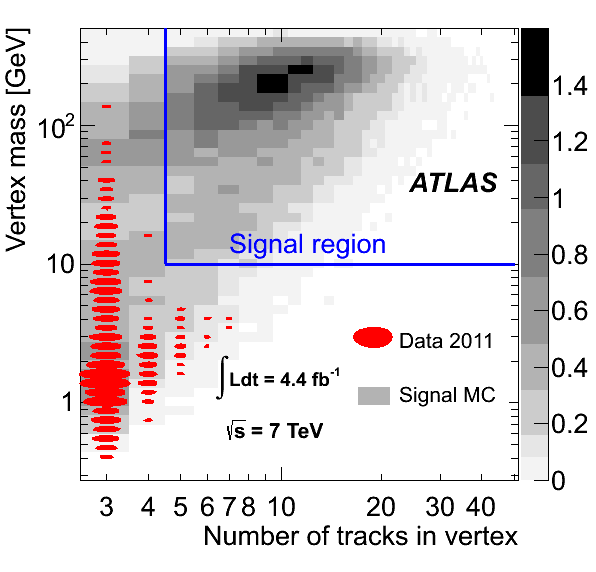}
\end{center}
\caption{Mass-Track multiplicity plane of the displaced vertex selected by the ATLAS search. One muon is also asked to point to the displaced vertex. 
The SUSY model considered has 700\GeV squarks, 500\GeV \ninoone and $\lambda'_{2ij}=0.3\times 10^{-5}$~\cite{ATLAS_DisplVert}.}
\label{fig:DisplVert}
\end{figure}

Reconstructing a displaced vertex with high mass and many tracks is undoubtedly a striking sign of new physics, as shown in Fig.~\ref{fig:DisplVert}. 
SUSY models with \ninoone LSP and very small RPV couplings can provide a plethora of possibilities: with leptons (electrons, muons or taus) attached to the displaced 
vertex or without leptons. While the latter case requires tracker and calorimeter information (cluster size, track multiplicity pointing to the cluster, ...), the former
relies on a dedicated tracking algorithm for non-standard displaced-vertex reconstruction. To date the only publicly available search interpreted in SUSY models 
is the one from ATLAS~\cite{ATLAS_DisplVert}. 
SUSY models with the simplified decay chain $\squark/\gluino \rightarrow \quark/\quark\aQuark + \ninoone \rightarrow \quark/\quark\aQuark + \mu \quark_{i} \quark_{j}$ 
are excluded for \ninoone lifetimes
below 1\m and squark (gluino) masses below 0.7 (0.9)\TeV. Other searches looking for long-lived neutral bosons decaying in two leptons~\cite{CMS_DisplVert} could be 
useful to reject some SUSY models -- though it is not yet done. In this case the displaced vertex is required to be at a distance of more than 5 standard deviations 
from the primary vertex in the transverse plane. Similar study is ongoing with two jets.

\begin{figure}[htbp]
\begin{center}
\includegraphics[width=\linewidth]{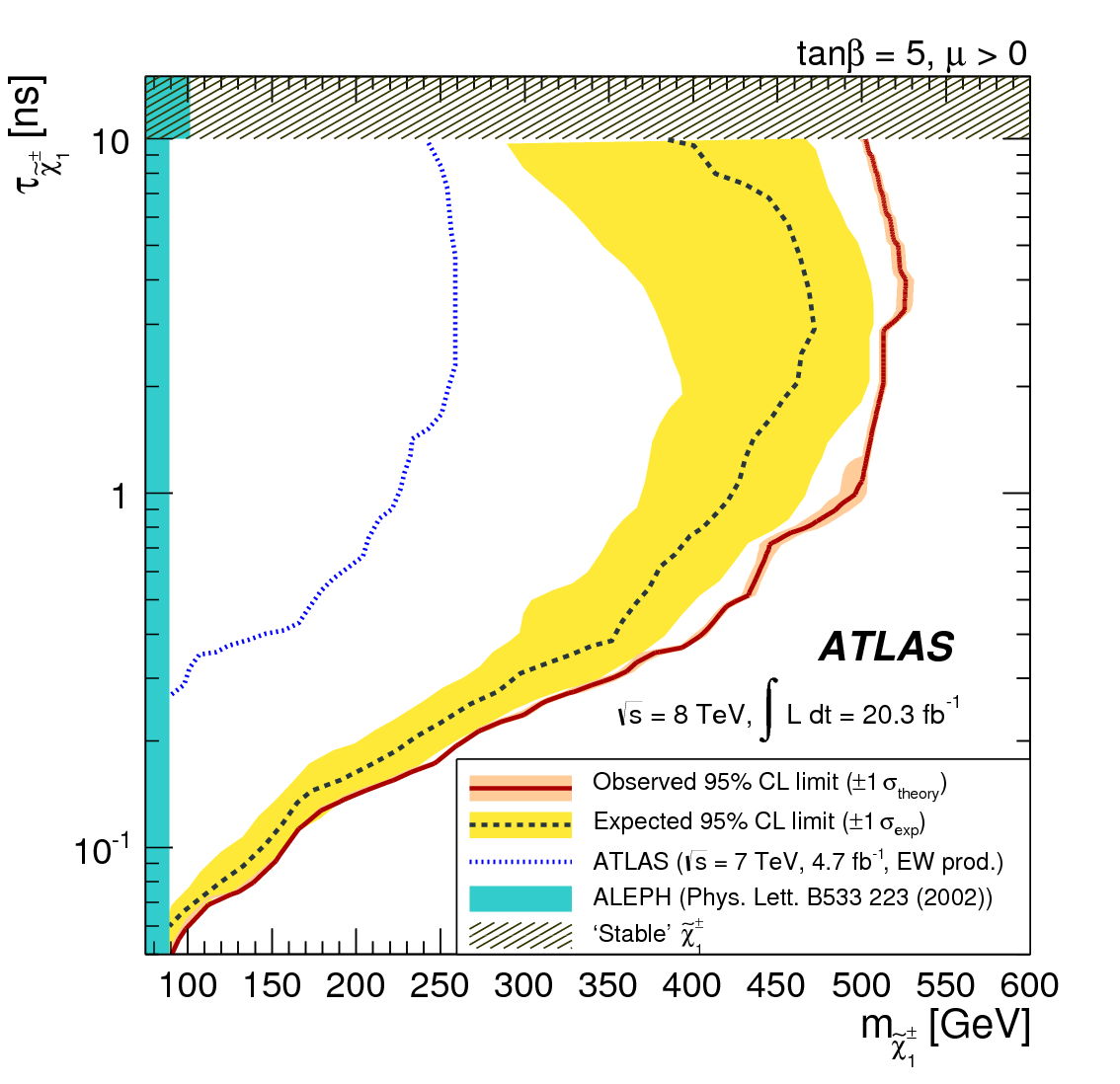}
\end{center}
\caption{Exclusion limit at 95\% CL for disappearing track searches in mass-lifetime plane of the \chipmone~\protect\cite{ATLAS_SUSY_DC1}.}
\label{fig:AMSB}
\end{figure}

AMSB~\cite{AMSB1,AMSB2} provides a well motivated case for metastable charginos since \chipmone and \ninoone are almost degenerate 
and $m_{\chipmone}-m_{\ninoone} \geq 140$\MeV. The situation is similar to scenario (b) of Fig.~\ref{fig:SUSY_EWKinos}. The chargino therefore decays 
after O(10\cm) to undetectable particles, a soft pion and the LSP. This will cause the chargino 
track to `disappear'. When produced directly (\chipone\chimone, \chipmone\ninoone) with an additional jet from initial 
state radiation to trigger the event, one (or two) tracks may have no/few associated hits in the outer region of the tracking system. The continuous TRT 
tracking of the outer part of the ATLAS inner detector gives sensitivity to this signature. With the additional requirement of a high-energetic isolated track, 
regions beyond the LEP limits can be excluded in the lifetime--mass plane of the chargino, as shown in Fig.~\ref{fig:AMSB}~\cite{ATLAS_SUSY_DC1}. Although originally 
motivated by AMSB, this result is largely model independent and fits also predictions in many ``unnatural" SUSY models~\cite{Unnatural1,Unnatural2,Unnatural3,Unnatural4}, 
where the chargino and the LSP are the only accessible sparticles at LHC.

If the gluino and the LSP are almost mass degenerate, the gluino lifetime could be long enough for it to hadronize in $R$-hadrons ($\gluino \qqbar$, $\gluino \quark\quark\quark'$) 
or $R$-gluino balls (\gluino \gluon)~\footnote{Similar reasoning also applies to squarks.}. These composite particles are detector-stable, highly ionizing, slowly moving 
(i.e. non-relativistic) and could change sign when they interact with the detector material. The signature can thus be a detector-stable charged particle, but also 
a charged particle turning neutral, or even a charged particle turning neutral and turning back charged. To fully explore all possibilities one needs to combine 
all possible detector measurements: $\beta\gamma$~\footnote{The variable $\beta$ is the particle velocity and $\gamma$ is the Lorentz boost.} from the pixel detector by 
inverting the Bethe-Bloch 
function, and $\beta$ from the calorimeter and muon spectrometer by measuring the arrival time in these devices. Together with the measure of the particle 
momentum $p$ in the tracker, or in the muon spectrometer, the composite particle mass $m=p/(\beta \gamma)$ can be inferred. 

Events are selected by dedicated slow muon or \MET triggers -- the latter is justified by the modest calorimeter energy depositions of the $R$-hadrons combined with  
high-energetic jets from initial state radiation. The background is evaluated by building templates for $p$, $\beta$ and $\beta\gamma$ in signal depopulated regions (like in 
Fig.~\ref{fig:DisplVert}, signal regions are generally background free). Non-colored particles can also be detector-stable and behave like heavy muons. Therefore, the
analysis techniques are similar to the ones used for colored particles. However, in that case, the best performing signal regions are the ones 
requiring two detector-stable slepton candidates. 

\begin{figure}[htbp]
\begin{center}
\includegraphics[width=\linewidth]{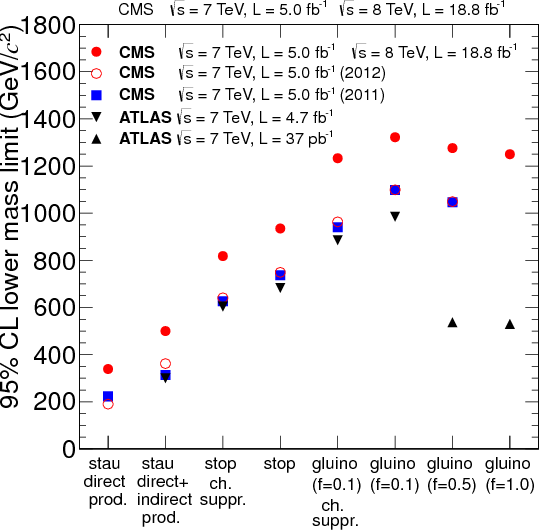}
\end{center}
\caption{Lower mass limit at 95\% CL on different scenarios for long-lived particles~\protect\cite{CMS_LLP}.}
\label{fig:HSCP}
\end{figure}

The current mass limits from detector-stable particles are presented in Fig.~\ref{fig:HSCP}~\cite{CMS_LLP} -- note that the
ATLAS results are still those from the 7\TeV run~\cite{ATLAS_LLP}. Because no Standard Model backgrounds exist, the limits obtained are generally higher than in the 
prompt-decay case. This is especially true for top and bottom squarks and staus -- where no limits on the direct production exist in the RPC prompt case. The gluino 
masses below 1.2\TeV are excluded independent of the hypotheses made for the interaction of the $R$-hadrons and $R$-gluino balls.

\begin{figure}[htbp]
\begin{center}
\includegraphics[width=\linewidth]{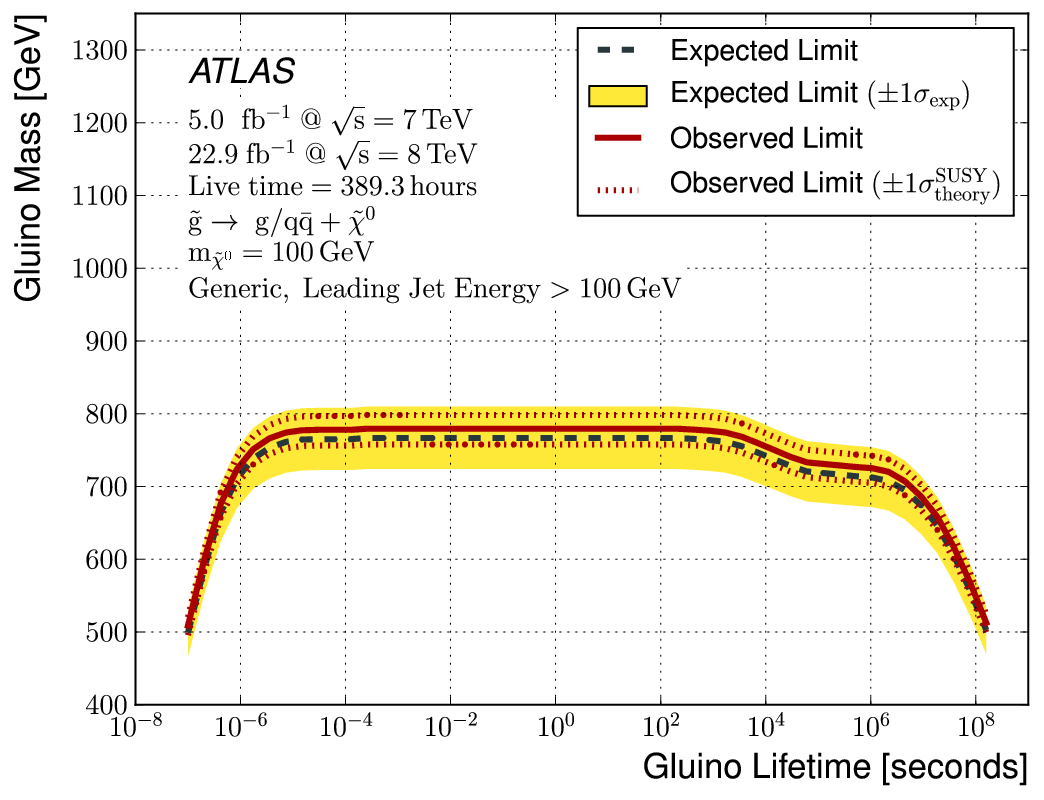}
\end{center}
\caption{Exclusion limit at 95\% CL for the stopped gluino searches in mass-lifetime plane of the gluino~\protect\cite{ATLAS_StoppedGluinos}.}
\label{fig:StoppedGluinos}
\end{figure}

A fraction of these slow-moving particles may come to rest within the detector volume and only decay later as 
$\gluino \rightarrow \qqbar \ninoone, g \ninoone$. A particular case is given when this happens in the calorimeter. The signature consists of a high energetic 
jet(s) in absence of collisions. The main background is then caused by calorimeter noise bursts, cosmic rays with high energy deposit or beam halo -- the leading background. 
Gluinos below 850\GeV are excluded for a gluino lifetime between 10\mus and 15~minutes, see Fig.~\ref{fig:StoppedGluinos}~\cite{ATLAS_StoppedGluinos}. This signature is 
generally present in unnatural SUSY models, where the gluino and the LSP are the only accessible sparticles at the LHC.

\subsection{Beyond MSSM}
\label{sec:Oth}

\begin{figure}[htbp]
\begin{center}
\includegraphics[width=\linewidth]{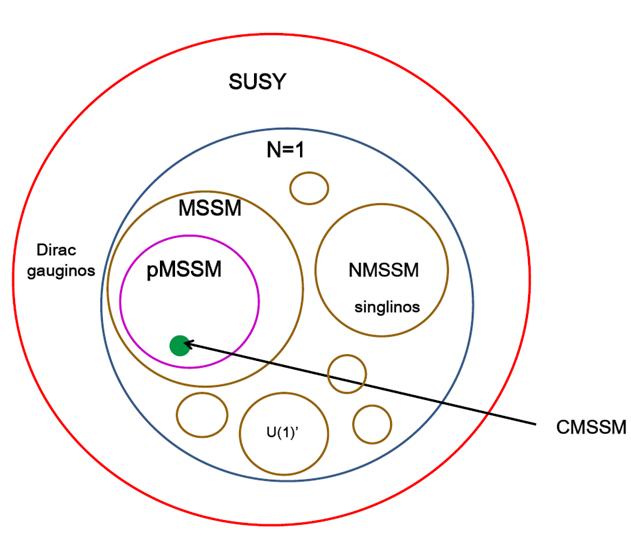}
\end{center}
\caption{SUSY Phase space and associated theories~\protect\cite{SUSY_PhaseSpace}.}
\label{fig:SUSY_PhaseSpace}
\end{figure}

The MSSM is firmly established since 30 years and serves as a basis for most of the SUSY searches at LHC. However, many possible extensions 
exist as shown in Fig.~\ref{fig:SUSY_PhaseSpace}. With mild departure from MSSM parsimony, they could explain the current absence of SUSY 
signals at LHC. In addition, they generally predict new signatures that could be searched for at LHC. We briefly review here the status of these beyond MSSM searches. 

The first category of models adds a gauge-singlet superfield to the MSSM. More specifically in the NMSSM~\cite{NMSSM}, two additional Higgs bosons and one 
neutralino (singlino, $\widetilde{S}$) 
are added to the MSSM. The extra Higgs are searched for as a peak around 10\GeV in the invariant di-muon mass~\cite{CMS_HiggsNMSSM}. The naturalness constraints 
from the 126\GeV Higgs mass are relaxed and singlino-like LSPs with very small couplings are possible -- changing the phenomenology of the SUSY EW sector. Because of 
this addition, these models predict final states with multi-leptons and \MET. Even if no dedicated searches exist yet at LHC, reinterpretations of present EWKino searches, 
presented in Sect.~\ref{sec:SUSY_EWK}, have already started to constrain NMSSM models~\cite{NMSSM_LHC}. Apart from NMSSM, another popular model is stealth SUSY~\cite{Stealth} where 
the invisible singlet and singlino are mass degenerate 
and light, which reduces drastically the amount of \MET -- all SUSY cascade decays end like $\widetilde{\text {S}}\rightarrow \text{S} (\rightarrow \text{jj}) \gravitino$ 
with a poorly boosted \gravitino. Experimental signatures comprise low \MET, displaced vertex and high-multiplicity final states including photons and a high number of \botq-jets. 
CMS already places a limit on stealth scenarios where S decays via a photon~\cite{CMS_Stealth}. Limits of 1.5\TeV on squarks initiating cascade decays are obtained. More refined 
searches are currently going on.

\begin{figure}[htbp]
\begin{center}
\includegraphics[width=\linewidth]{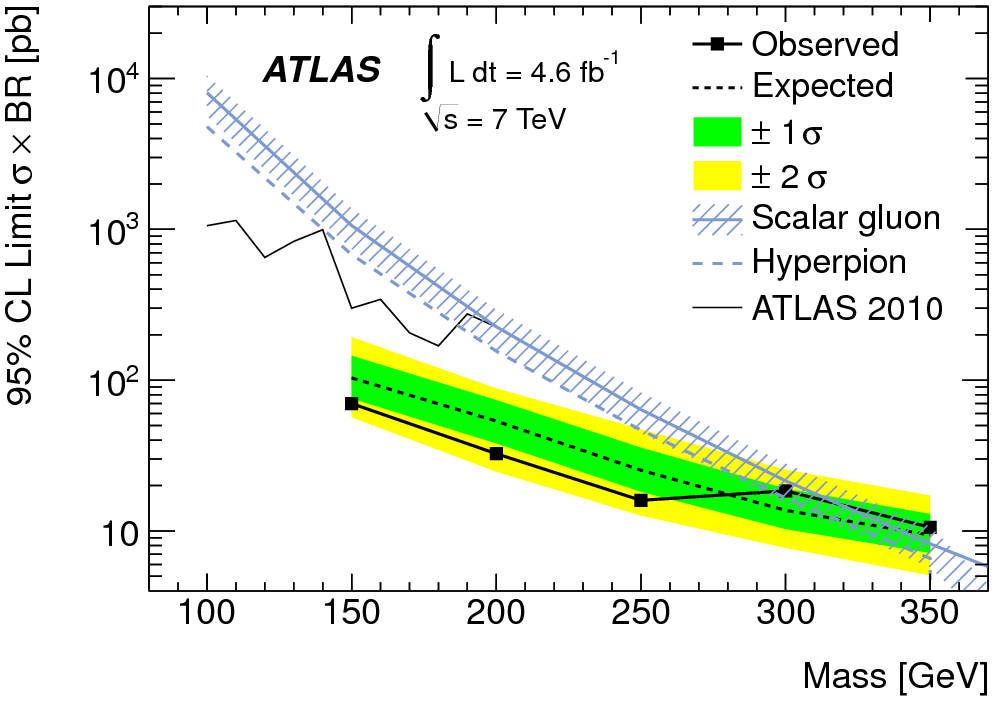}
\end{center}
\caption{Exclusion limits at 95\% CL on the sgluon pair production cross section as a function of the sgluon mass~\protect\cite{ATLAS_ScalarGluons}.}
\label{fig:ScalarGluon}
\end{figure}

The second category of models postulates the gluino to be of Dirac type instead of Majorana as in the MSSM~\footnote{It can be also the case for other gauginos but 
only the gluino is considered since LHC limits are generally quite strong on the gluino mass.}. This happens in theories which extend 
the $R$-parity concept to a continuous symmetry (MRSSM~\cite{MRSSM}), hybrid N=1/N=2 model~\cite{Hybrid} and Supersoft 
SUSY (SSSM~\cite{SSSM}). For the two 
first models, a new particle (the sgluon) completes the MSSM multiplet composed of gluons and gluinos. In all models, the 
constraint on the gluino mass in the natural spectrum is relaxed since the radiative corrections are truncated. Because of that, gluino-gluino cross sections 
are expected to be lower than in the MSSM, weakening the current constraint. The sgluon provides also new signatures to search for. Above a mass of 350\GeV, sgluons 
dominantly decay in two tops. They can for example be searched for by requiring two same-sign leptons (no public 
results are presently available). At lower mass, sgluons decay in two gluons, giving a pair of two-jet resonances with equal mass. A unique pairing of 
the four highest energetic jets is achieved for each event by minimizing the pairwise separation. A peak in the dijet mass distribution is then searched 
for, while the shape and the normalization of the multi-jet background are estimated by a data-driven method. Sgluons are excluded for masses below 300\GeV; see 
Fig.~\ref{fig:ScalarGluon}~\cite{ATLAS_ScalarGluons}. It is interesting to note that this signature, unique in BSM models at LHC, is limited at 
low mass by the multijet energy trigger threshold and therefore strong limits are already obtained with the first run at 7\TeV in 2010.

\section{SUSY Prospects at LHC beyond the first run}
\label{sec:Prospects}

The LHC program is approved until 2022. In this program, the center-of-mass energy will be 13\TeV at the restart and it
should reach gradually 14\TeV. Expected luminosities are $\sim$100\fbinv at the end of 2017 and $\sim$300\fbinv in 2022. A possible extension, called High-Luminosity-LHC (HL-LHC), 
is planned from 2024 to 2030-2035 and could deliver ultimately $\sim$3000\fbinv. The average number of pile-up events per bunch crossing is expected to rise up 
to 140 in 2030. Expected discovery sensitivities and exclusion limits have recently been extracted by ATLAS and CMS for benchmark processes of 
plain vanilla MSSM, i.e. $R$-parity conservation and \ninoone LSP~\cite{Snowmass2013,ECFA2013}.

With the increase in beam energy in 2015, the strong SUSY cross sections are greatly enhanced, opening a new phase space to explore with 
already a low integrated luminosity recorded (1-10\fbinv). A few examples are given here for sparticles at the energy frontier. The cross section for 
gluino-gluino production (with squarks decoupled) is enhanced by more than a factor of 20 for gluino masses of around 1.3\TeV. The cross section for the 
pair production of squarks of the first/second generation as well as for third-generation 
squarks, rises by about a factor of ten for squark masses of 800\GeV. 
With 300\fbinv, mass degenerate squarks (of the first and second generation) and gluinos of up to 2.7\TeV could be discovered
with $5\sigma$ significance. For higher gluino (squarks) masses, e.g.~3.5\TeV, squark (gluino) masses could be discovered up to 2.5 (2)\TeV. 
With the HL-LHC, a gain of about 
300\GeV in the mass range is expected. Finally the top squark could be discovered up to masses of about 800 (1000)\GeV with 300 (3000)\fbinv, 
assuming a 100\% branching ratio to a top quark and a \ninoone.

Searches for electroweak particles like neutralinos and charginos benefit more from the large expected integrated luminosity than for the increase in $\sqrt{s}$ due to 
their low cross sections. By the end of the HL-LHC, 
neutralinos (\ninotwo) and charginos (\chipmone) decaying to Z\ninoone and W\ninoone respectively, as in scenario (a) of Fig.~\ref{fig:SUSY_EWKinos}, can be 
discovered up to a mass of 700~GeV for \ninoone masses of up to 200\GeV with a 5$\sigma$ significance. In case of no signal, the exclusion
limits are about 200 to 300\GeV higher. Some sensitivity is also expected to the compressed and more natural scenario (c) of Fig.~\ref{fig:SUSY_EWKinos} where 
the \ninoone is higgsino-like.

In summary, the full harvest of the LHC, including the HL-LHC, could explore the largest part of the most 
interesting weak-scale SUSY phase space, which will remain a hot topic to be tested for at least the next two decades.

\section{Conclusions}
\label{sec:Conclu}

ATLAS and CMS, the two general-purpose LHC experiments, have developed a coherent and ambitious program to search for new 
particles at the energy frontier, O(0.1-1\TeV) with 25\fbinv of proton-proton collision data with center-of-mass energies of 7 and 8\TeV. 
These efforts were successful: a Higgs boson of 126\GeV mass was discovered after two years of running, closing the list of 
Standard Model particles to be found. 

This discovery fits with expectations from the minimal realisation of N=1 SUSY, a.k.a weak-scale SUSY or MSSM. However, this model also predicts new particles at the energy frontier that could solve the gauge hierarchy problem of the 
Standard Model, i.e. the quadratic divergence of the Higgs mass at higher energy. Assuming that $R$-parity is conserved, the plain vanilla 
SUSY solution predicts a `natural' or low fine-tuning mass spectrum composed of gluino masses around 1\TeV, top squark and left-handed 
bottom squark masses around 500\GeV  and chargino/neutralino masses below 500\GeV. All other SUSY particles could be of much higher mass. In this framework, the 
two favored LSP candidates could be the lightest neutralino or the gravitino. The ATLAS and CMS experiments have probed, by direct searches, the uncharted 
heart of the MSSM spectrum, attracting high attention from the community. To date, these dedicated searches, mainly based on the presence of multijets 
and \MET (but not only), have not revealed any sign of new physics. 

Not all results from Run~1 are currently available but the following general conclusions could be drawn: the gluino mass, governed by 
only one SUSY parameter, $M_3$, could be excluded below 1\TeV irrespective of the SUSY mass spectrum in between the gluino and the LSP, and the nature of the LSP. 
This conclusion applies well for open SUSY spectra, i.e. mass difference between gluino and the LSP above O(500\GeV). But the excluded gluino mass region should be lowered to 600\GeV when 
considering more and more compressed spectra, because the jet and \MET softening decreases the acceptance -- the presence of 
isolated lepton(s) can 
partially correct for that. 

Constraints on squarks of first and second generation are generally softer and less general, since mass degeneracy 
between families is often assumed. A strong focus was put on the third squark generation (top and bottom squark) because of their particular 
position in the natural spectrum. When top and bottom squarks are directly produced, final states are generally less complex than for the gluino and squarks of first 
and second generation and composed of multiple \botq-jets and lepton(s). Dedicated searches dramatically shrink the allowed window, 
but they are (presently) unsuccessful. As an illustration, holes in the top squark searches are presently located near the top mass funnel, $m_{\sTop}=m_{\topq}+m_{\ninoone}$, 
for $m_{\ninoone}>$100\GeV, at low mass difference between top squark and \ninoone or in very intricate top squark decay chains. 

The weak SUSY sector (charginos, neutralinos and sleptons) is also probed extensively at LHC. Because of lower cross section, the 
Standard Model background is only reducible considering multi-lepton final states. These leptons are provided by the leptonic decay of the 
\W, \Z, Higgs and/or sleptons. Compared with the other sectors no strong general conclusions are drawn (yet) because of the high number of 
possible final states and the complexity of the sector governed by around 10 parameters. Nonetheless, constraints are generally always 
going beyond the LEP ones, and biting the natural spectrum in many cases.

Therefore it is fair to say that, even if not all 8\TeV results are currently available, plain vanilla MSSM is under high pressure. More 
definitive conclusions will come when all these results will be interpreted with a full scan of the main 19-20 MSSM parameters (some assumptions are made on the 
other 105-19 parameters), which will happen in 2014. Meanwhile, more focus has been put on searches for long-lived particles, $R$-parity violating scenarios 
and new theoretical ideas (Stealth SUSY, 
Dirac gauginos, extra gauge-singlet superfield) that provide many striking signatures, generally background free and relying more on detector performance. 
Such scenarios are in most cases compatible with the absence of experimental evidence for plain vanilla MSSM. A huge number of possibilities exists and the 
most important ones have been covered (or are worked on), with presently no sign of SUSY. It is interesting 
to note that the gluino mass is constrained to be above 1\TeV in the models that have been considered.

All these direct searches for new particles have been complemented by indirect searches that we briefly mention for completeness. At LHC, the main improvement 
comes from the measurement of the B$_{\text s}^0 \rightarrow \mu^+ \mu^-$ branching ratio, where the leading SUSY contributions involve SUSY Higgses (A$^0$ and H$^0$) in 
penguin diagrams. However, a good agreement with the Standard Model was found~\cite{LHCb_Bsmumu,CMS_Bsmumu}, and no irreducible limit exists on SUSY models.
Other indirect evidence for new physics could be found when searching for the flavor-changing decay of the top quark like $\topq \rightarrow \charm \Higgs$, 
which is very strongly suppressed in the Standard Model, BR=O(10$^{-15}$). Contributions from SUSY Higgses in virtual loops of the decay 
amplitude can enhance the cross section significantly, by factors up to nine orders of magnitude. Nevertheless, up to now, no evidence for 
this decay has been found when considering $\Higgs\rightarrow WW,ZZ,\tau\tau$ leptonic decay~\cite{CMS_3l_GMSB} and $\Higgs\rightarrow \gamma \gamma$ decay~\cite{ATLAS_tcHgamgam}.

A new phase of exploration will start in 2015 with the restart of the LHC at higher energy. But the present situation after the first run could well fit an 
Ernest Rutherford quote `An alleged scientific discovery has no merit unless it can be explained to a barmaid': we can now happily discuss with the barmaid 
of the SM bar, but we could not yet find the door of the SUSY bar!

\begin{acknowledgements}
We would like to thank our colleagues in CMS and ATLAS who helped us producing the combined plots.
\end{acknowledgements}

\bibliographystyle{spphys}       
\bibliography{SUSY_Exp2}   

\end{document}